\documentclass[apj,iop]{emulateapj}
\usepackage{lineno}

\usepackage{natbib}
\usepackage{longtable}
\usepackage{graphicx}
\usepackage[caption=false]{subfig}
\usepackage{float}
\usepackage{epstopdf}
\usepackage{amsmath}
\usepackage{appendix}
\usepackage{hyperref}
\usepackage{booktabs}
\usepackage{scrextend}
\usepackage{xcolor}
\usepackage{CJKutf8}

\newcommand{\HII}{H{\footnotesize II}}

\newcommand{\NII}{[N~{\footnotesize II}]}

\newcommand{\OIII}{[O~{\footnotesize III}]}

\newcommand{\ha}{H$\alpha$}
\newcommand{\hbeta}{H$\beta$}

\newcommand{\msun}{$M_\odot$}

\shorttitle{Chandra and HST Observations of Dwarf Galaxies}
\shortauthors{Latimer et al.}

\begin{document}

\title{A {\it Chandra} and {\it HST} View of {\it WISE}-Selected AGN Candidates in Dwarf Galaxies}

\author{Colin J. Latimer}
\affil{eXtreme Gravity Institute, Department of Physics, Montana State University, Bozeman, MT 59717, USA}
\email{colin.latimer@montana.edu}

\author{Amy E. Reines}
\affil{eXtreme Gravity Institute, Department of Physics, Montana State University, Bozeman, MT 59717, USA}

\author{Kevin N. Hainline}
\affil{Steward Observatory, University of Arizona, 933 North Cherry Avenue, Tucson, AZ 85721, USA}

\author{Jenny E. Greene}
\affil{Department Astrophysical Sciences, Princeton University, Princeton, NJ 08544, USA}

\author{Daniel Stern}
\affil{Jet Propulsion Laboratory, California Institute of Technology, 4800 Oak Grove Drive, Mail Stop 169-221, Pasadena, CA 91109, USA}

\begin{abstract}

Reliably identifying active galactic nuclei (AGNs) in dwarf galaxies is key to understanding black hole demographics at low masses and constraining models for black hole seed formation. Here we present \textit{Chandra X-ray Observatory} observations of eleven dwarf galaxies that were chosen as AGN candidates using \textit{Wide-field Infrared Survey Explorer} (\textit{WISE}) mid-infrared (mid-IR) color-color selection. \textit{Hubble Space Telescope} images are also presented for ten of the galaxies.
Based on Sloan Digital Sky Survey spectroscopy, six galaxies in our sample have optical evidence for hosting AGNs and five are classified as star-forming.  
We detect X-ray point sources with luminosities above that expected from X-ray binaries in the nuclei of five of the six galaxies with optical evidence of AGNs. However, the X-ray emission from these AGNs is generally much lower than expected based on AGN scaling relations with infrared and optical tracers.
We do not find compelling evidence for AGNs in the five optically-selected star-forming galaxies despite having red mid-IR colors. Only two are detected in X-rays and their properties are consistent with stellar-mass X-ray binaries. 
Based on this multiwavelength study, we conclude that two-color mid-IR AGN diagnostics at the resolution of {\it WISE} cannot be used to reliably select AGNs in optically-star-forming dwarf galaxies.  Future observations in the infrared with the {\it James Webb Space Telescope} offer a promising path forward.

\end{abstract}

\section{Introduction}\label{sec:intro}

The vast majority of massive galaxies play host to supermassive black holes (BHs) at their cores. However, we do not know how these monster BHs came to be. Ideally, we would directly observe galaxies in the early Universe and the ``seed" BHs they harbor, but these BHs currently remain out of observational reach.  They are simply too distant and faint to detect with existing facilities \citep[e.g.,][]{volonterireines2016,vitoetal2018}.
Instead, we turn to nearby dwarf galaxies. These galaxies are relatively low mass ($M_\star \lesssim 10^{9.5}$~\msun) and may host BHs of similar mass to those in early-universe galaxies {\citep[for reviews, see][]{reinescomastri16,mezcua17,greene20}}. We can use the occupation fraction of massive BHs in these proxy galaxies as well as scaling relations at low mass to help distinguish between different theories concerning the formation of seed BHs such as remnants from Pop III stars, stellar mergers in compact star clusters, or direct collapse of protogalactic gas {\citep{volonteri10,vanWassenhove10,greene12,ricarte18}.}

There is now growing evidence for the existence of BHs in dwarf galaxies, with detailed studies of single galaxies to large-scale surveys {\citep[e.g.][]{reines11,reines13,reines14, reines20,baldassare15,baldassare16,baldassare17,baldassare18,baldassare20,mezcua18,mezcua19,mezcua20, schramm13,moran14,lemons15,hainline16,pardo16,dickey19,nguyen19,latimer19,schutte19,birchall20,lupi20}}. Multiple studies have tried to make progress identifying massive BHs in the mid-infrared (mid-IR) using data from the \textit{Wide-field Infrared Survey Explorer} (\textit{WISE}) \citep{wright10} by extrapolating active galactic nuclei (AGN) diagnostics \citep[e.g.][]{jarrett11,stern12} to low-mass galaxies \citep{satyapal14,sartori15,marleau17}. 
These studies find that the IR-selected AGN fraction seems to increase at low galaxy masses, a puzzling conclusion that contradicts studies at other wavelengths.
Motivated by these results, as well as earlier work (e.g., with {\it WISE}) that revealed extreme star-forming dwarf galaxies could produce very red mid-IR colors 
\citep{hirashita04,reines08,griffith11,izotov11,izotov14,remyruyer15,oconnor16},
\citet{hainline16} revisited mid-IR AGN selection techniques applied to dwarf galaxies.

\begin{deluxetable*}{ccccccccrcc}
\tabletypesize{\footnotesize}
\tablecaption{Sample of Dwarf Galaxies with Mid-IR Selected Candidate AGNs}
\tablewidth{0pt}
\tablehead{
\colhead{ID} & \colhead{BPT} & \colhead{NSAID} & \colhead {RGG ID} & \colhead{SDSS Name}  & \colhead{$N_{\rm H}$} & \colhead{$z$} & \colhead{$r_{50}$} &  \colhead{$M_g$} & \colhead{$g-r$} & \colhead{log $M_\star$/\msun} \\
\colhead{ }   & \colhead{Class.} & \colhead{ } & \colhead{ } & \colhead{ } & \colhead{($10^{20}$ cm$^{-2}$)} & \colhead{ } & \colhead{(kpc)}  & \colhead{(mag)} & \colhead{(mag)} & \colhead{ } \\
\colhead{(1)} & \colhead{(2)} & \colhead{(3)} & \colhead{(4)} & \colhead{(5)} & \colhead{(6)}  & \colhead{(7)} & \colhead{(8)} & \colhead{(9)} & \colhead{(10)} & \colhead{(11)} }
\startdata
1  	    	 & AGN & 68765  & 3       & J032224.64+401119.8   &    13.60  & 0.02679 & 1.0 & $-18.7$  & 0.467 & 9.4 \\
2\footnote{\label{broadline}Broad-line AGNs \citep{baldassare17}}           & AGN & 10779  & 9       & J090613.75+561015.5   &    2.83  & 0.04697 & 1.7 & $-19.0$  & 0.400 & 9.4 \\
3\footref{broadline}            & AGN & 125318 & 11      & J095418.15+471725.1   &    1.03   & 0.03284 & 2.0 & $-18.7$  & 0.438 & 9.1 \\
4   		 & AGN & 113566 & 18      & J114359.58+244251.7   &    2.09   & 0.04992 & 1.3 & $-18.7$  & 0.647 & 9.5 \\
5   		 & AGN & 104527 & 24      & J133245.62+263449.3   &    0.99   & 0.04693 & 1.2 & $-19.1$  & 0.272 & 9.4\\
6\footref{broadline}          & Comp. & 79874  & 119     & J152637.36+065941.6   &    3.47   & 0.03829 & 0.8 & $-18.7$  & 0.248 & 9.3 \\
7 	    	 & SF & 6205   & \nodata & J005904.10+010004.2   &    3.06   & 0.01743 & 0.9 & $-18.3$  & 0.211 & 8.7 \\
8            & SF & 98135  & \nodata & J154748.99+220303.2   &    4.64   & 0.03154 & 0.7 & $-18.1$  & $-0.459$ & 8.2 \\
9            & SF & 57649  & \nodata & J160135.95+311353.7   &    2.72   & 0.03085 & 1.3 & $-18.1$  & 0.109 & 8.6 \\
10            & SF & 4610 & \nodata & J173501.25+570308.8  &     3.44   & 0.04797 & 0.9 & $-20.5$  & $-1.050$ & 8.8 \\
11            & SF & 151888 & \nodata & J233244.60$-$005847.9 &    4.00  & 0.02437 & 1.3 & $-18.9$  & 0.344 & 9.3
\enddata
\tablecomments{Column 1: Identification number used in this paper.
Column 2: Classification of the galaxy as AGN, composite (Comp.) or star-forming (SF) from the \NII/\ha~BPT diagram (left panel in Figure \ref{fig:bpt})
Column 3: Identification number in the NSA.
Column 4: Identification number assigned by \cite{reines13}.
Column 5: SDSS name.
Column 6: Galactic neutral hydrogen column density \citep{dickey90}\footnote{Retrieved via \url{https://cxc.harvard.edu/toolkit/colden.jsp}}.  
Column 7: redshift, specifically the \texttt{zdist} parameter from the NSA.
Column 8: Petrosian 50\% light radius.  
Column 9: absolute $g$-band magnitude corrected for foreground Galactic extinction.
Column 10: $g-r$ color.  
Column 11: log galaxy stellar mass.
The values given in columns 7-11 are from the NSA and we assume $h=0.73$. }
\label{tab:sample}
\end{deluxetable*}

\cite{hainline16} find that dwarf galaxies with the reddest mid-IR colors have the youngest stellar populations and highest star formation rates, demonstrating that dwarf galaxies can heat dust in a way that can mimic luminous AGNs. In particular, they show that using a single
$W1-W2$ color cut to select candidate AGNs leads to severe contamination from dwarf starburst galaxies. 
While this mid-IR color selection often fails for the dwarf population, it appears to be quite robust for more luminous galaxies \citep{stern12}.
From \citet{hainline16}, it was not clear whether employing the two-color AGN selection of \citet{jarrett11} would suffer from contamination as well.
They therefore presented a sample of candidate AGNs in dwarf galaxies falling within the color-color AGN selection box of \cite{jarrett11} rather than employing a single color cut \citep{stern12}. 
Overall, their results are more in line with findings at optical wavelengths; that is, no enhanced active fraction at low galaxy masses is seen. A subsequent theoretical investigation by \cite{satyapal18} also showed that some extreme starbursts with high ionization parameters or gas densities could mimic AGNs in mid-IR color space.

In this paper, we observationally probe the efficacy of the mid-IR \textit{WISE} color-color selection box of \cite{jarrett11} in dwarf galaxies using high resolution X-ray and optical/near-IR {(NIR)} observations with {\it Chandra} and {\it HST}. Specifically, we seek to understand whether this AGN selection technique can effectively distinguish between BH activity and star formation when applied to low-mass galaxies.

\section{Sample Selection} \label{sec:samp}

The dwarf galaxies studied here come from the sample of \citet{hainline16} that have mid-IR colors suggestive of AGNs.  \citet{hainline16} start with a sample of ${\sim}18,000$ dwarf galaxies ($M_{*} \lesssim 3 \times 10^{9}$~\msun) in the  
NASA-Sloan Atlas (v0\_1\_2)\footnote{\url{http://nsatlas.org/}} with significant detections in the first three bands of the ALLWISE data release \citep[see Section 2 in][]{hainline16}. From this, they identify 41 galaxies with $W1-W2$ vs.\ $W2-W3$ colors that fall within the \cite{jarrett11} \textit{WISE} AGN selection box. Eleven of these galaxies have 
optical emission-line measurements (both redshifts and fluxes) and signal-to-noise ratios ${>}5$ in all four \textit{WISE} bands. 
We obtained new \textit{Chandra X-ray Observatory} and \textit{Hubble Space Telescope} (\textit{HST}) observations for seven of these dwarf galaxies (Proposal ID: 20700286; PI: Reines) and archival {\it Chandra} and {\it HST} observations for three additional galaxies (Proposal ID: 16700103; PI: Reines); the latter are broad-line AGNs/Composites from the \citet{reines13} sample and the {\it Chandra }X-ray $+$ {\it HST} {ultraviolet (UV; $\sim 2700$ \AA)} observations have previously been presented by \citet{baldassare17}. The final galaxy had an archival \textit{Chandra} observation, but not {\it HST} imaging.

\begin{figure}[h!]
\centering
\includegraphics[width=0.48\textwidth]{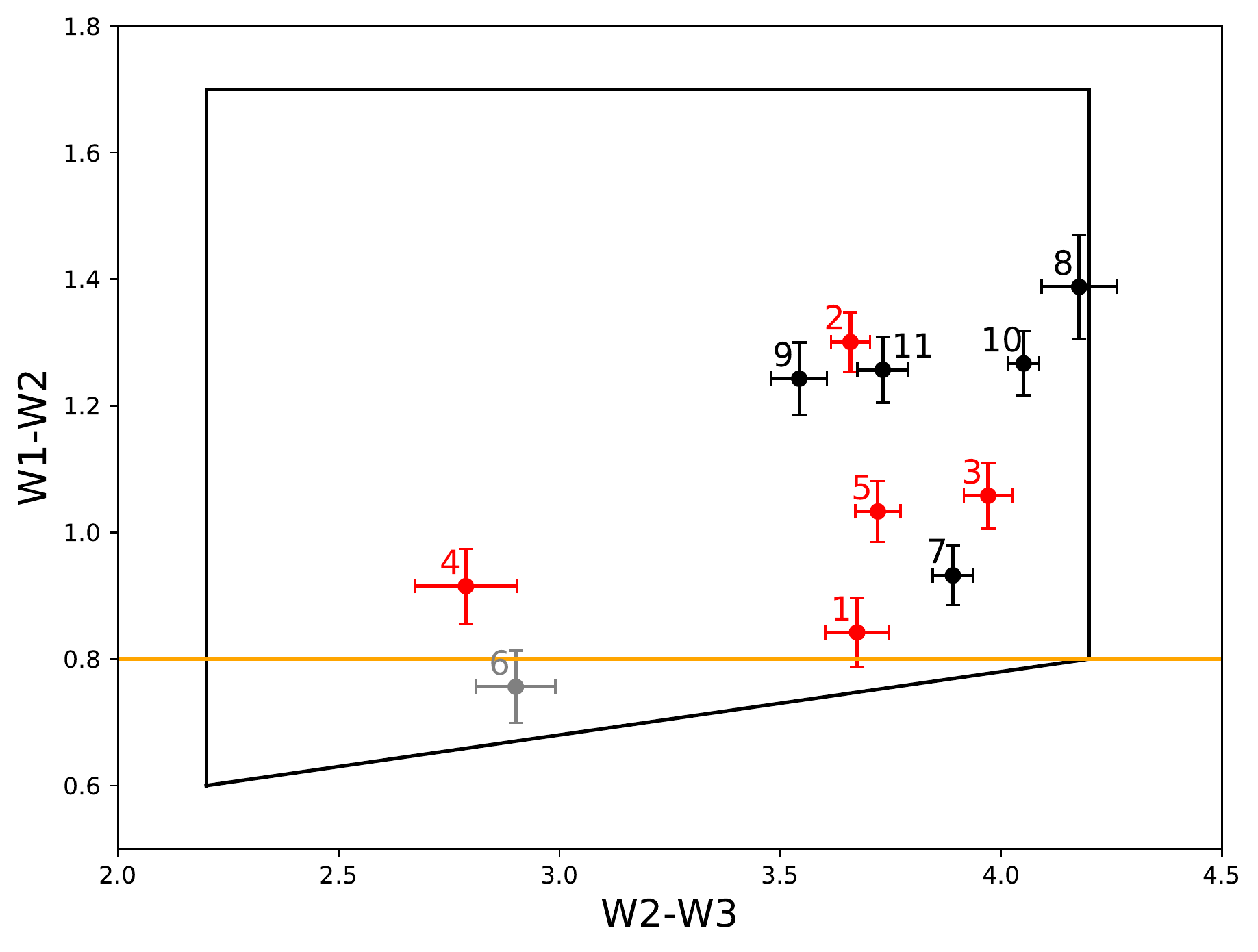}
\caption{\textit{WISE} color-color diagram for the dwarf galaxies in our sample. The galaxies classified as AGN, composite, and star-forming are in red, grey, and black, respectively, using the \NII/\ha~BPT diagram 
(left panel in Figure \ref{fig:bpt}). 
In black, we plot the AGN selection box from \cite{jarrett11}, and in orange we show the W1-W$2 > 0.8$ selection criterion from \cite{stern12}. 
}
\label{fig:wisecolor}
\end{figure}

\begin{figure*}
\centering
\includegraphics[width=\textwidth]{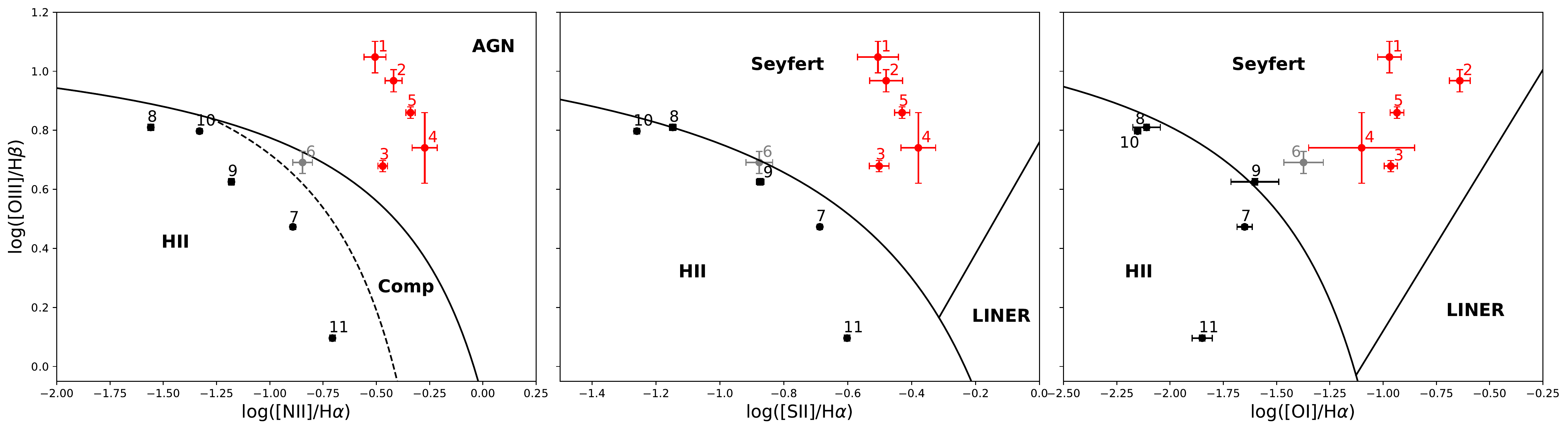}
\caption{Optical narrow emission line diagnostic diagrams for our galaxies. Regions are separated according to the classification scheme in \cite{kewley06}. The galaxies falling in the AGN, composite, and star-forming regions in the leftmost diagram are in red, grey, and black, respectively.}
\label{fig:bpt}
\end{figure*}

Properties of the eleven dwarf galaxies in our sample are summarized in Table \ref{tab:sample}, and we plot our galaxies on the \textit{WISE} color-color diagram in Figure \ref{fig:wisecolor} (see Table \ref{tab:wisemags} for magnitudes). Five of our galaxies are classified as AGNs, one as Composite, and five as star-forming using the \NII/\ha~BPT diagram (left panel in Figure \ref{fig:bpt}). The line fluxes used for the ratios in Figure \ref{fig:bpt} are measured from Sloan Digital Sky Survey (SDSS) spectra, as described in \cite{reines13}. Three-color \textit{HST} images of our galaxies are shown in Figure \ref{fig:rgb}, except in the case of ID 10 for which we use Dark Energy Camera Legacy Survey (DECaLS) imaging.

\begin{deluxetable}{cccccc}
\tabletypesize{\footnotesize}
\tablecaption{\textit{WISE} magnitudes}
\tablewidth{0pt}
\tablehead{
\colhead{ID} & \colhead{W1} & \colhead{W2}  & \colhead{W3} & \colhead{W1-W2} & \colhead{W2-W3} \\
\colhead{ } & \colhead{(mag)} & \colhead{(mag)} & \colhead{(mag)} & \colhead{(mag)} & \colhead{(mag)} \\
\colhead{(1)} & \colhead{(2)} & \colhead{(3)} & \colhead{(4)} & \colhead{(5)} & \colhead{(6)} }
\startdata
1  	    	 & 14.01   & 13.17 & 9.50   & 0.84 & 3.67   \\
2            & 13.40   & 12.10 & 8.44   & 1.30 & 3.66   \\
3            & 13.86   & 12.80 & 8.83   & 1.06 & 3.97   \\
4    		 & 14.28   & 13.36 & 10.57  & 0.92 & 2.79   \\
5    		 & 13.51   & 12.48 & 8.76   & 1.03 & 3.72   \\
6            & 14.12   & 13.37 & 10.46  & 0.76 & 2.90   \\  
7 	    	 & 13.31   & 12.38 & 8.49   & 0.93 & 3.89   \\
8 	    	 & 15.58   & 14.20 & 10.02  & 1.39 & 4.18   \\
9 	    	 & 14.53   & 13.28 & 9.74   & 1.24 & 3.54   \\
10           & 13.04   & 11.77 & 7.72   & 1.27 & 4.05   \\
11           & 13.76   & 12.50 & 8.77   & 1.26 & 3.73   
\enddata
\label{tab:wisemags}
\end{deluxetable}

\begin{figure*}
\captionsetup[subfigure]{labelformat=empty}
{\includegraphics[width=0.25\textwidth]{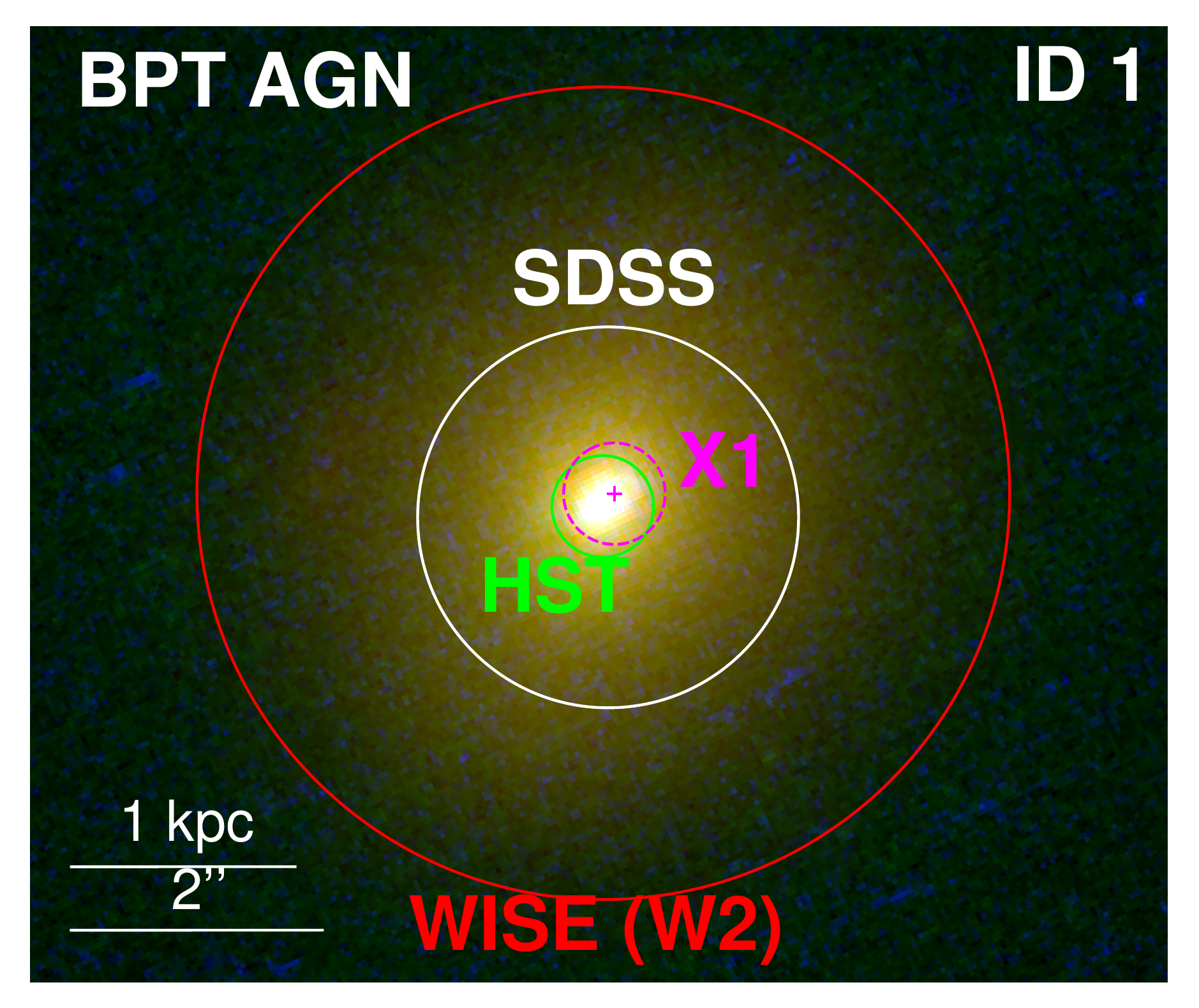}}\hfill
{\includegraphics[width=0.25\textwidth]{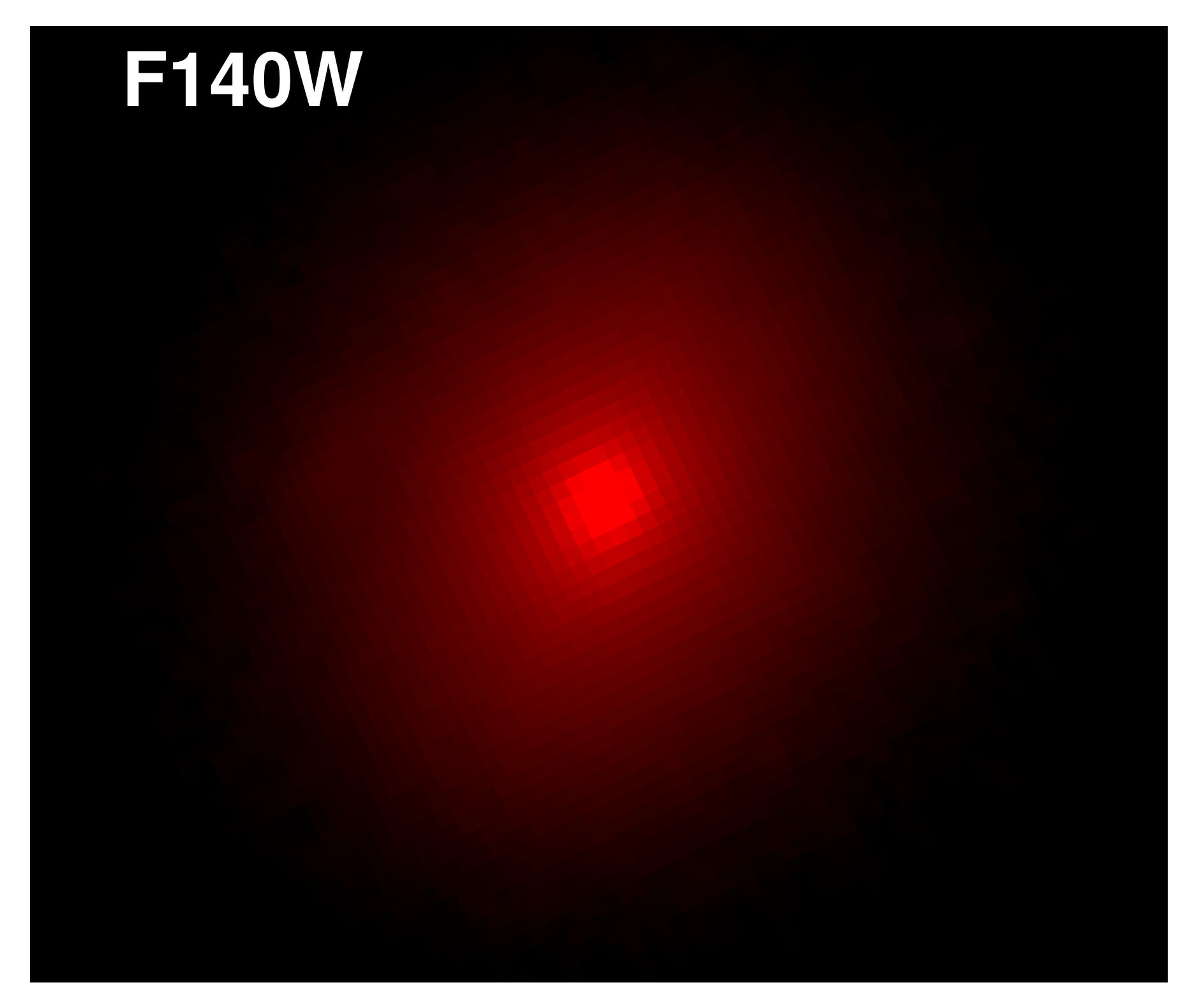}}\hfill
{\includegraphics[width=0.25\textwidth]{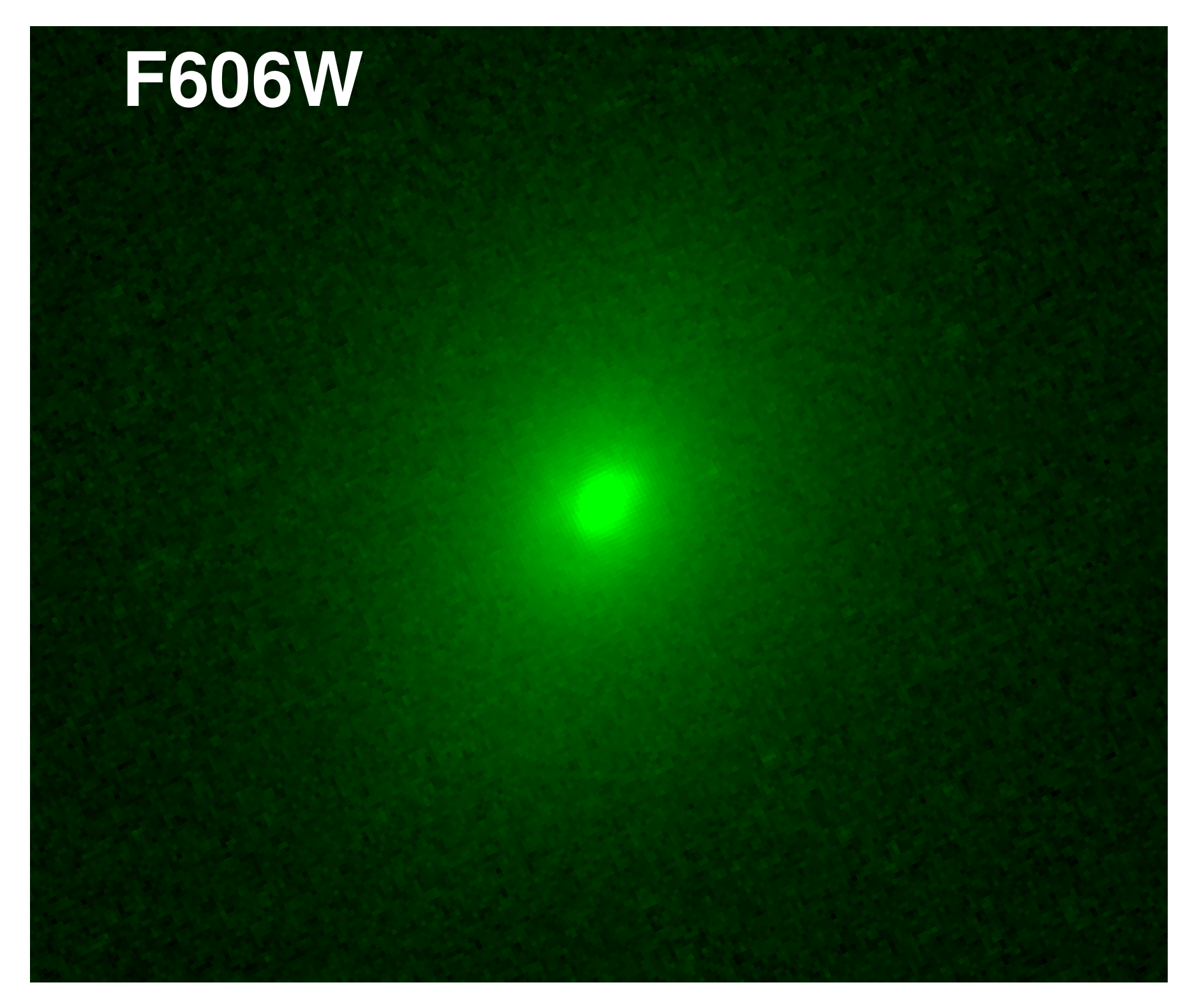}}\hfill
{\includegraphics[width=0.25\textwidth]{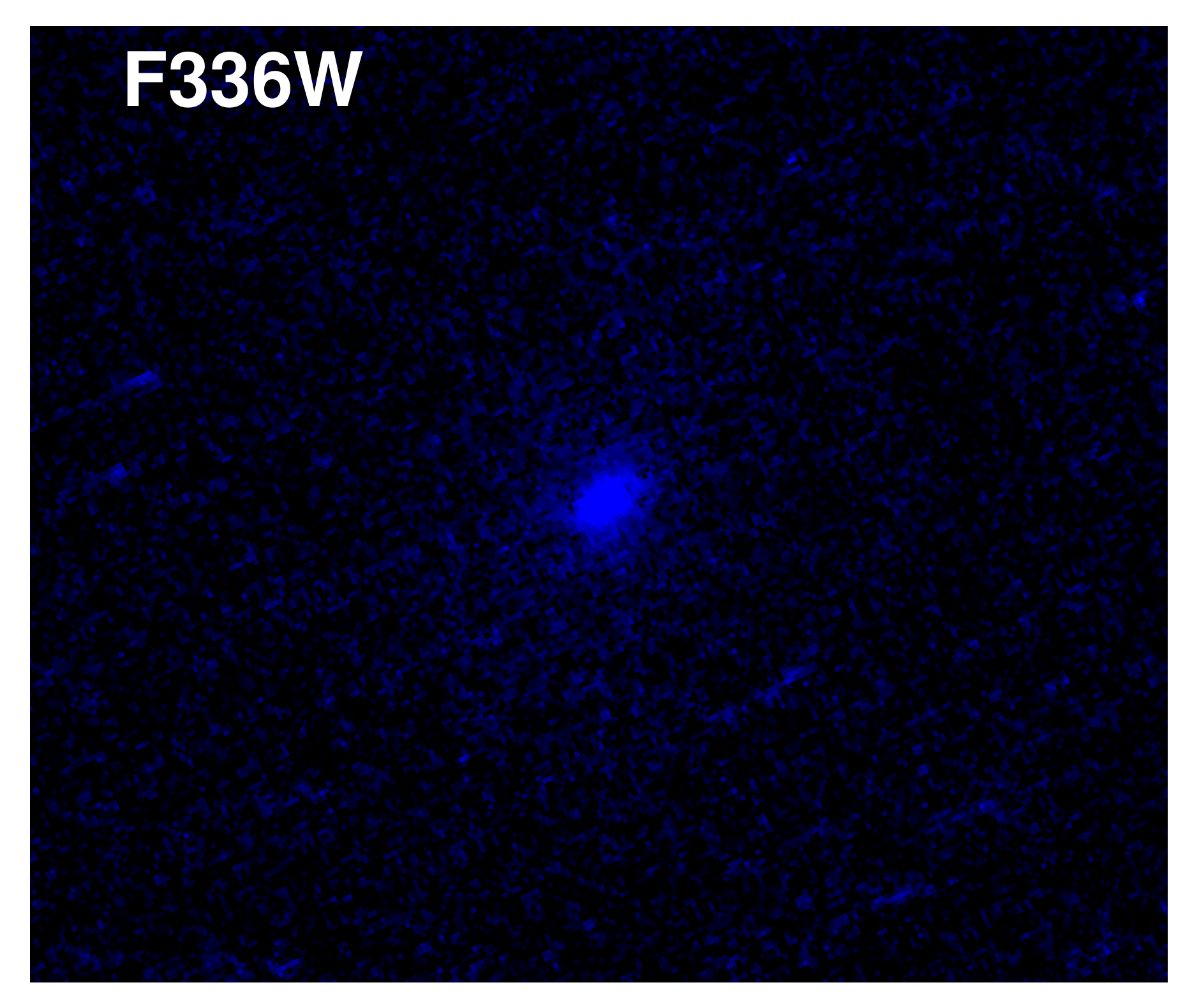}}\hfill

{\includegraphics[width=0.25\textwidth]{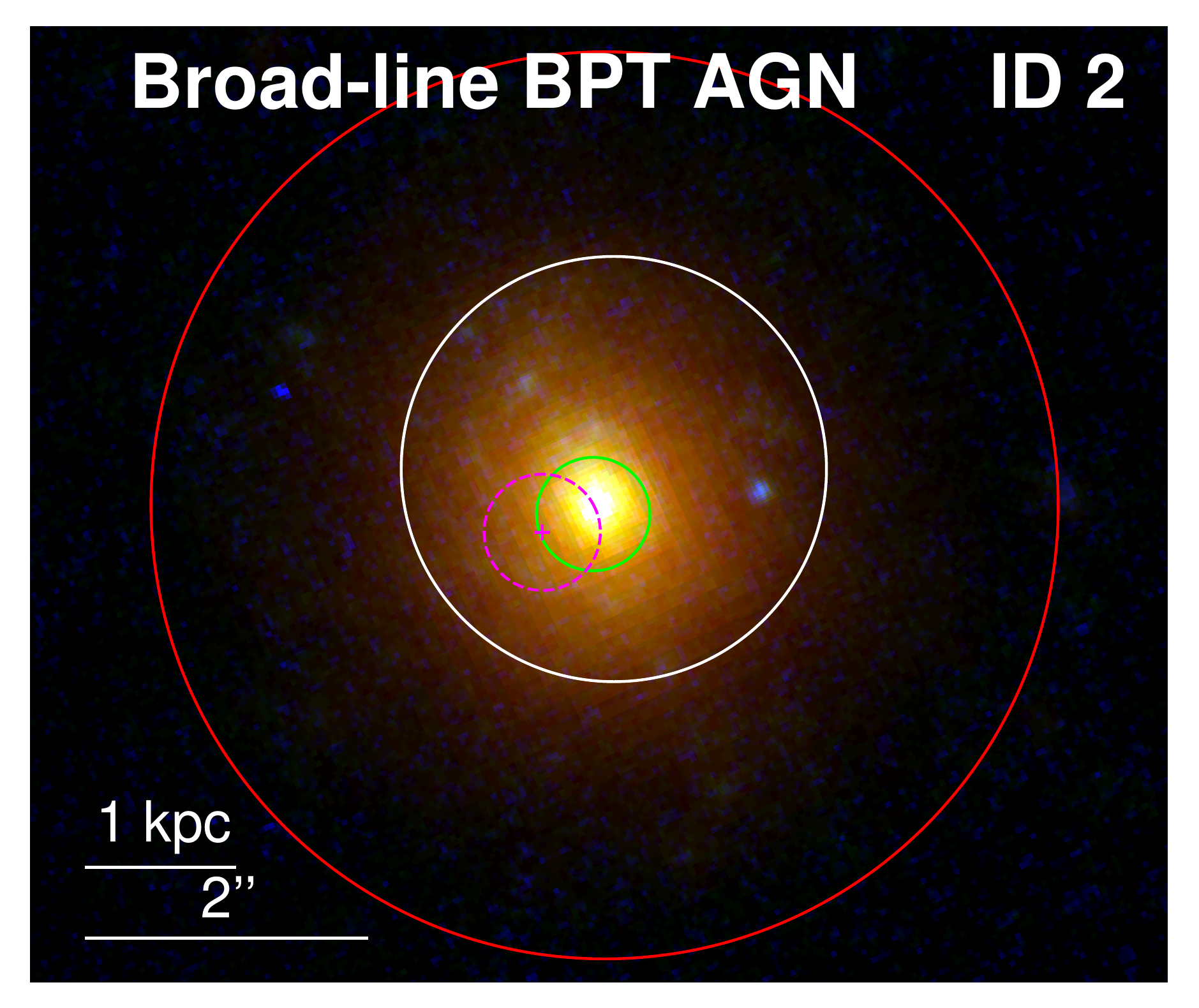}}\hfill
{\includegraphics[width=0.25\textwidth]{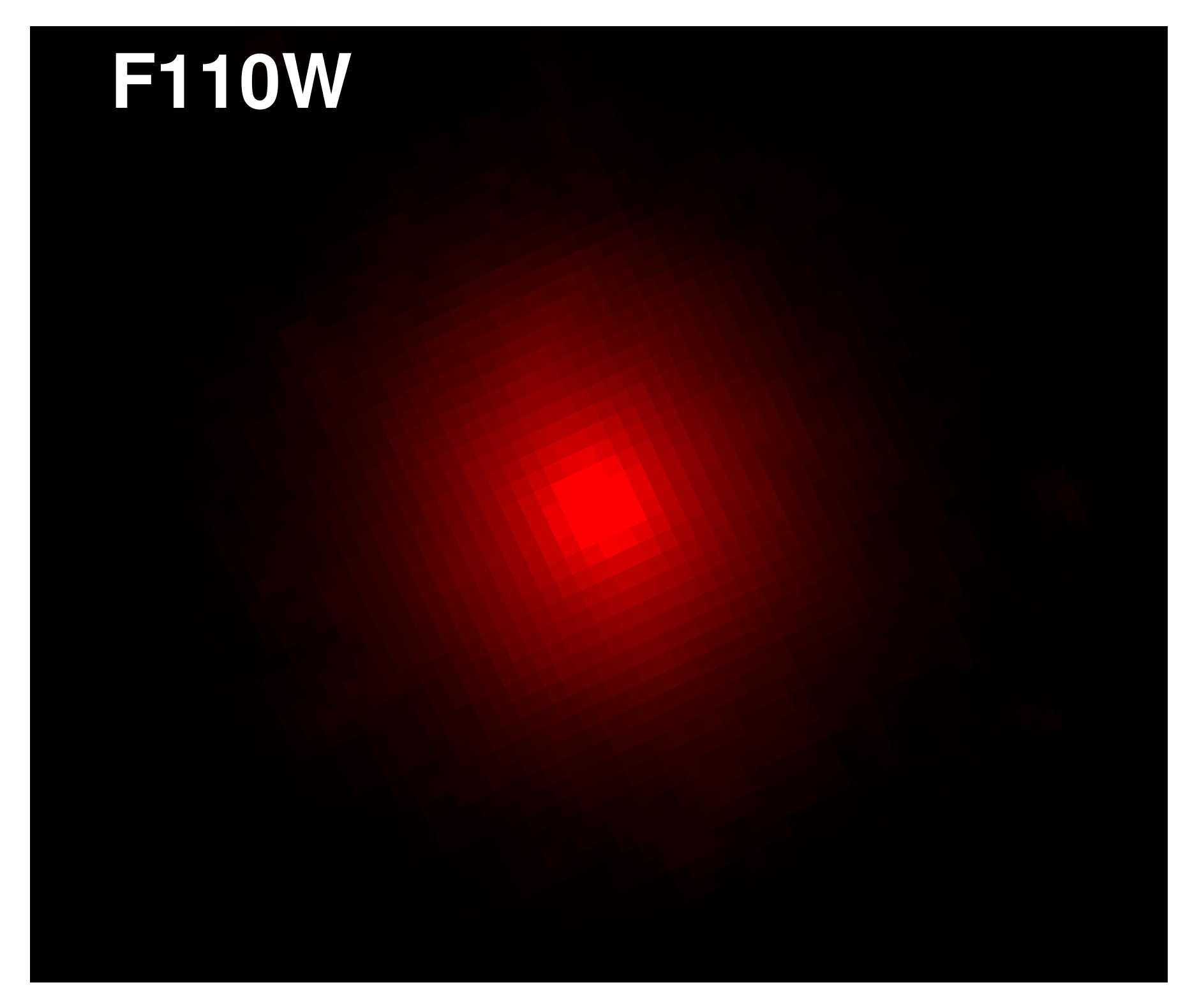}}\hfill
{\includegraphics[width=0.25\textwidth]{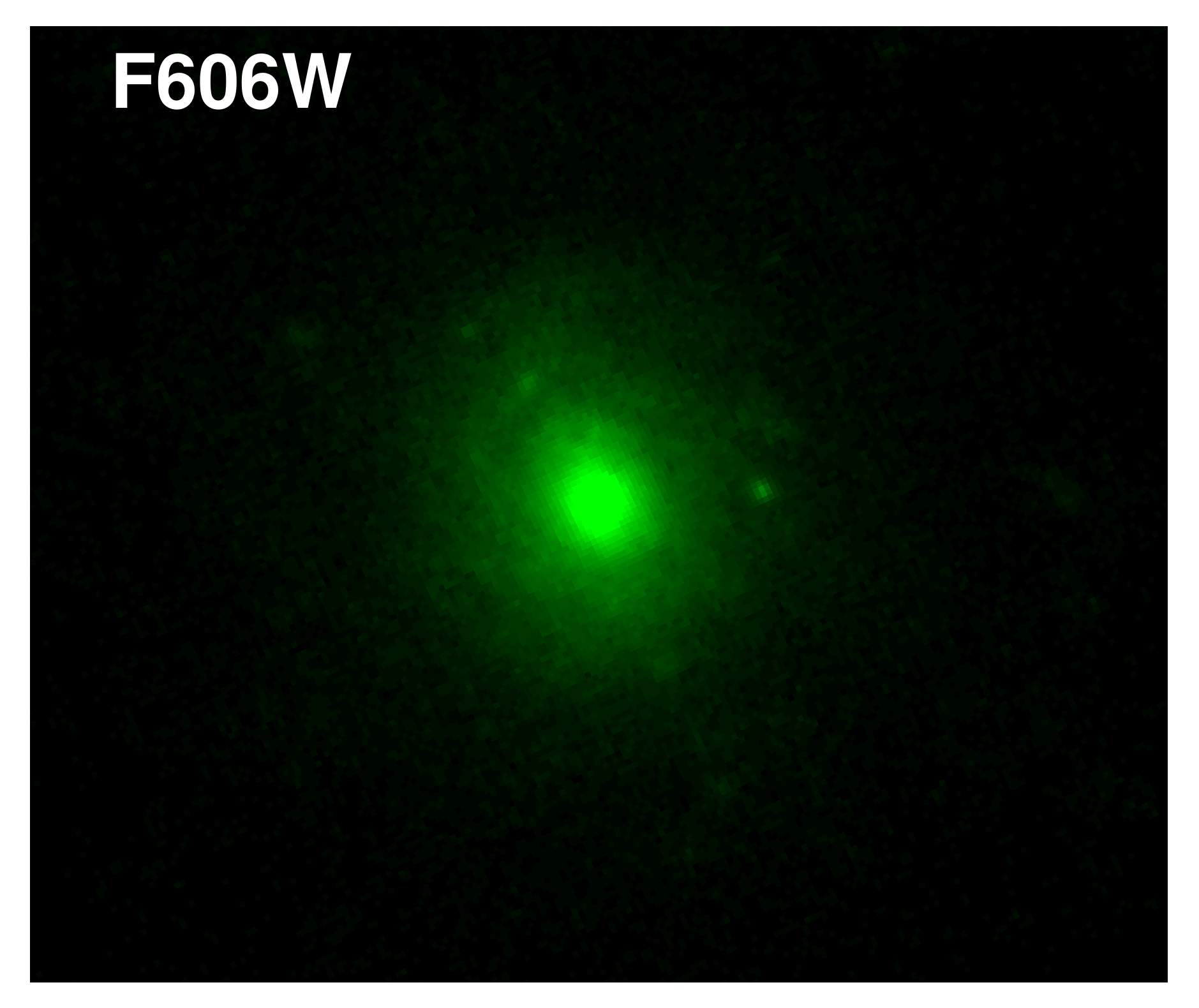}}\hfill
{\includegraphics[width=0.25\textwidth]{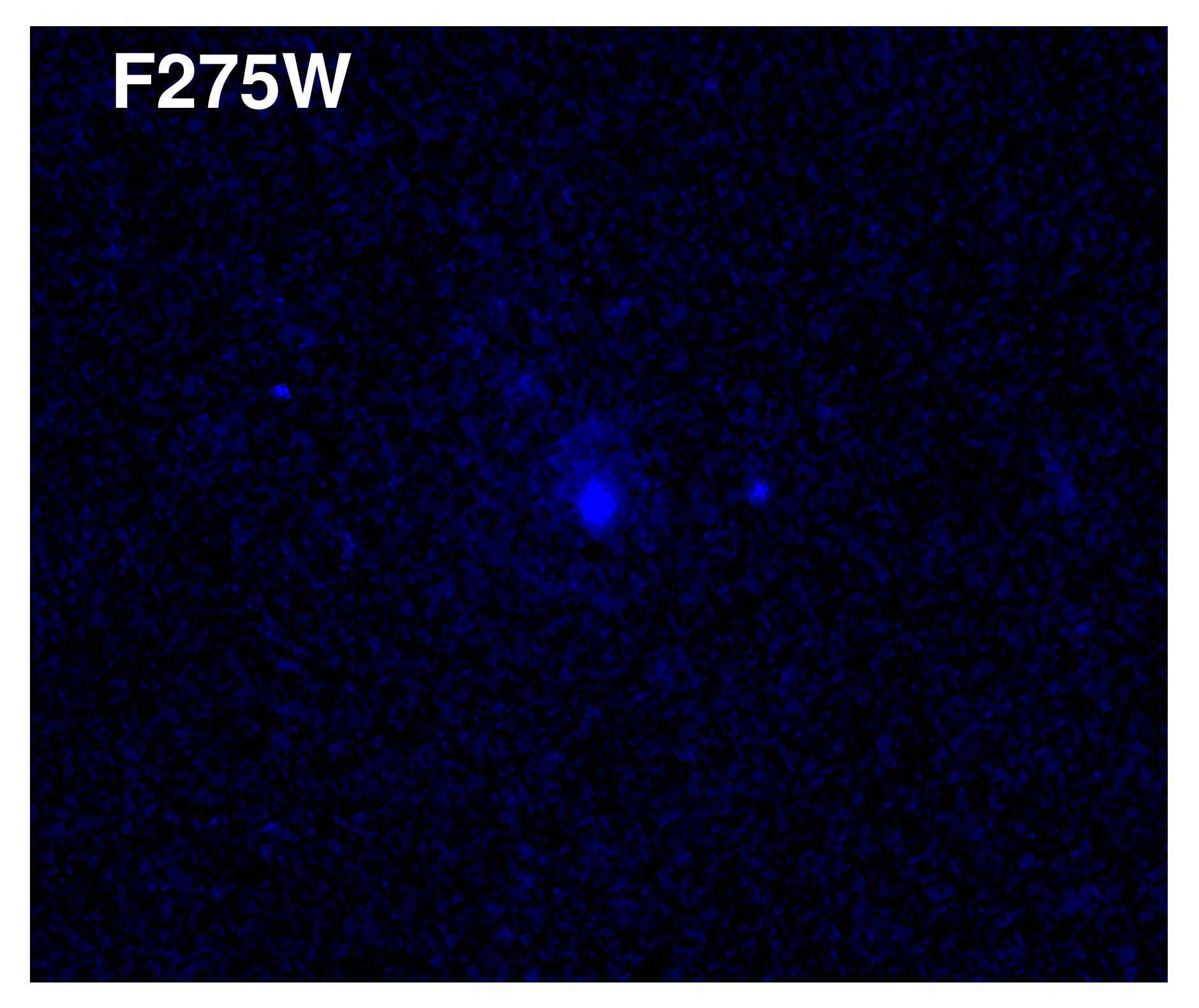}}\hfill

{\includegraphics[width=0.25\textwidth]{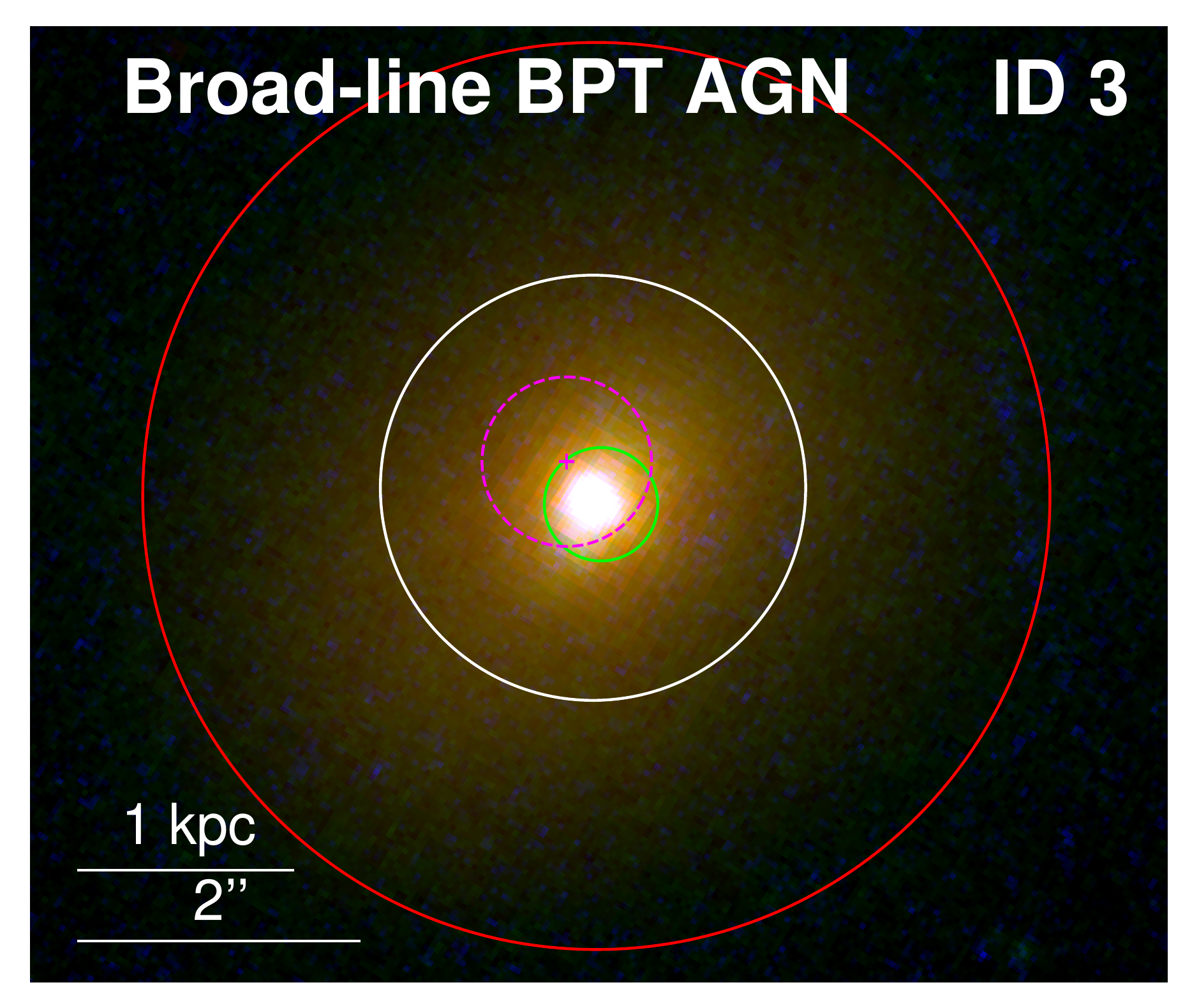}}\hfill
{\includegraphics[width=0.25\textwidth]{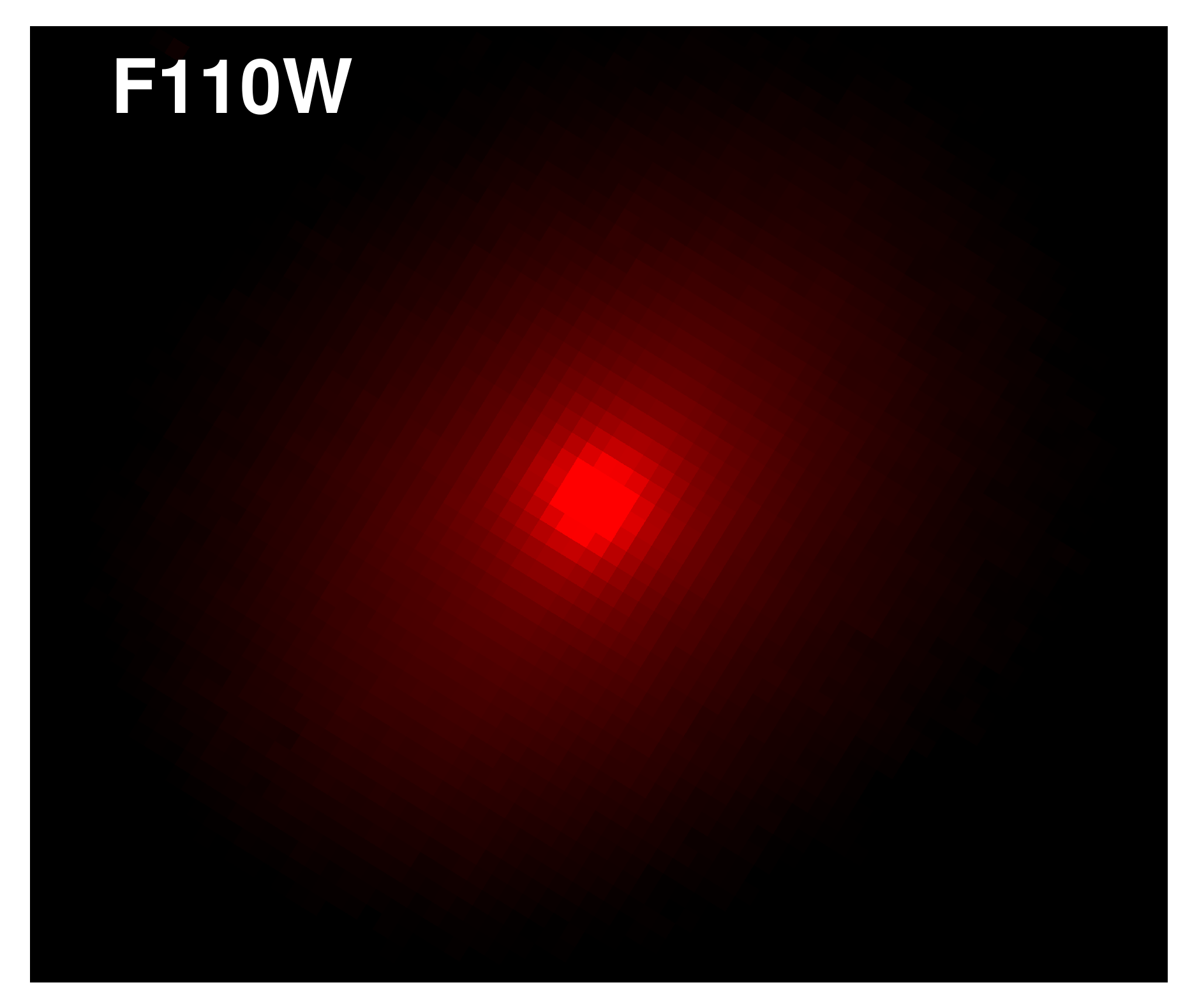}}\hfill
{\includegraphics[width=0.25\textwidth]{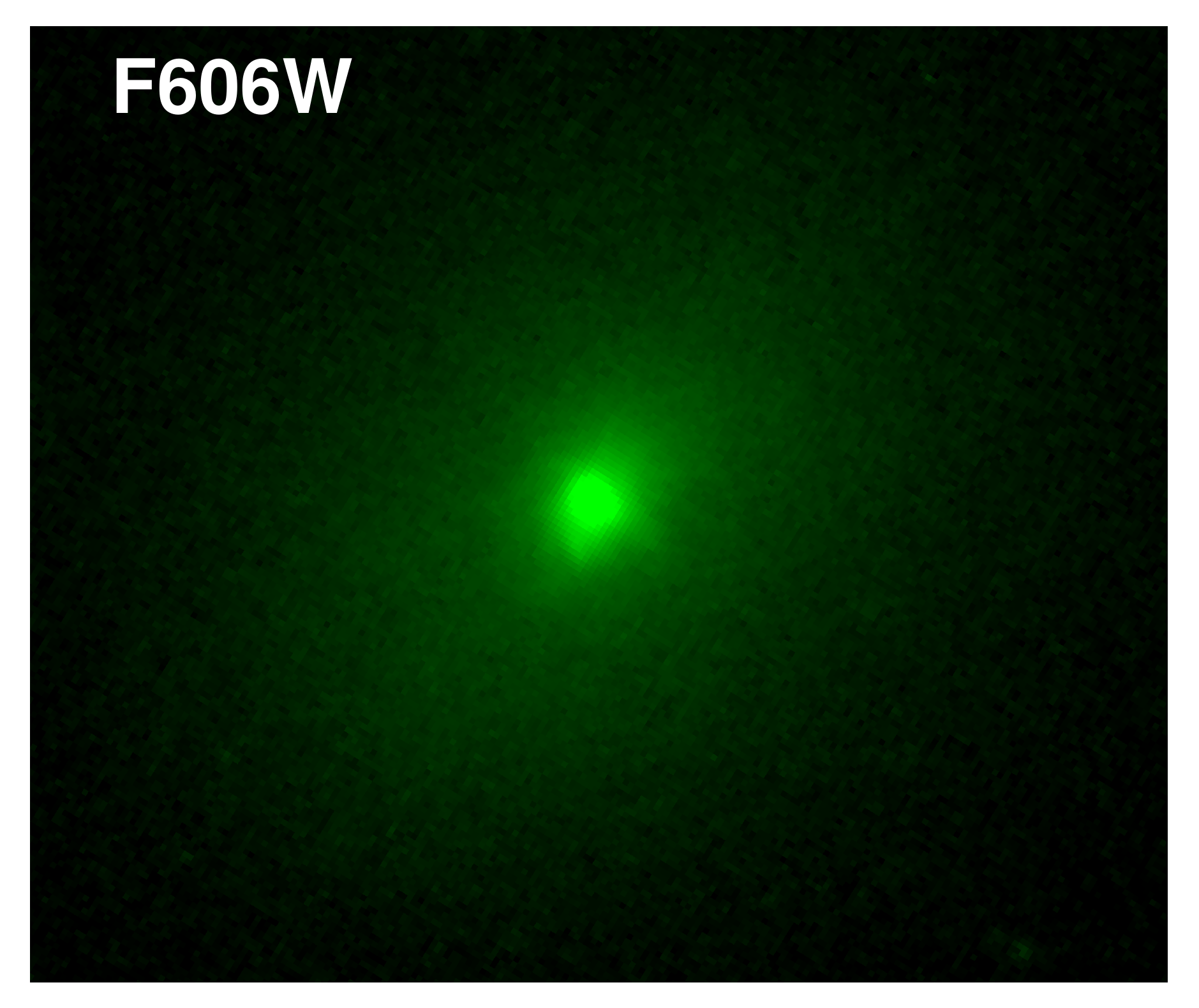}}\hfill
{\includegraphics[width=0.25\textwidth]{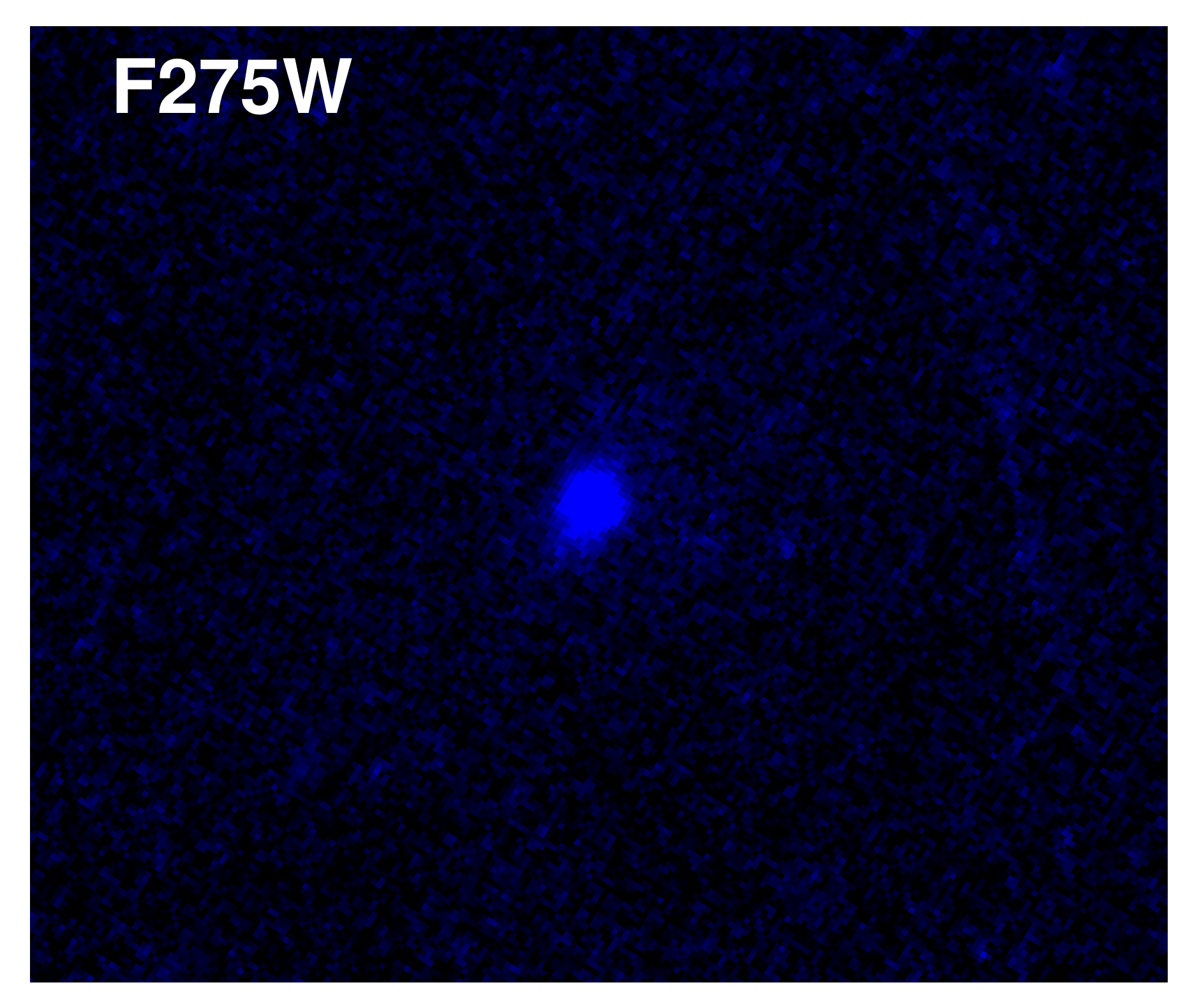}}\hfill

{\includegraphics[width=0.25\textwidth]{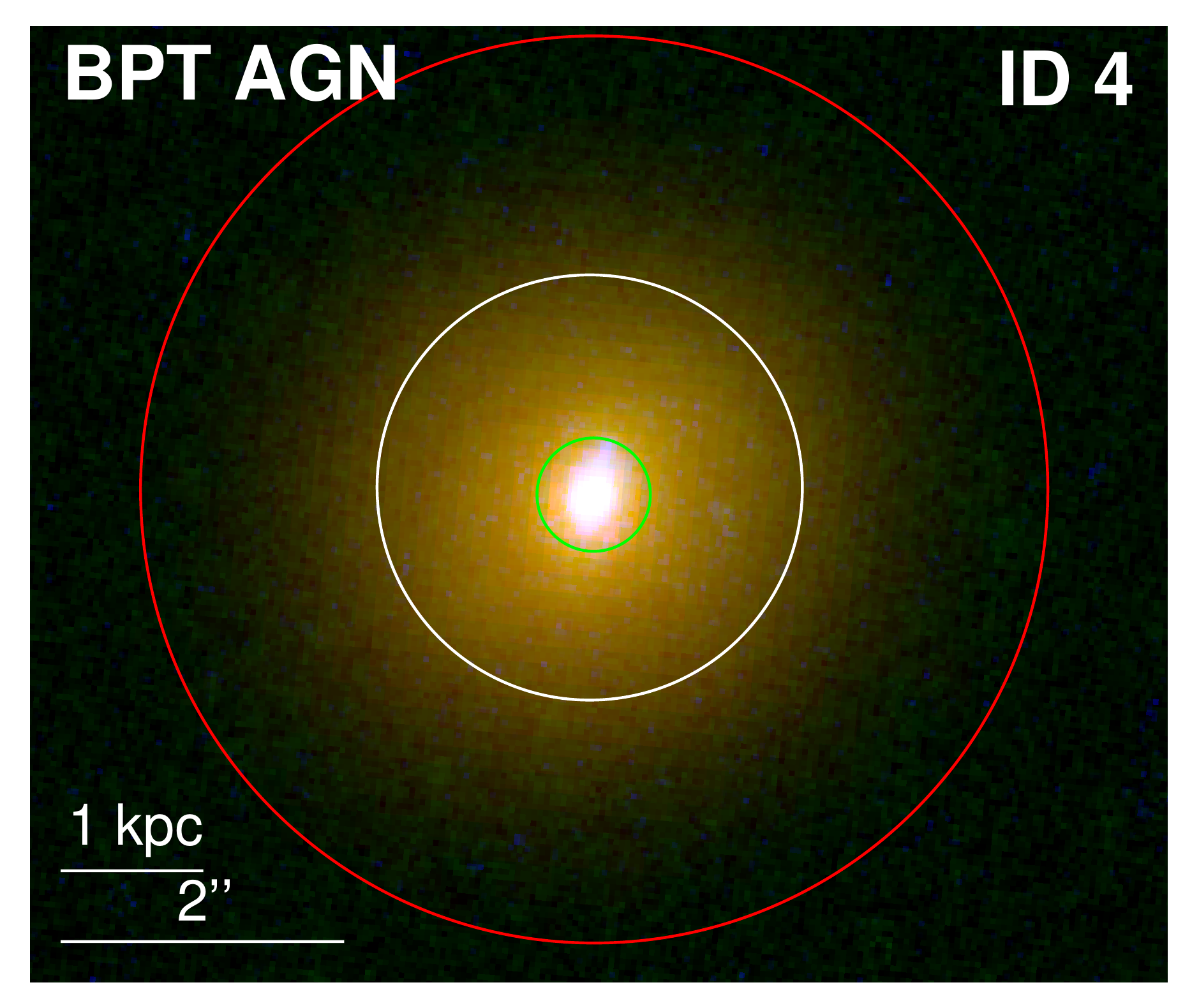}}\hfill
{\includegraphics[width=0.25\textwidth]{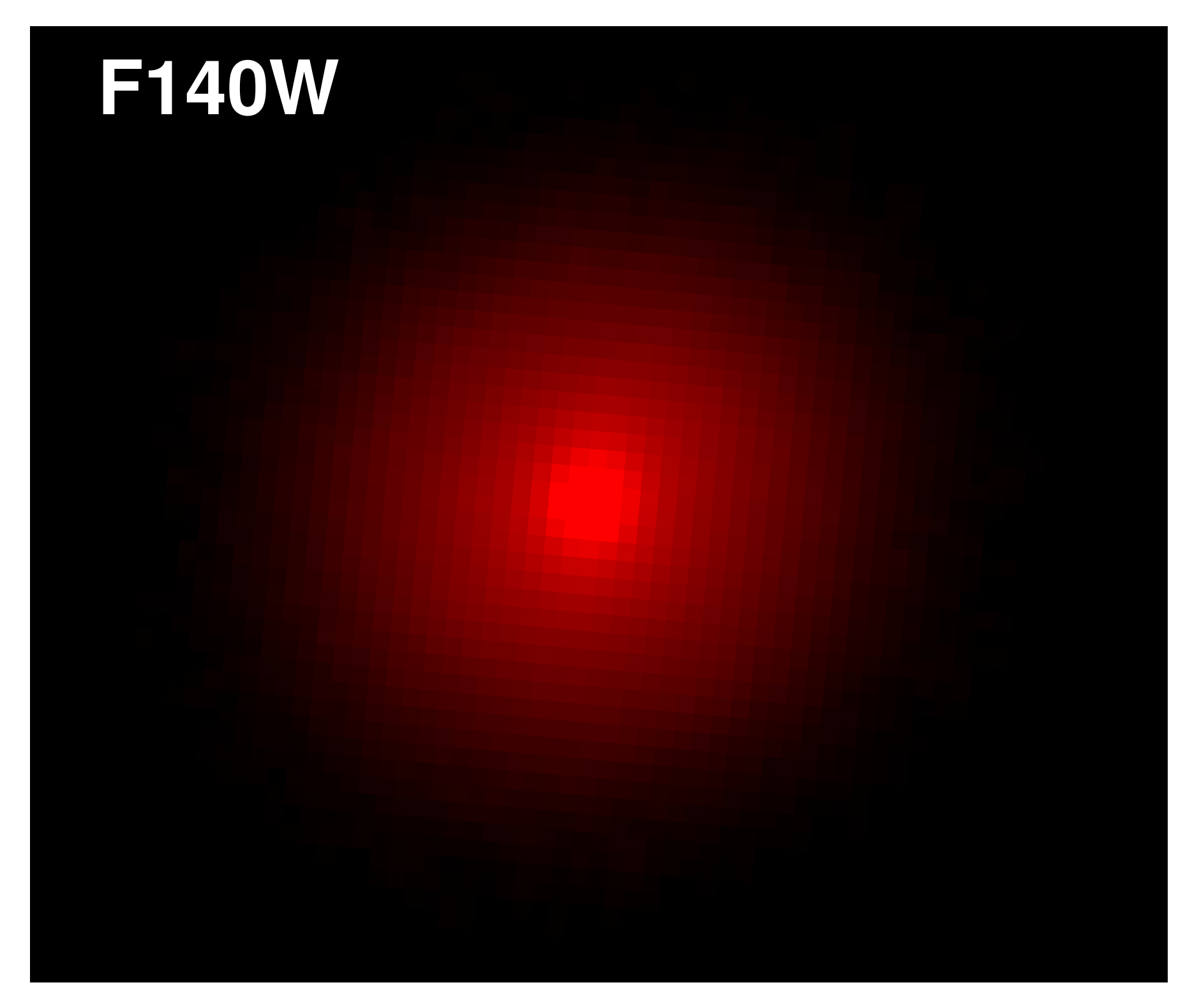}}\hfill
{\includegraphics[width=0.25\textwidth]{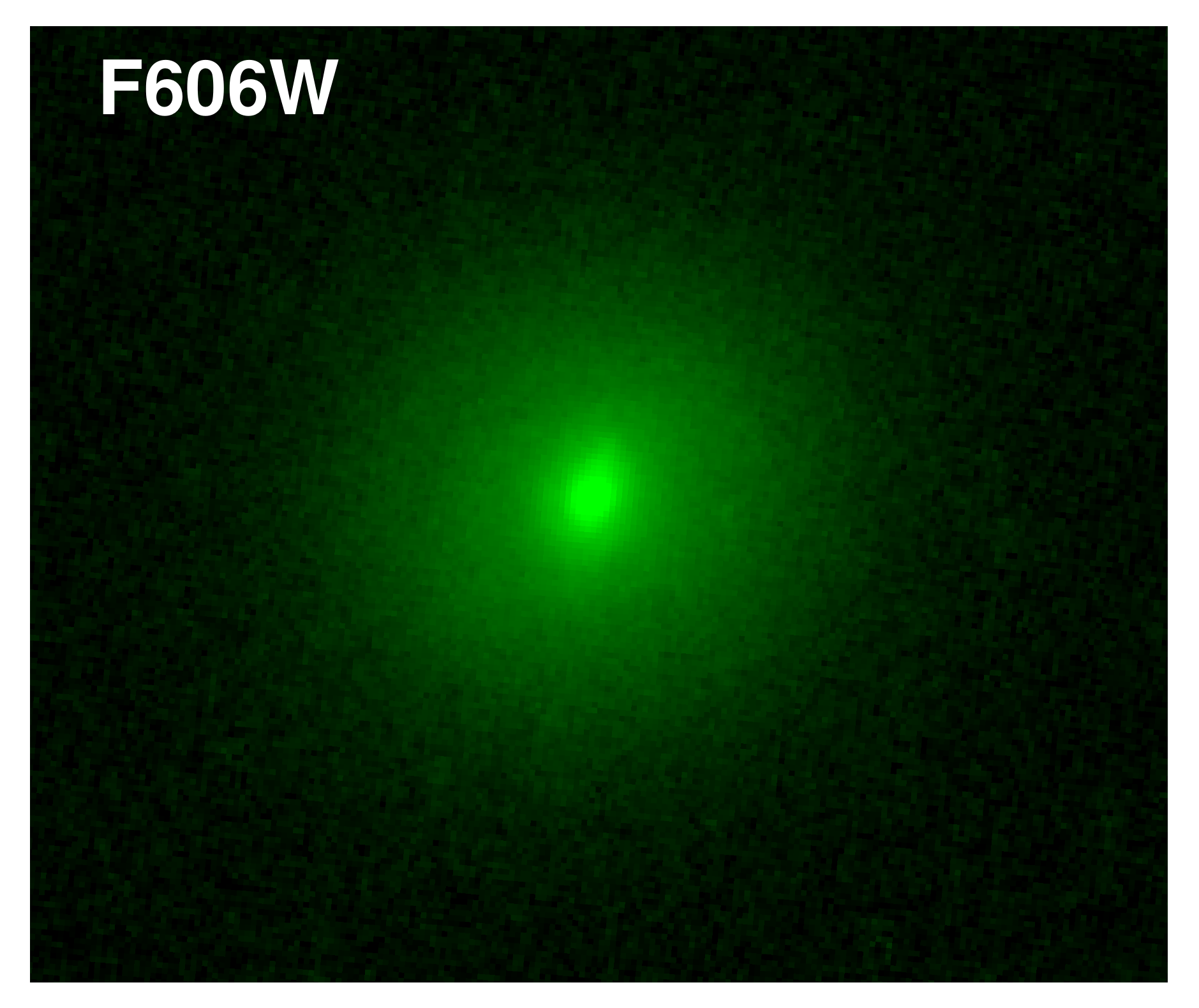}}\hfill
{\includegraphics[width=0.25\textwidth]{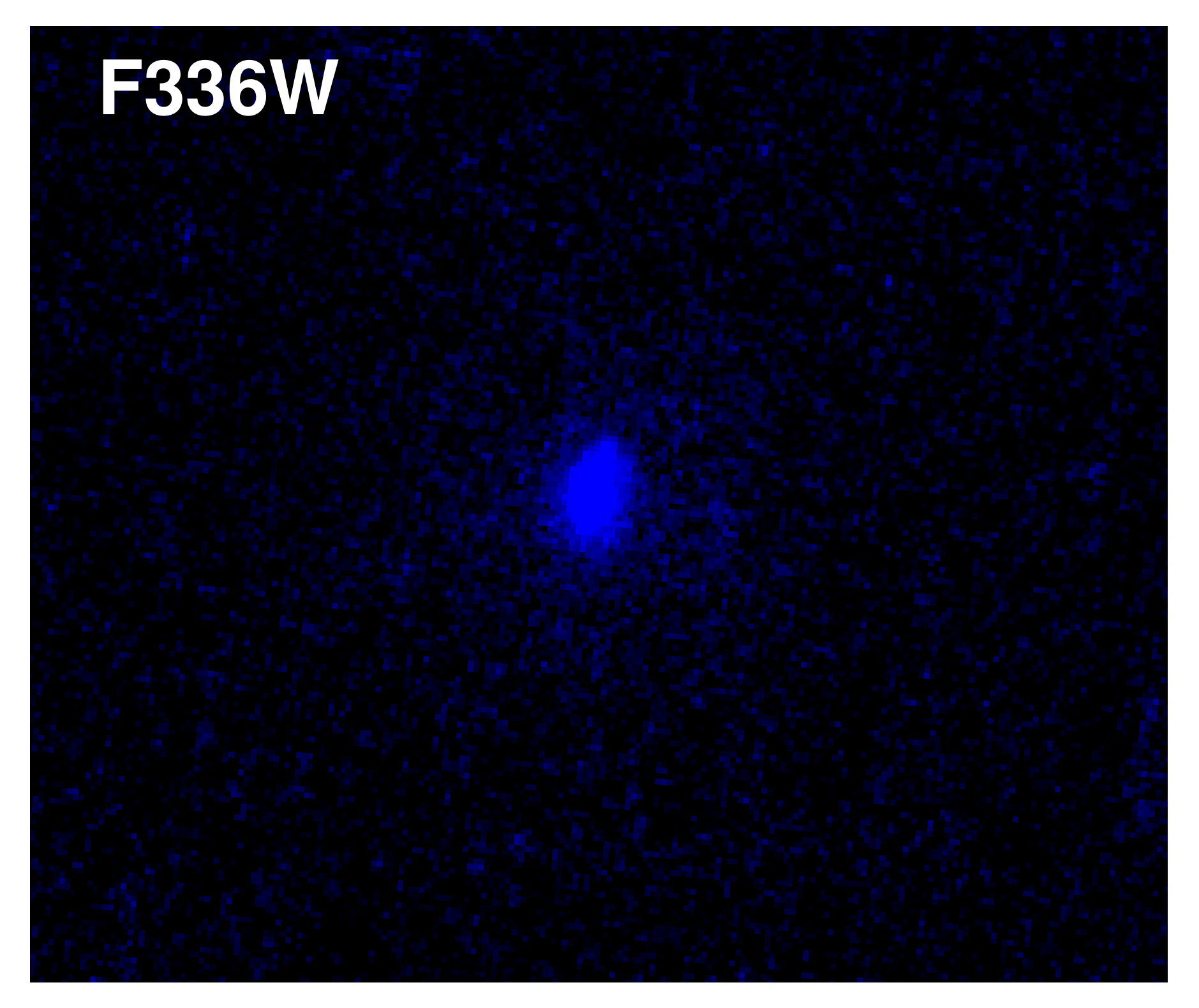}}\hfill

{\includegraphics[width=0.25\textwidth]{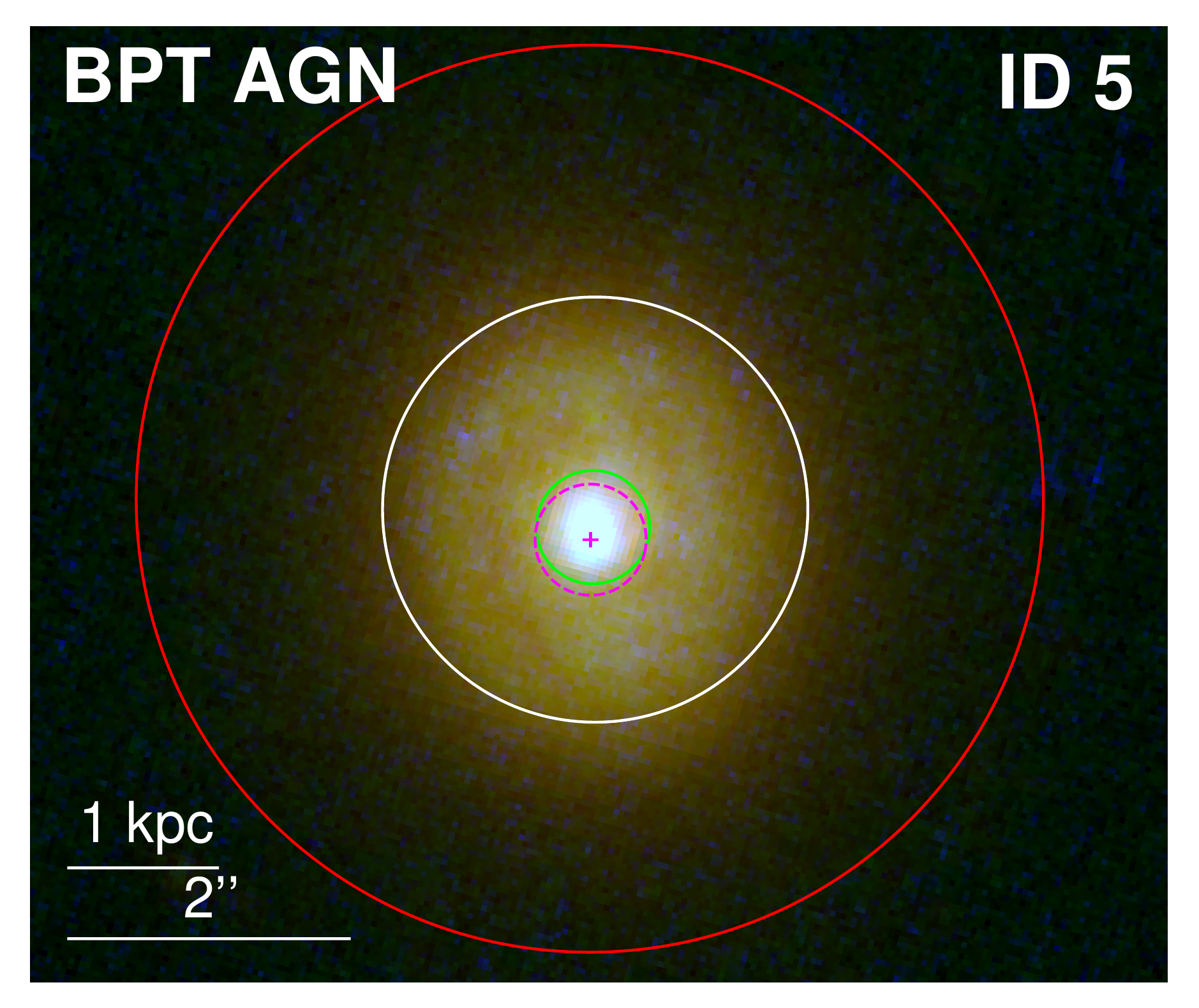}}\hfill
{\includegraphics[width=0.25\textwidth]{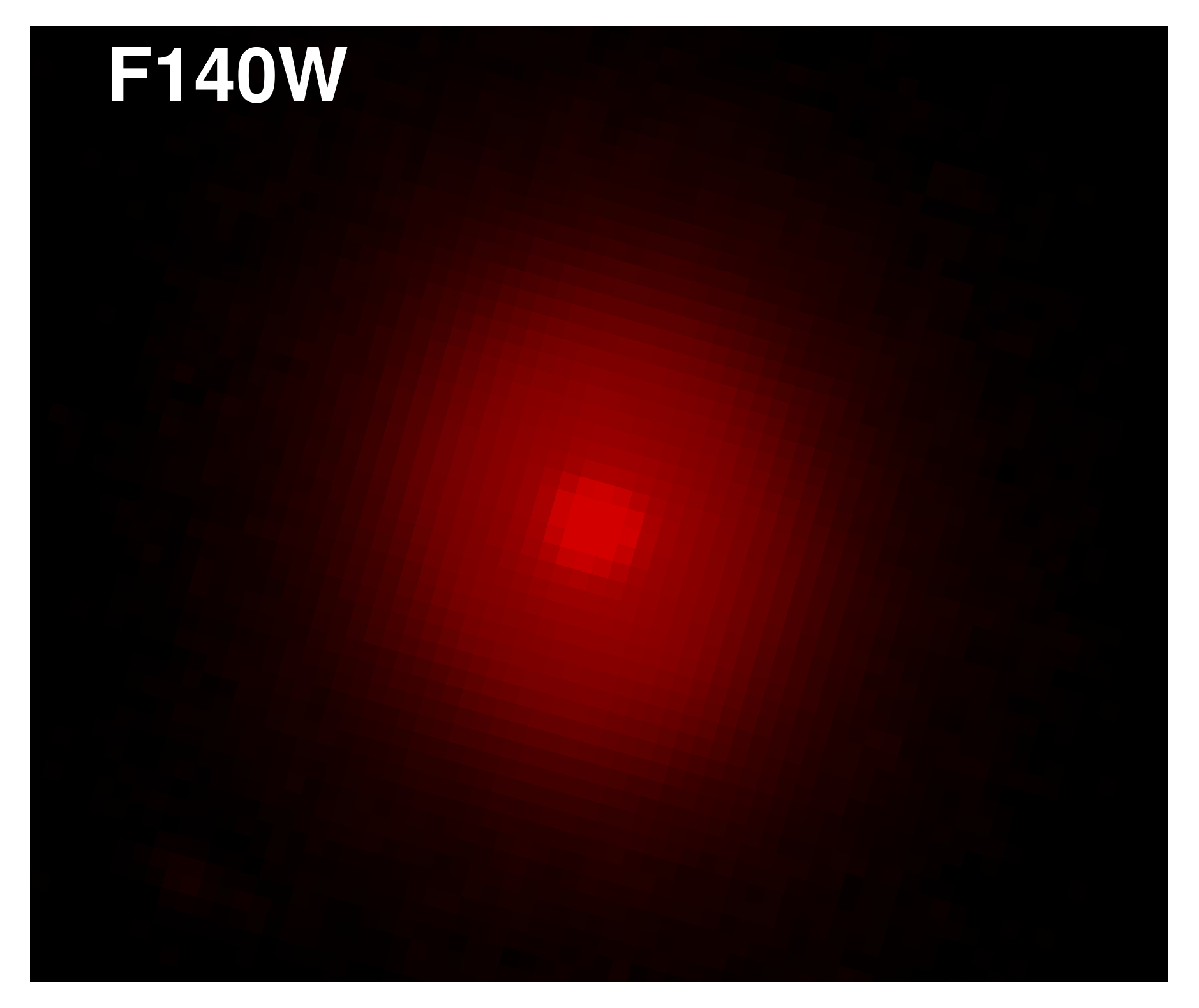}}\hfill
{\includegraphics[width=0.25\textwidth]{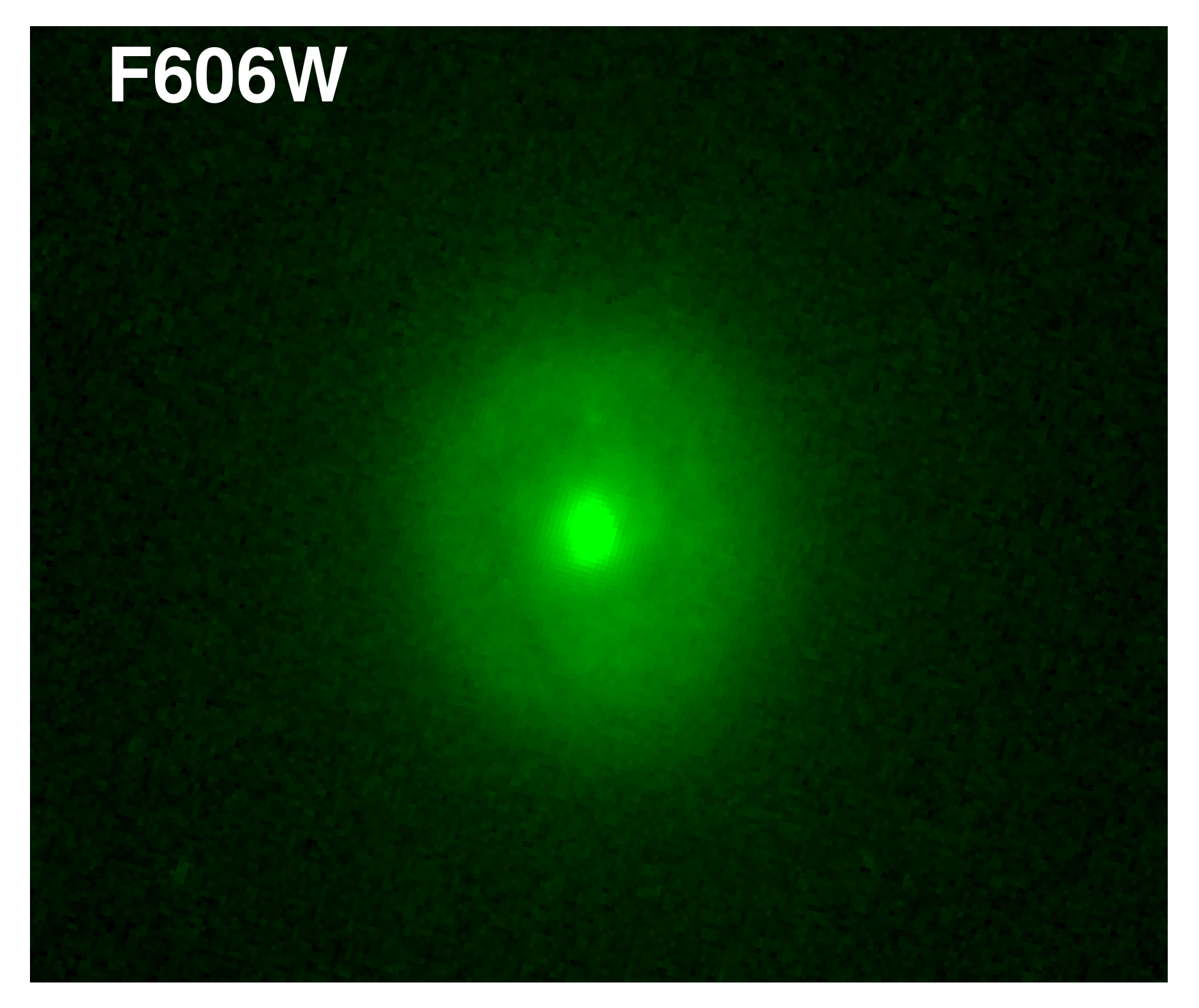}}\hfill
{\includegraphics[width=0.25\textwidth]{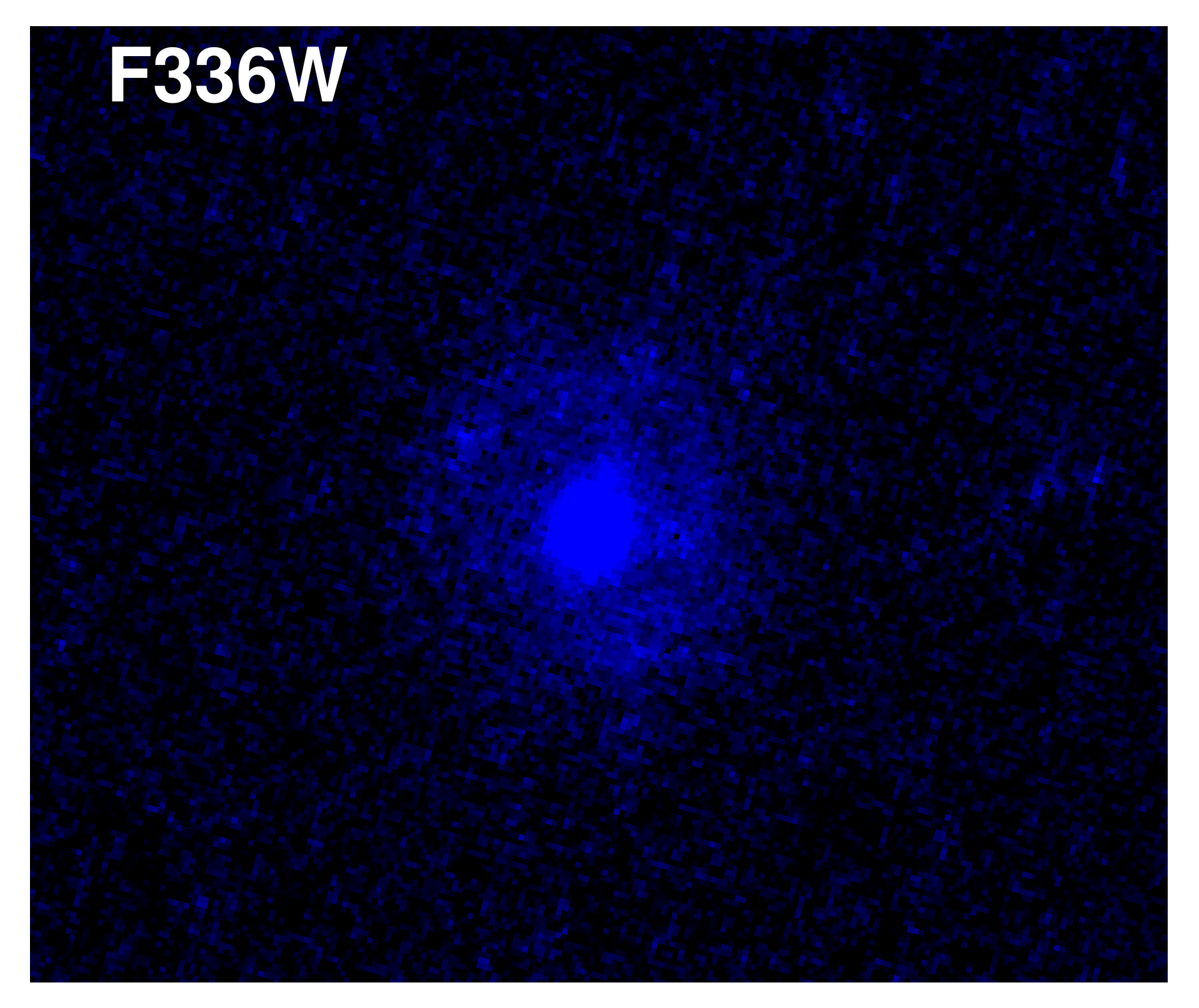}}\hfill

{\includegraphics[width=0.25\textwidth]{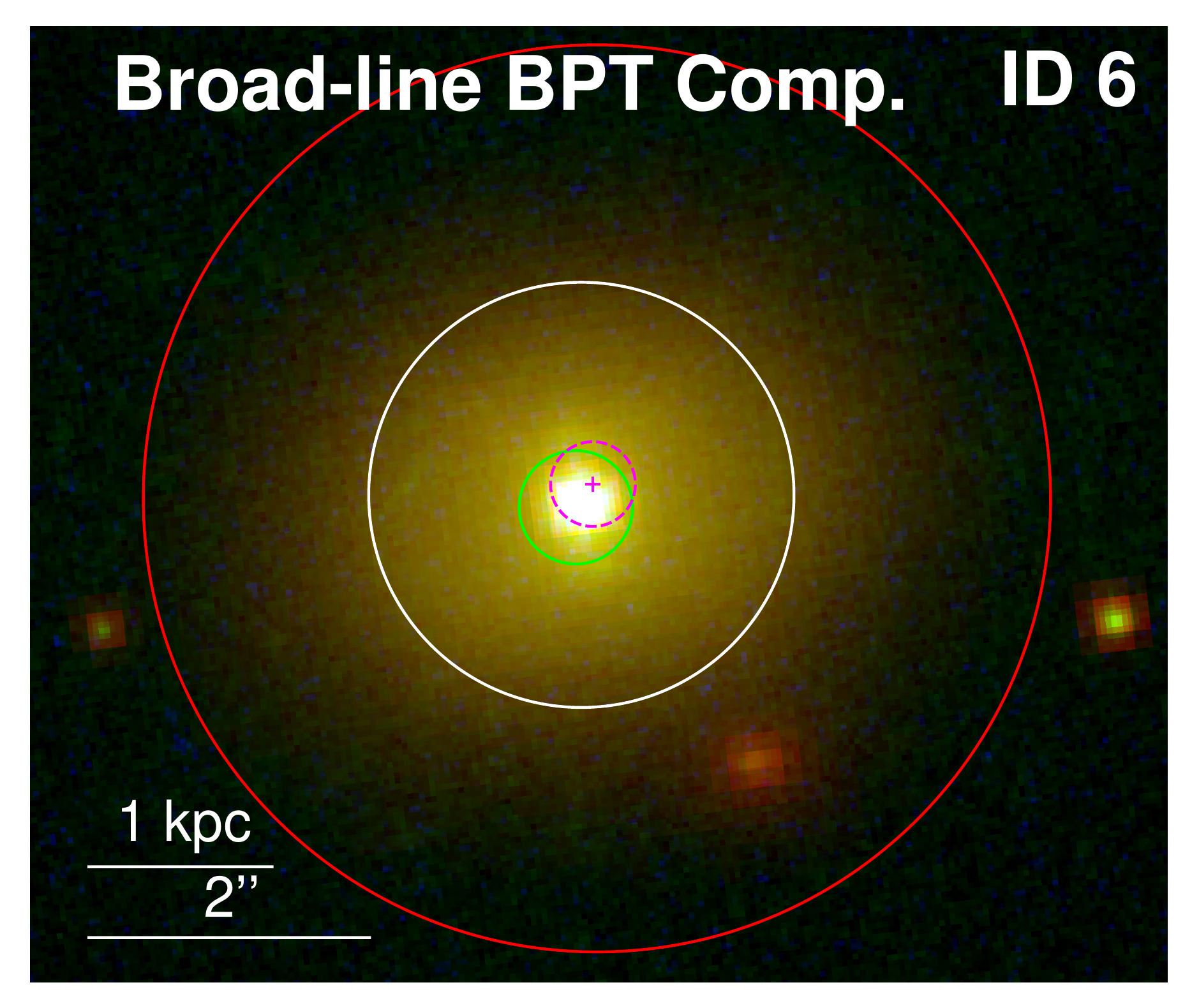}}\hfill
{\includegraphics[width=0.25\textwidth]{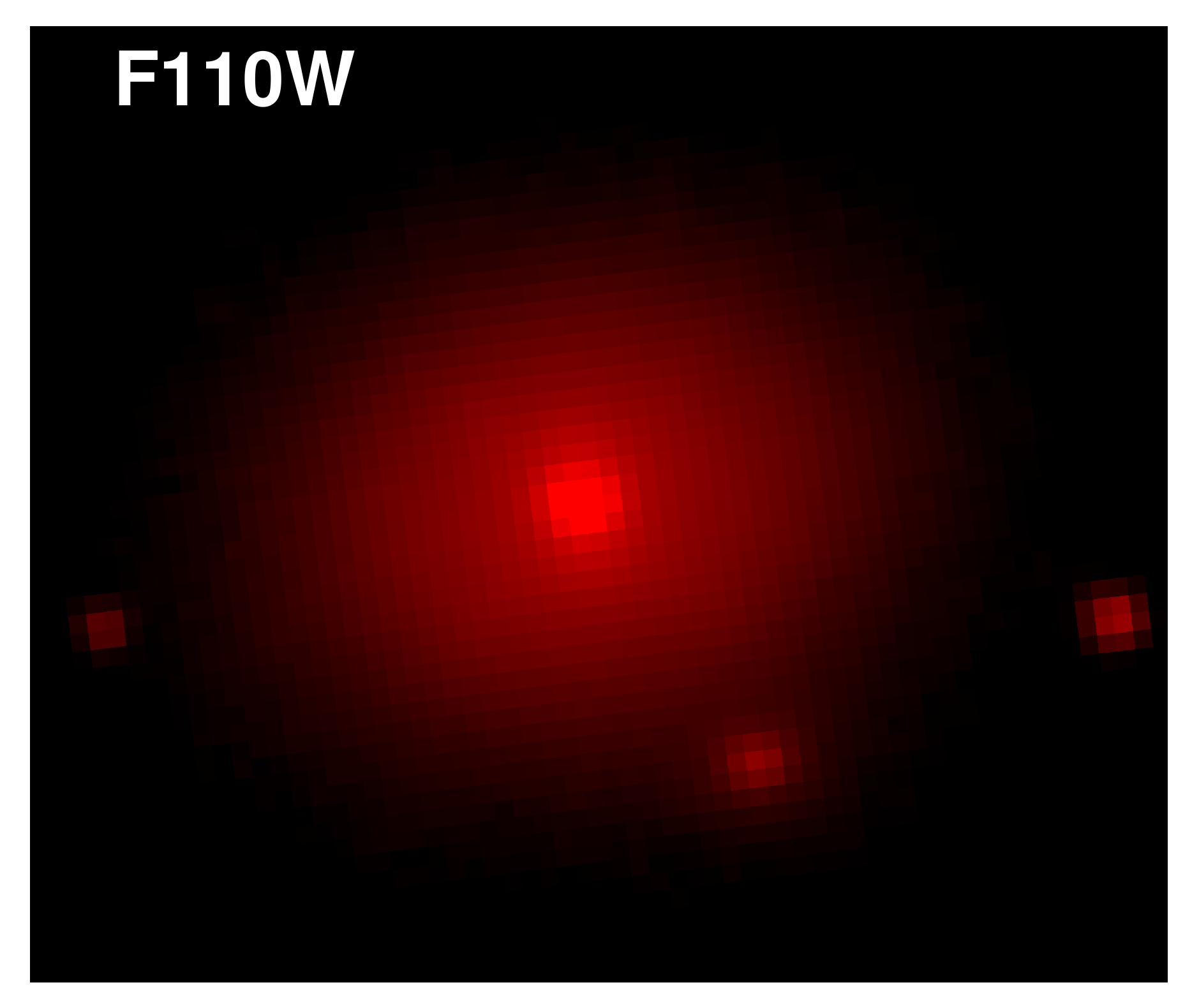}}\hfill
{\includegraphics[width=0.25\textwidth]{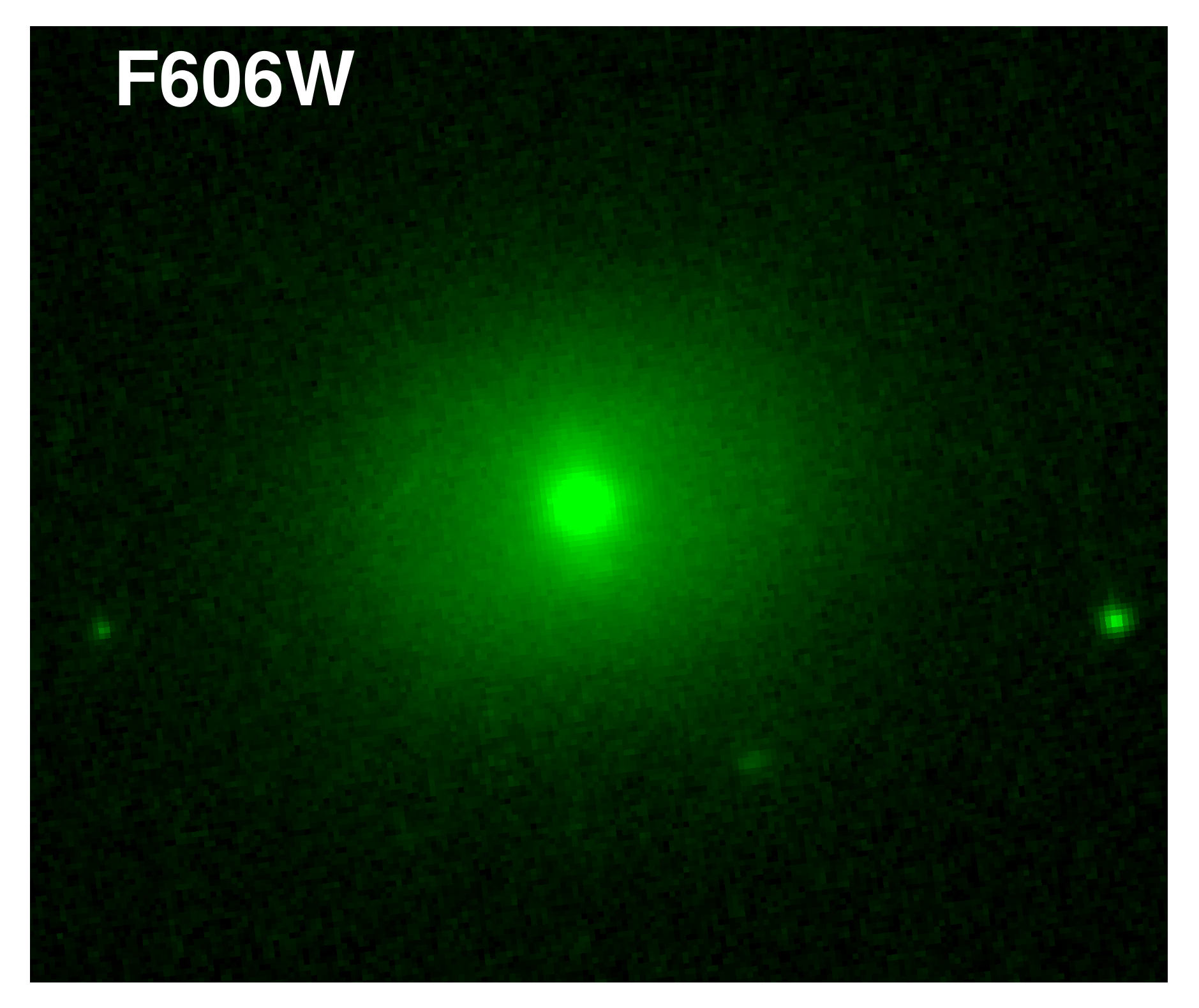}}\hfill
{\includegraphics[width=0.25\textwidth]{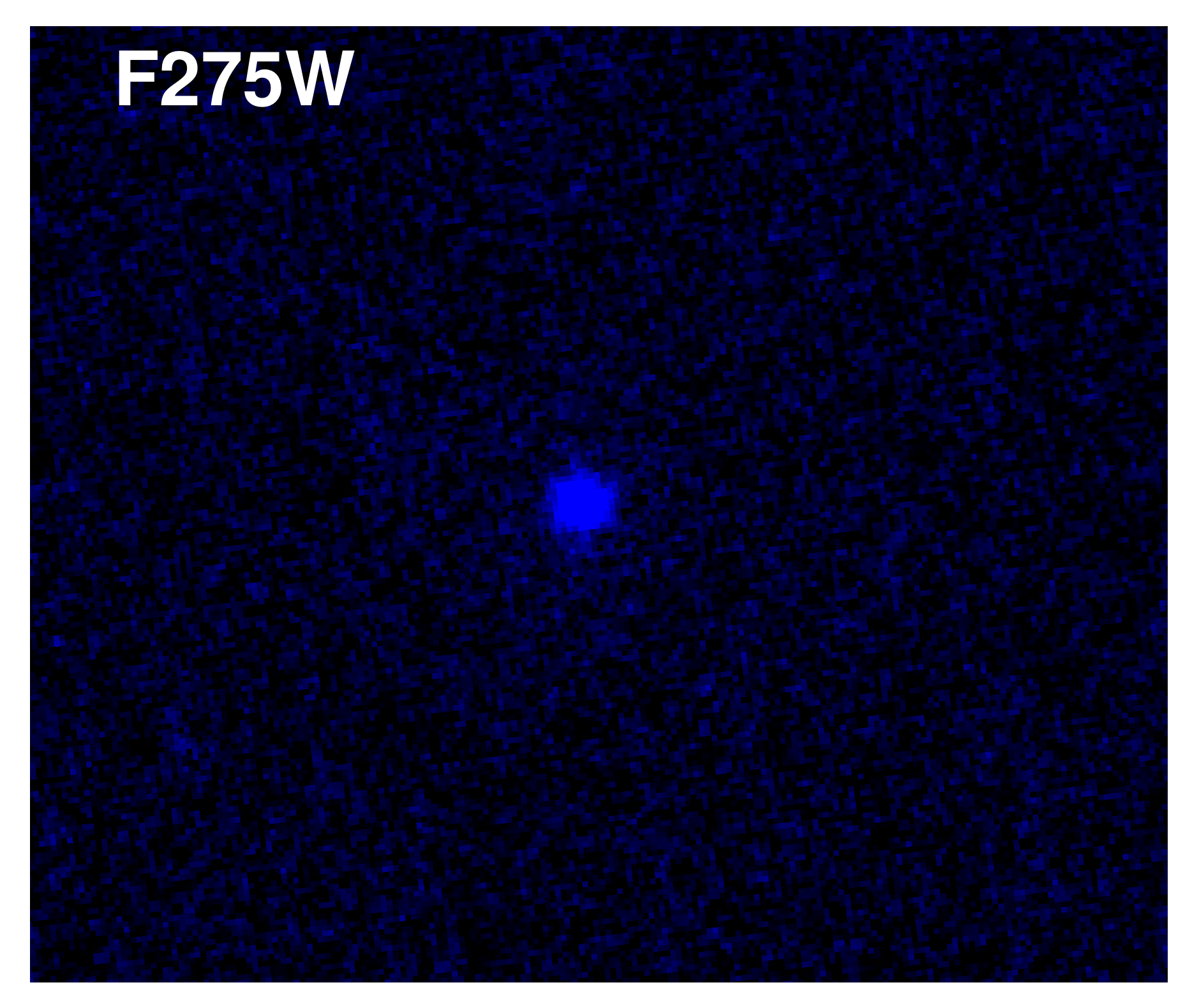}}\hfill

\caption{Three-color \textit{HST} images of our galaxies. Red shows the NIR image (F140W for IDs 1,4-5,7-9,11, F110W for IDs 2-3,6), green shows the optical band (F606W), and blue shows the $U$/UV band (F336W for IDs 1,4-5,7-9,11, F275W for IDs 2-3,6). We overlay green circles of radii 0\farcs4 at the peak of the {NIR} emission in the {\it HST} images, as well as the positions and 95\% positional uncertainties of the detected X-ray sources in {magenta}. The SDSS 3\arcsec\ spectroscopic apertures are shown as white circles and the resolution of the \textit{WISE} W2-band (6.4\arcsec) is in red. The text in the upper left corner denotes the BPT classification of the galaxy. Note that for ID 10, \textit{HST} imaging was unavailable, so DECaLS imaging was used instead, with red, green, and blue as the $z$, $r$, and $g$ bands, respectively.
} 
\label{fig:rgb}
\end{figure*}

\begin{figure*}
\ContinuedFloat

{\includegraphics[width=0.25\textwidth]{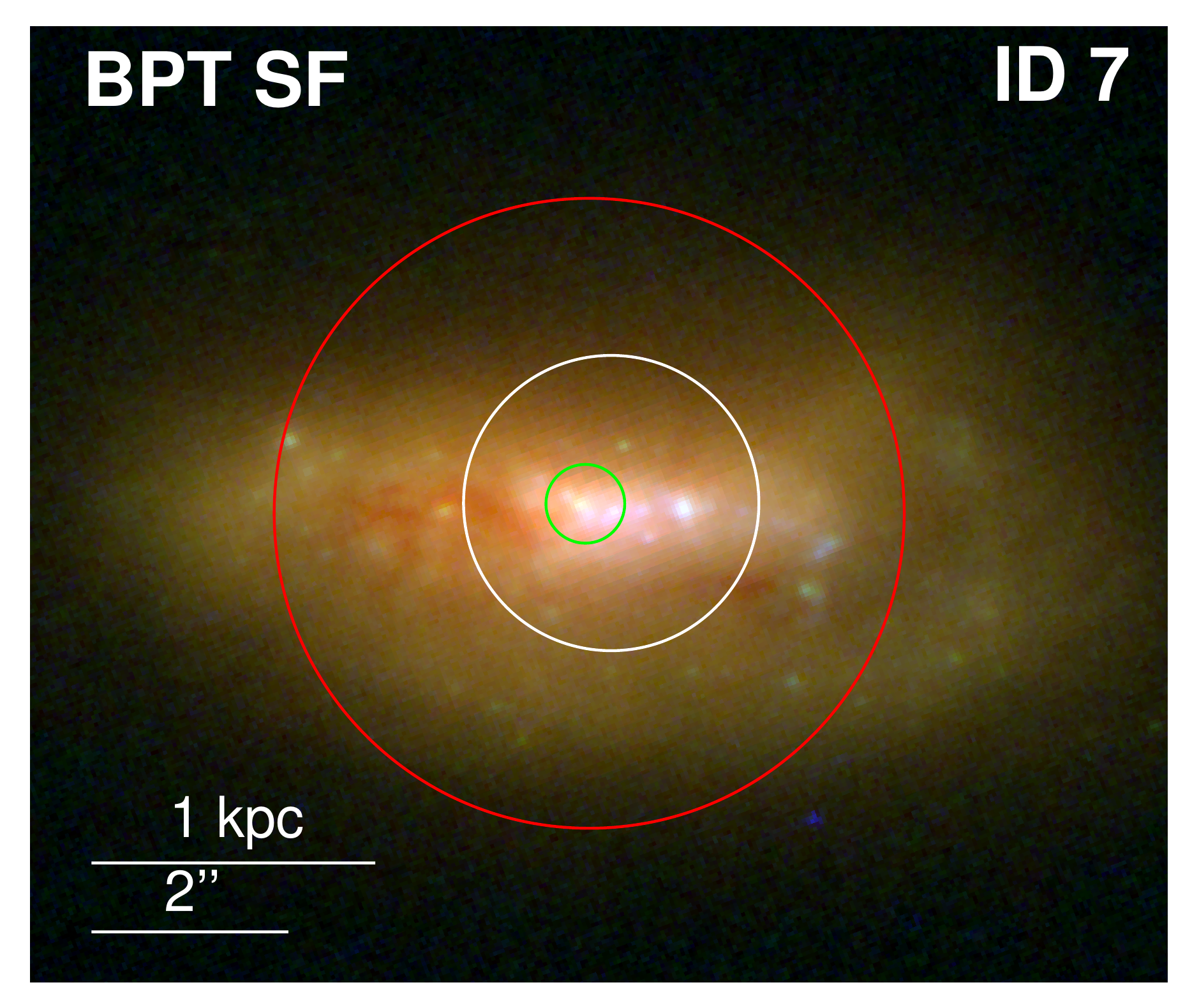}}\hfill
{\includegraphics[width=0.25\textwidth]{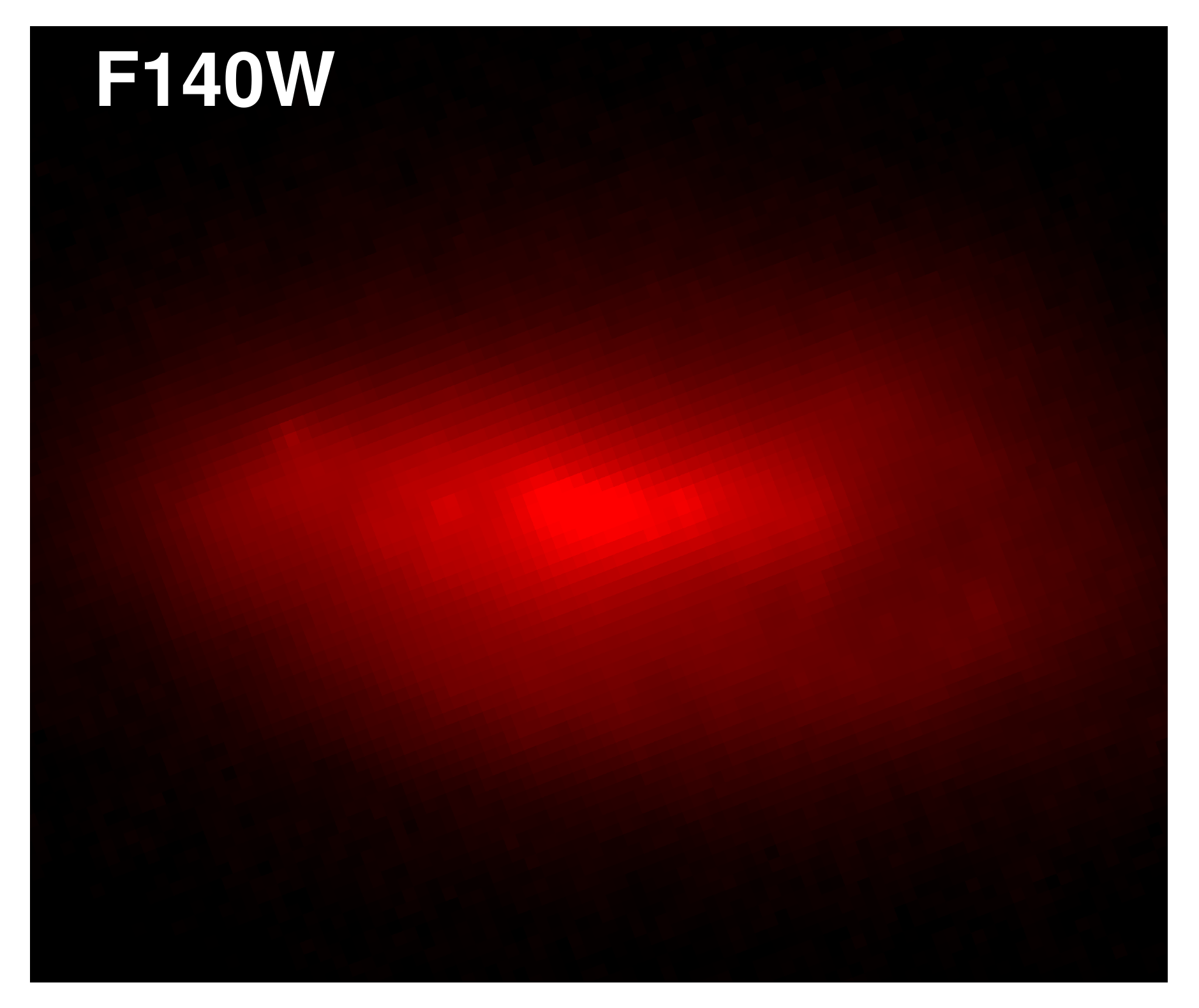}}\hfill
{\includegraphics[width=0.25\textwidth]{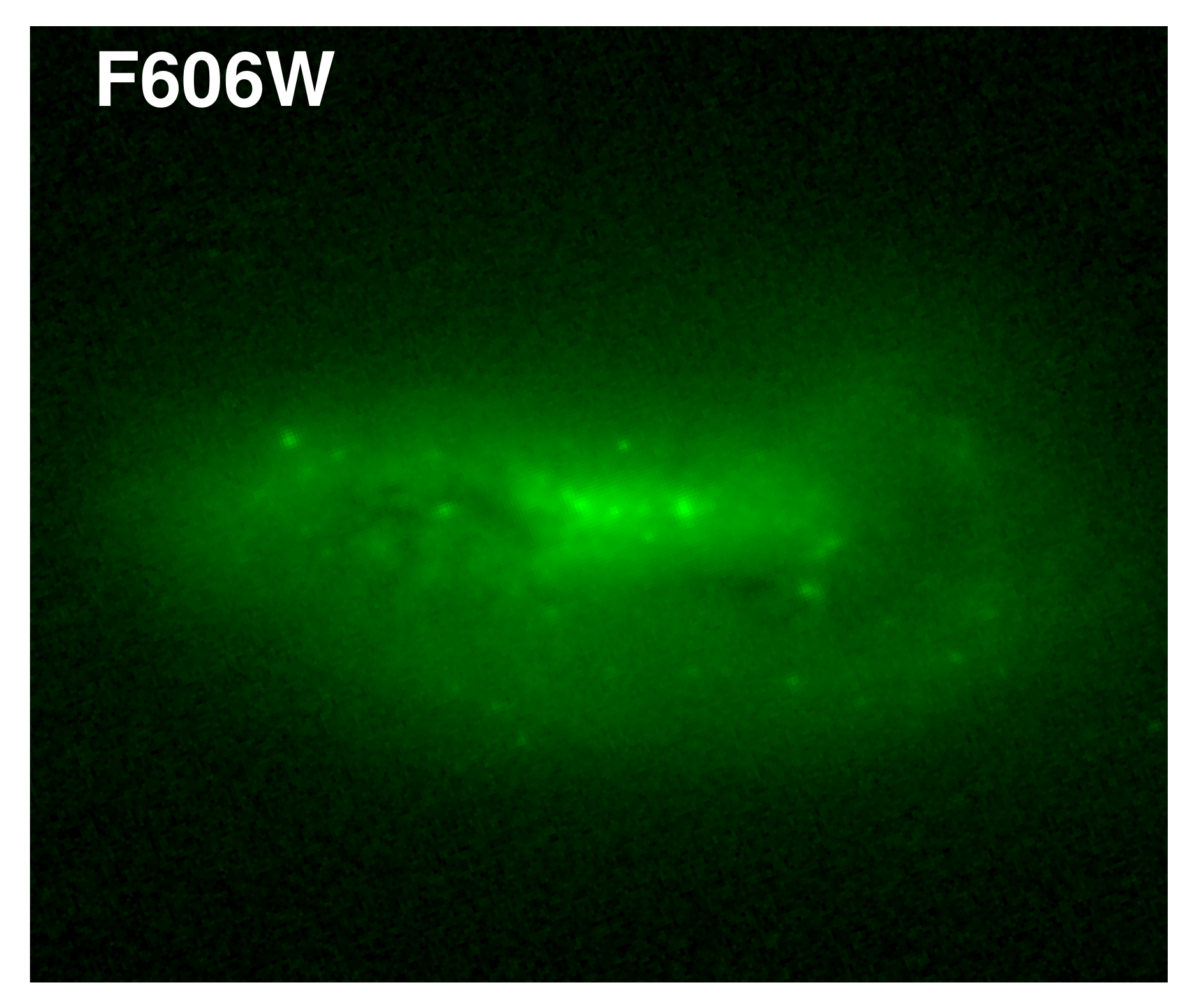}}\hfill
{\includegraphics[width=0.25\textwidth]{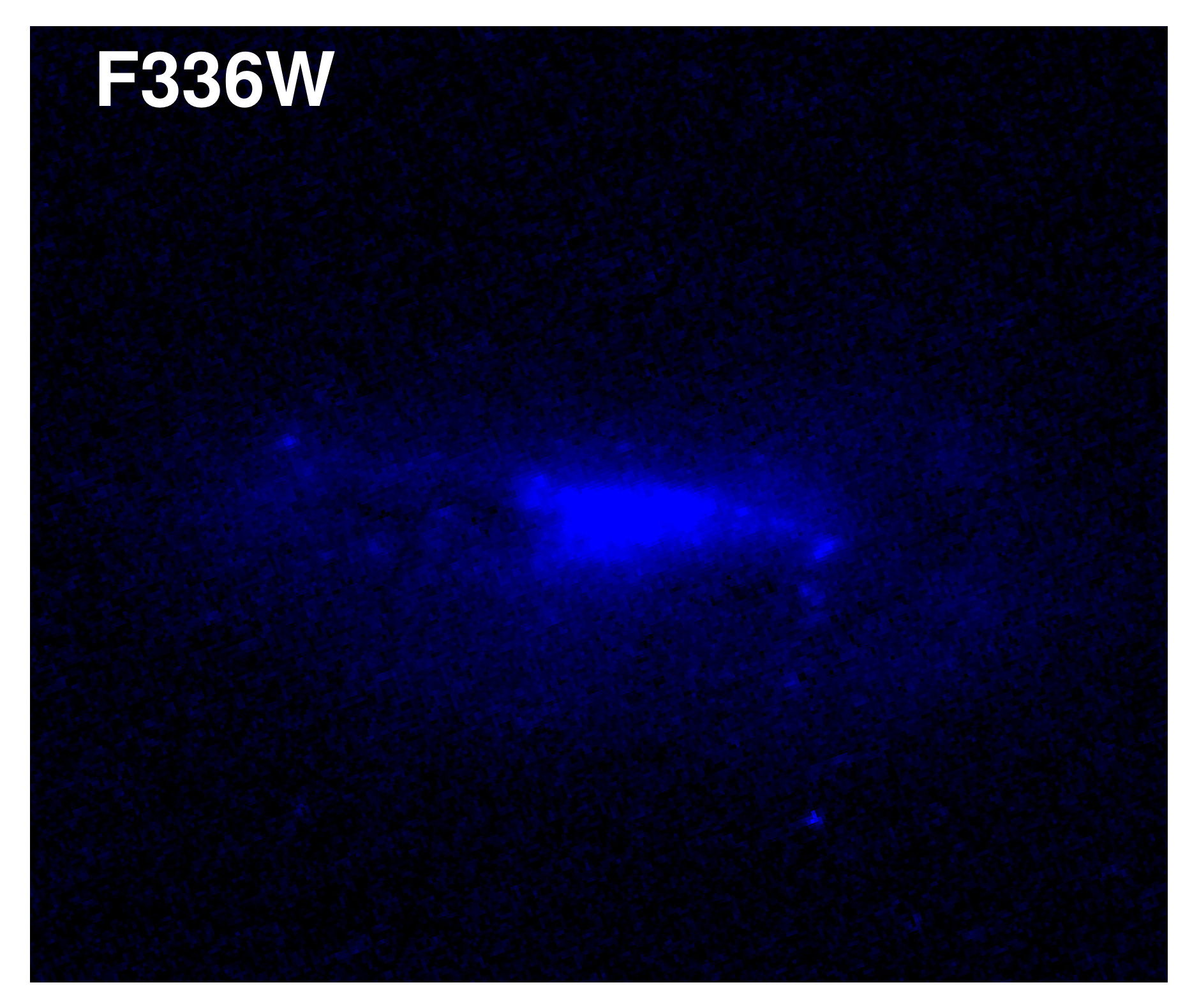}}\hfill

{\includegraphics[width=0.25\textwidth]{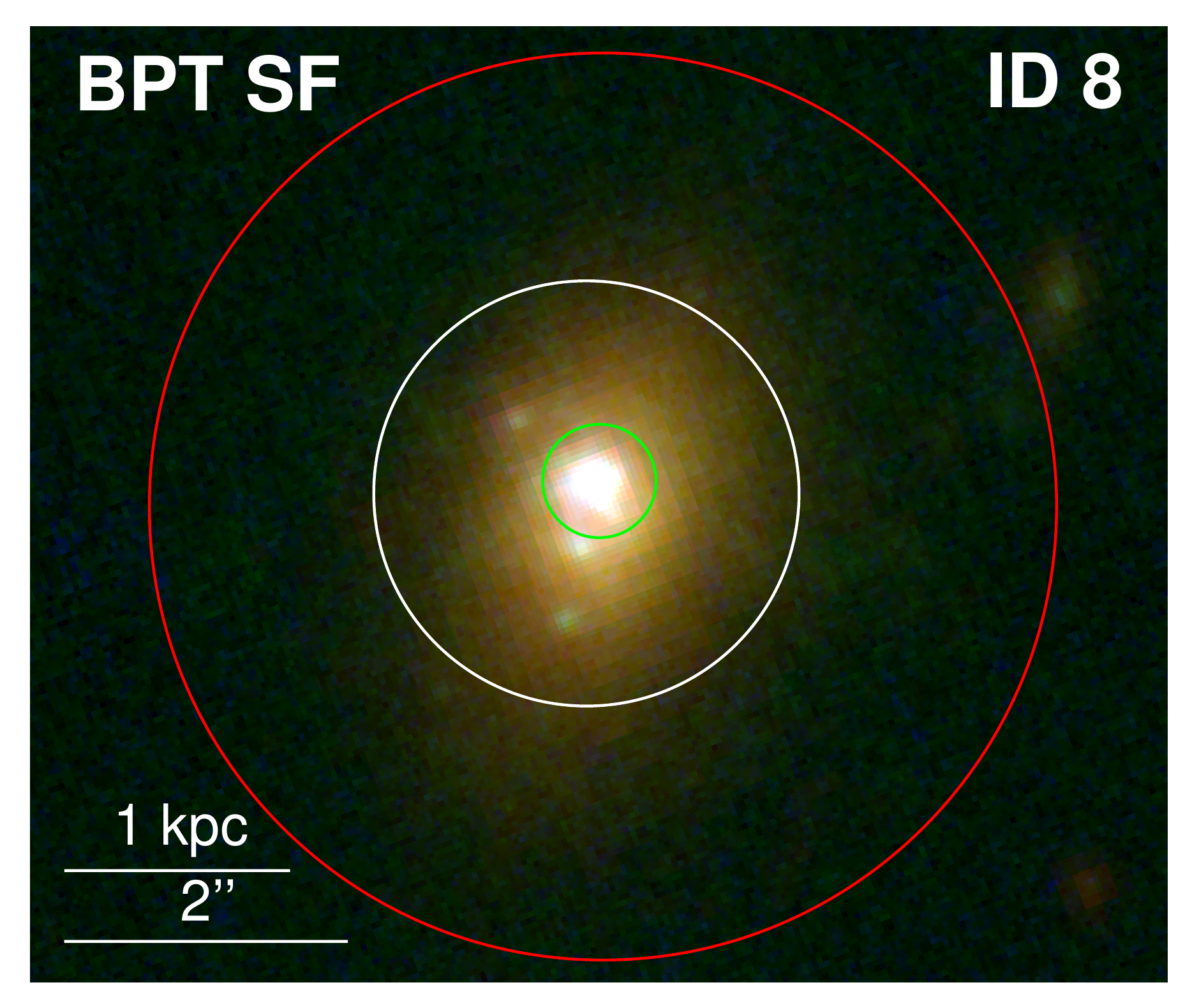}}\hfill
{\includegraphics[width=0.25\textwidth]{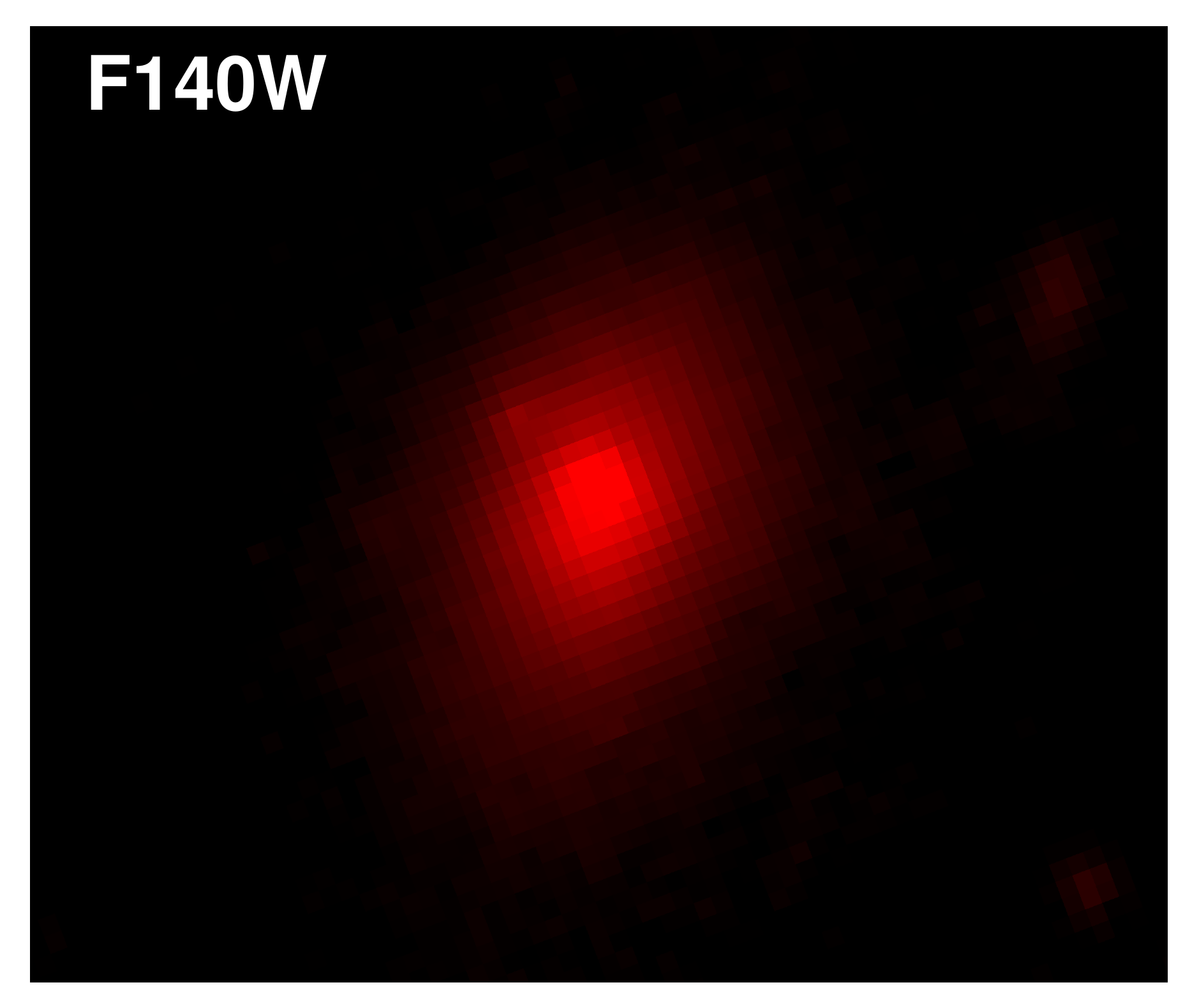}}\hfill
{\includegraphics[width=0.25\textwidth]{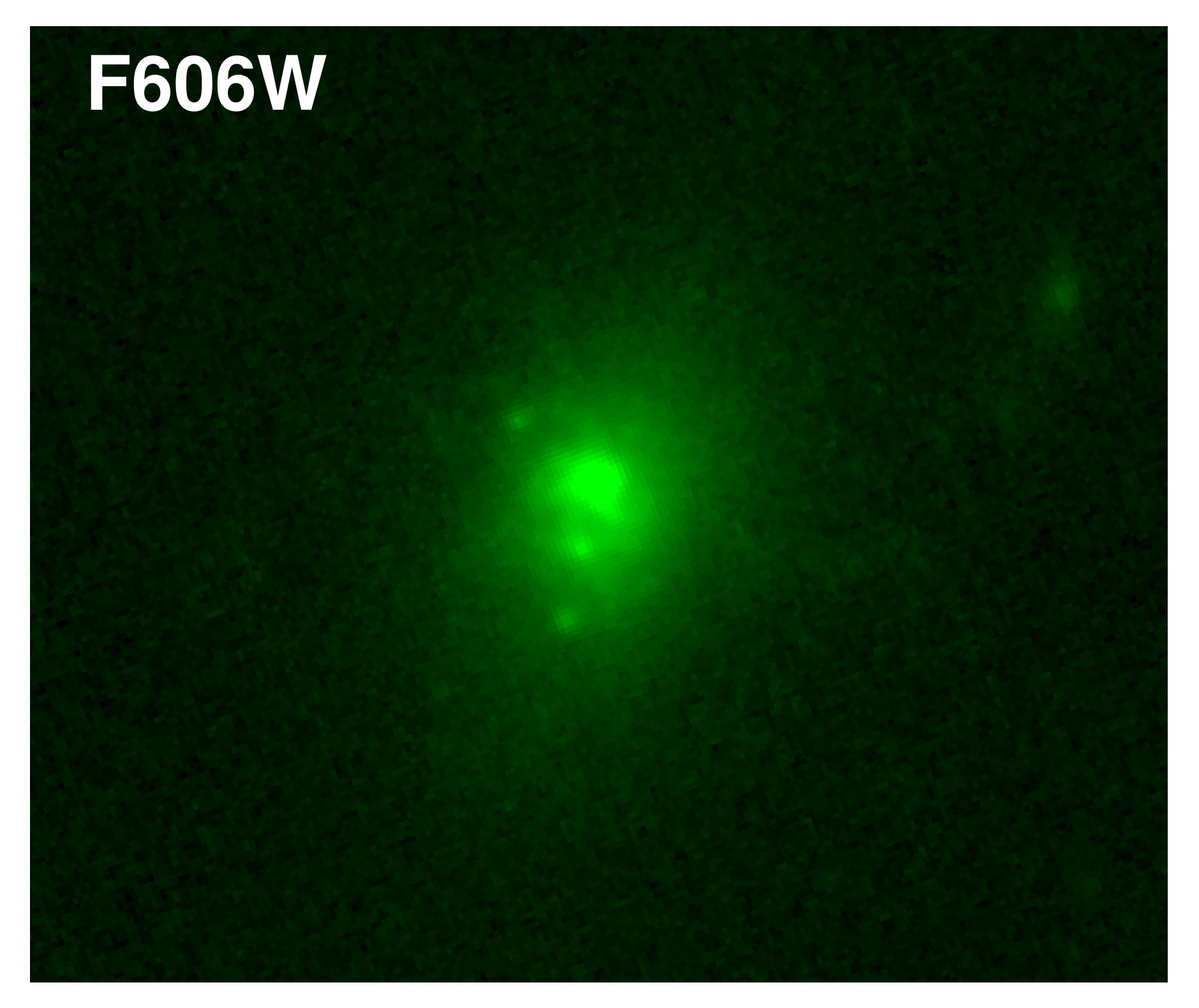}}\hfill
{\includegraphics[width=0.25\textwidth]{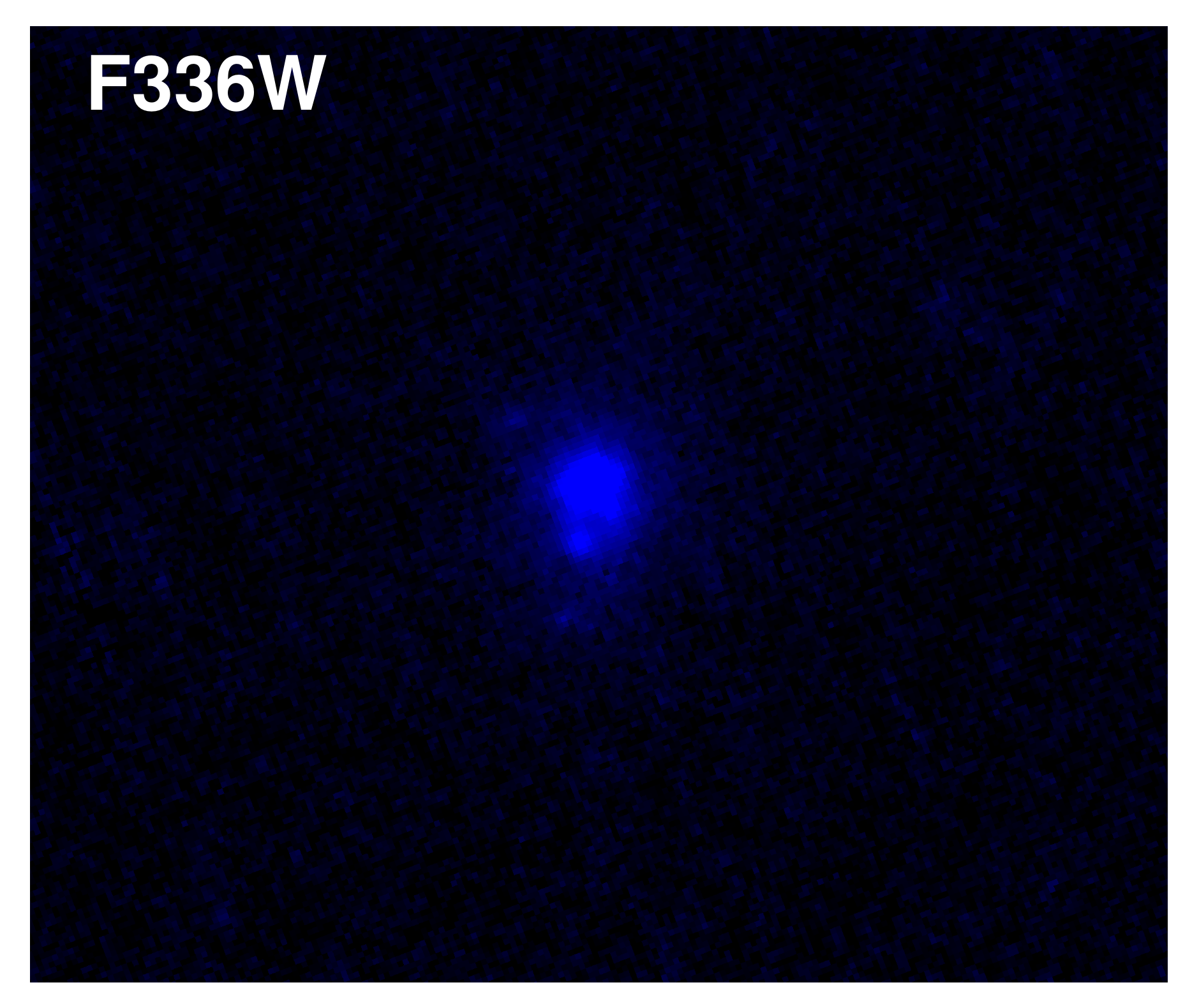}}\hfill

{\includegraphics[width=0.25\textwidth]{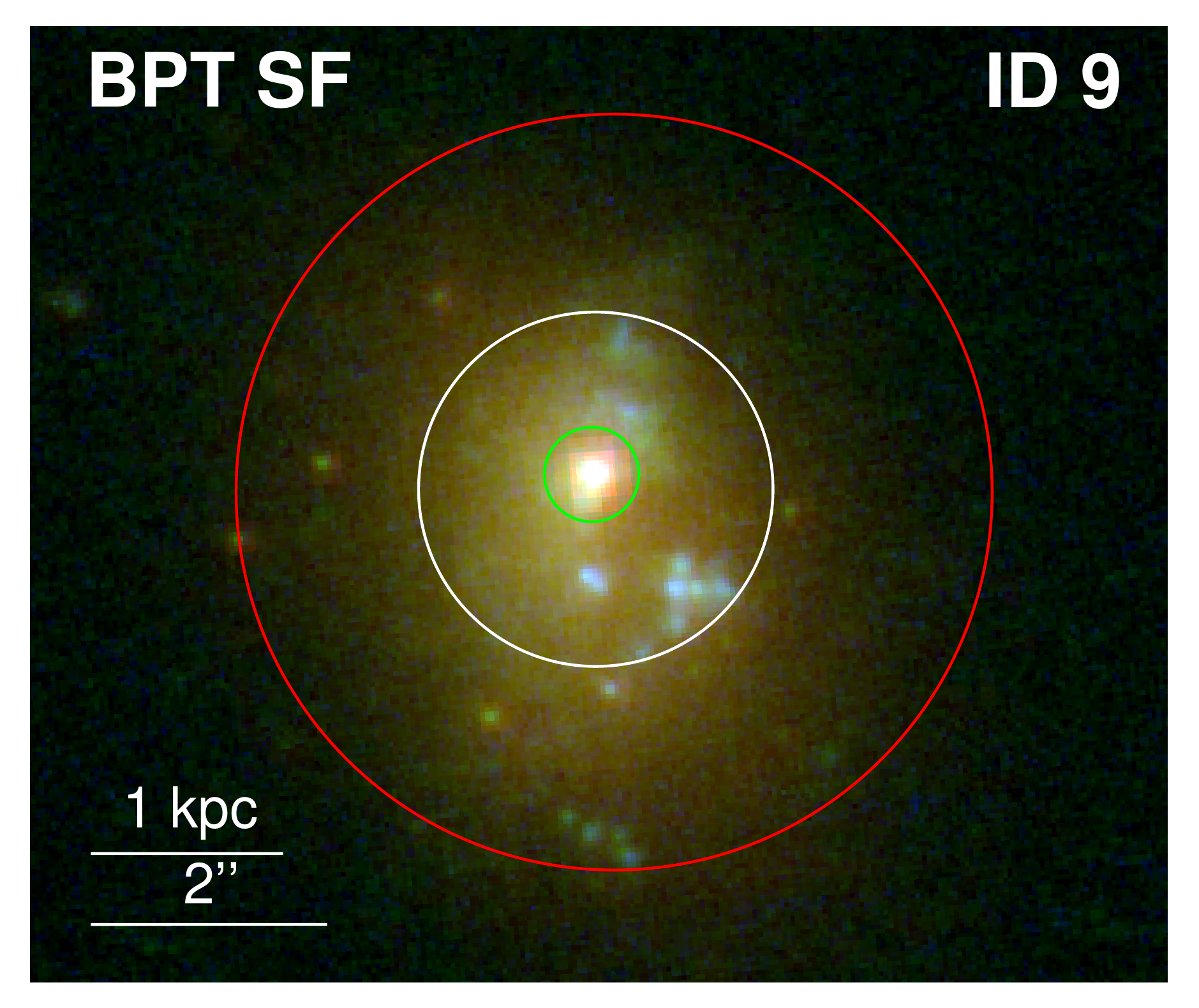}}\hfill
{\includegraphics[width=0.25\textwidth]{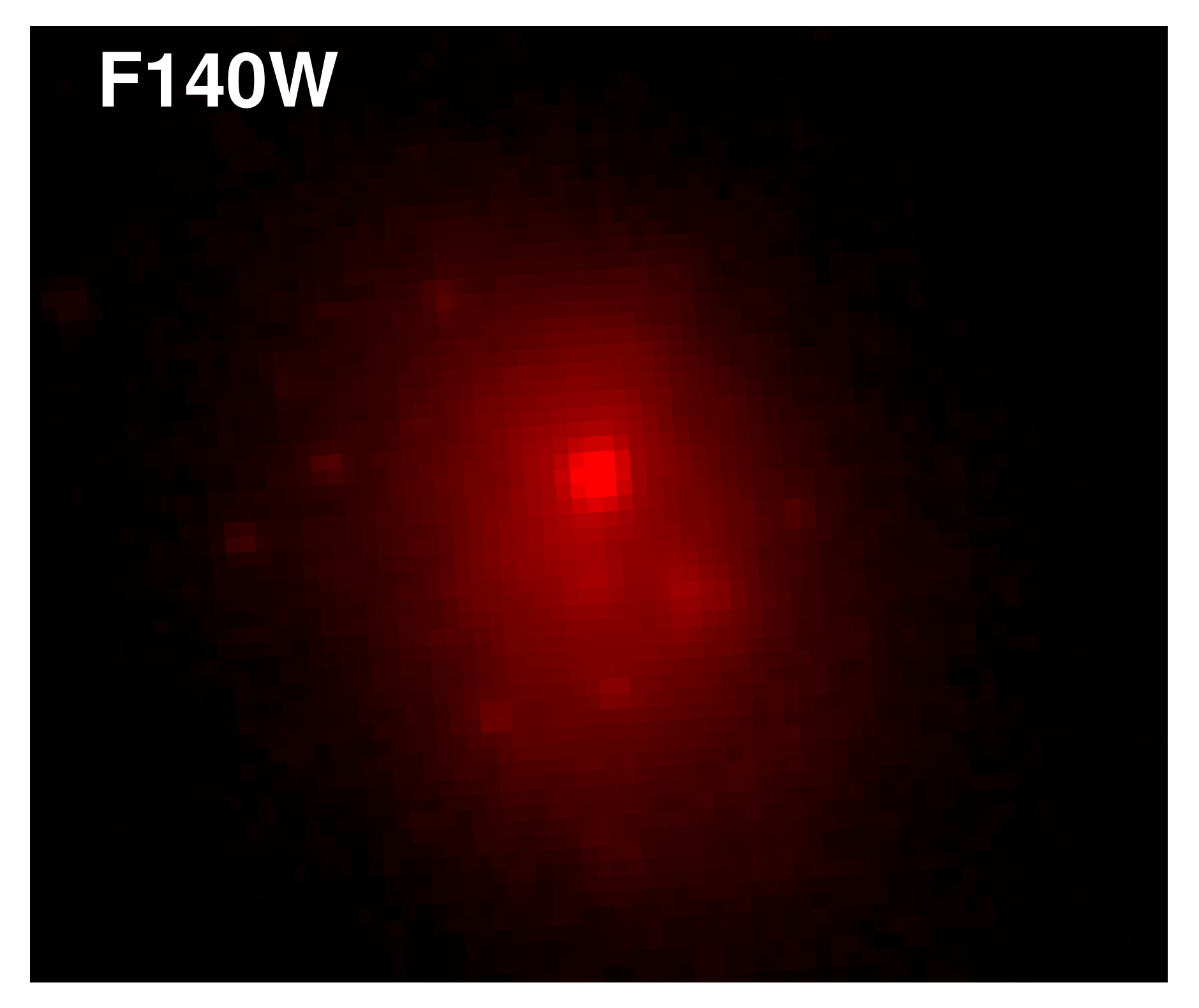}}\hfill
{\includegraphics[width=0.25\textwidth]{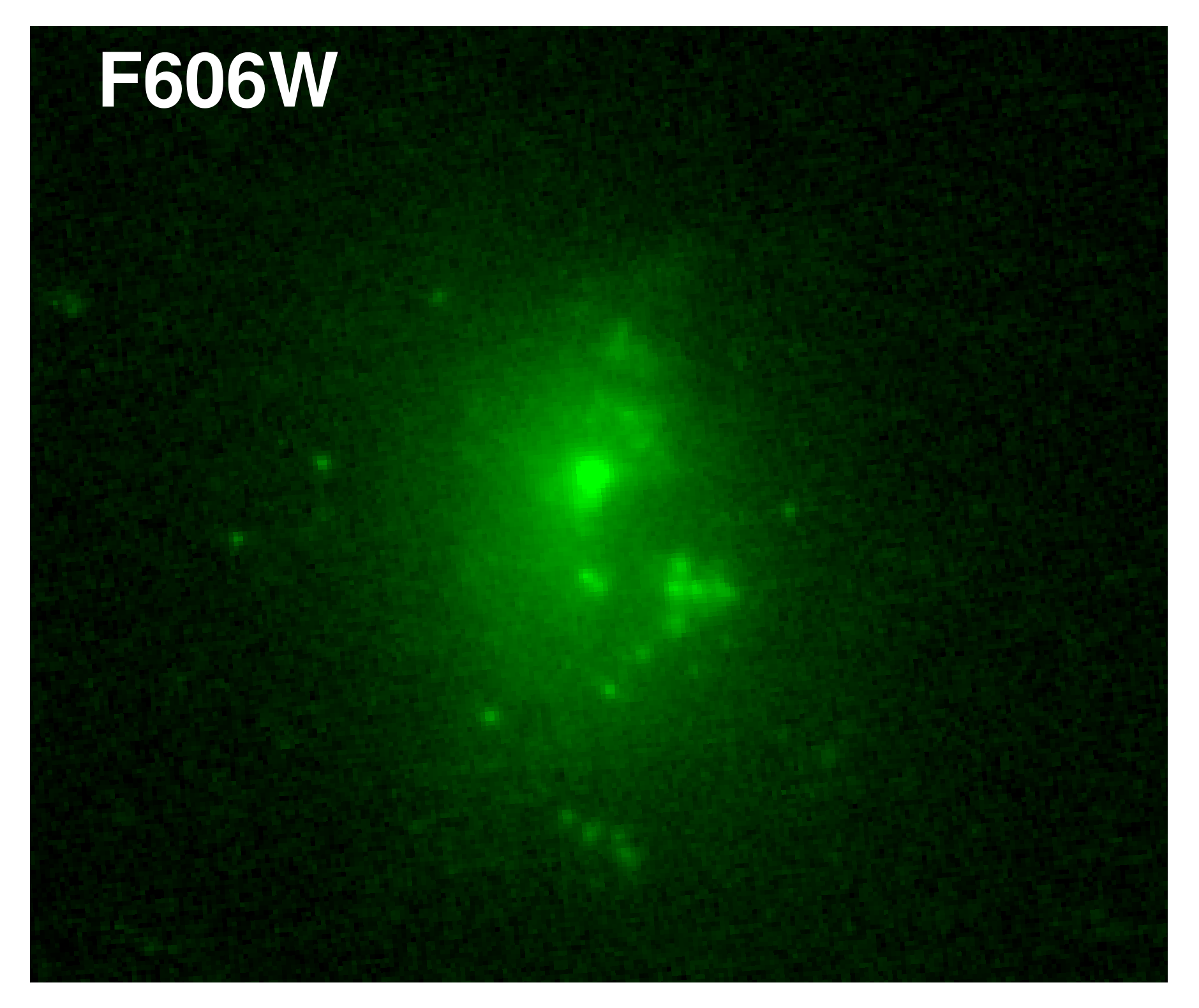}}\hfill
{\includegraphics[width=0.25\textwidth]{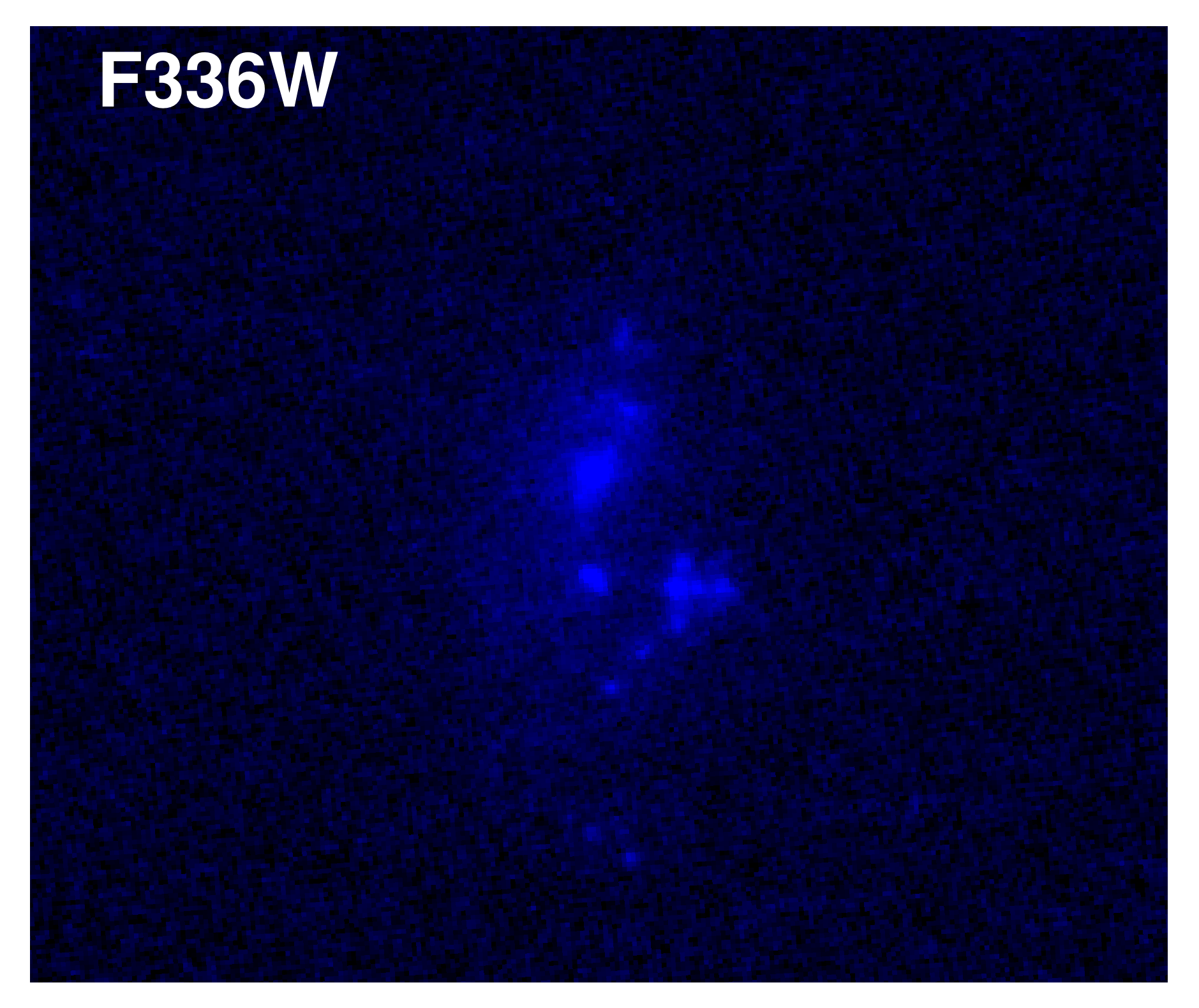}}\hfill

{\includegraphics[width=0.25\textwidth]{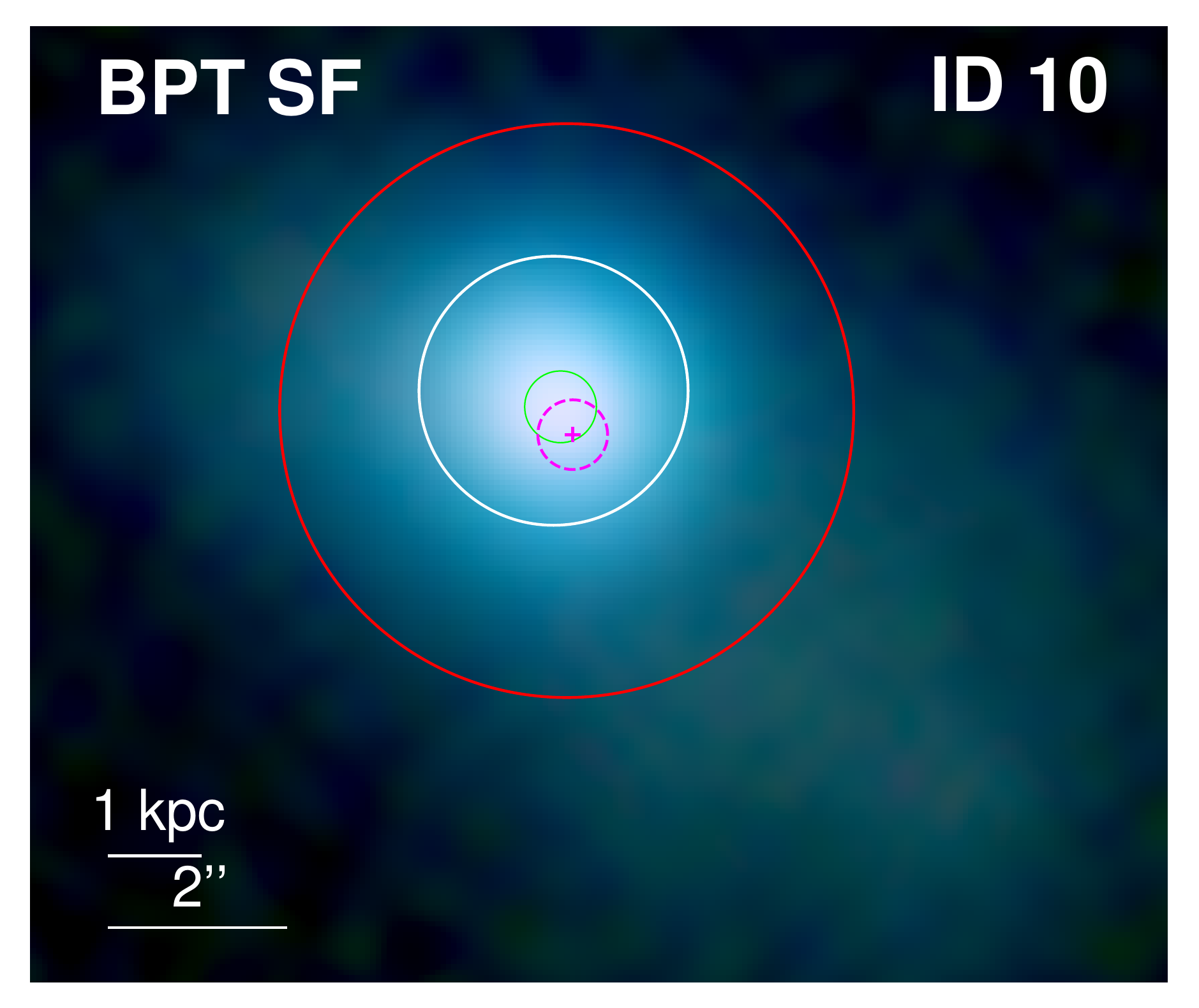}}\hfill
{\includegraphics[width=0.25\textwidth]{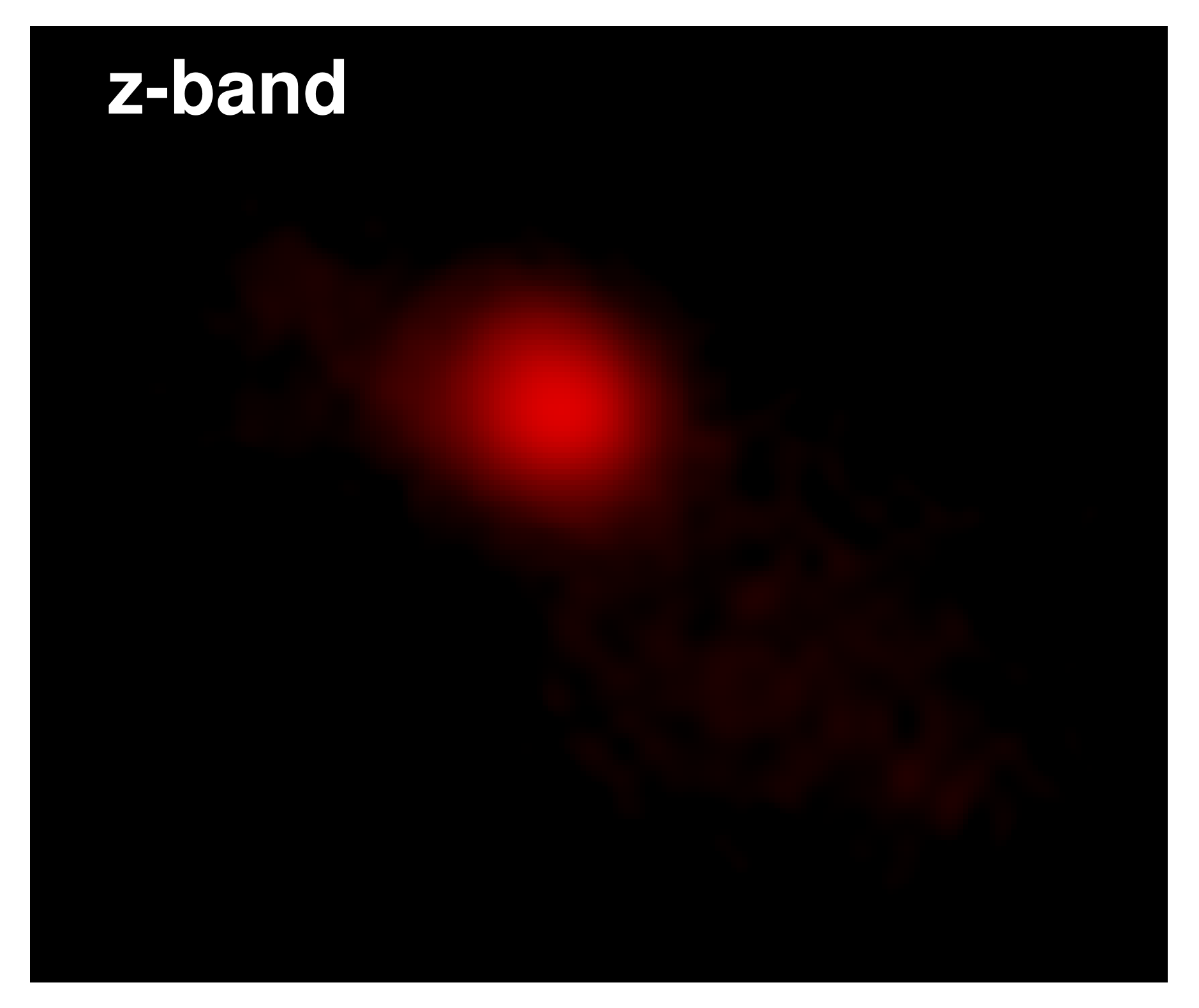}}\hfill
{\includegraphics[width=0.25\textwidth]{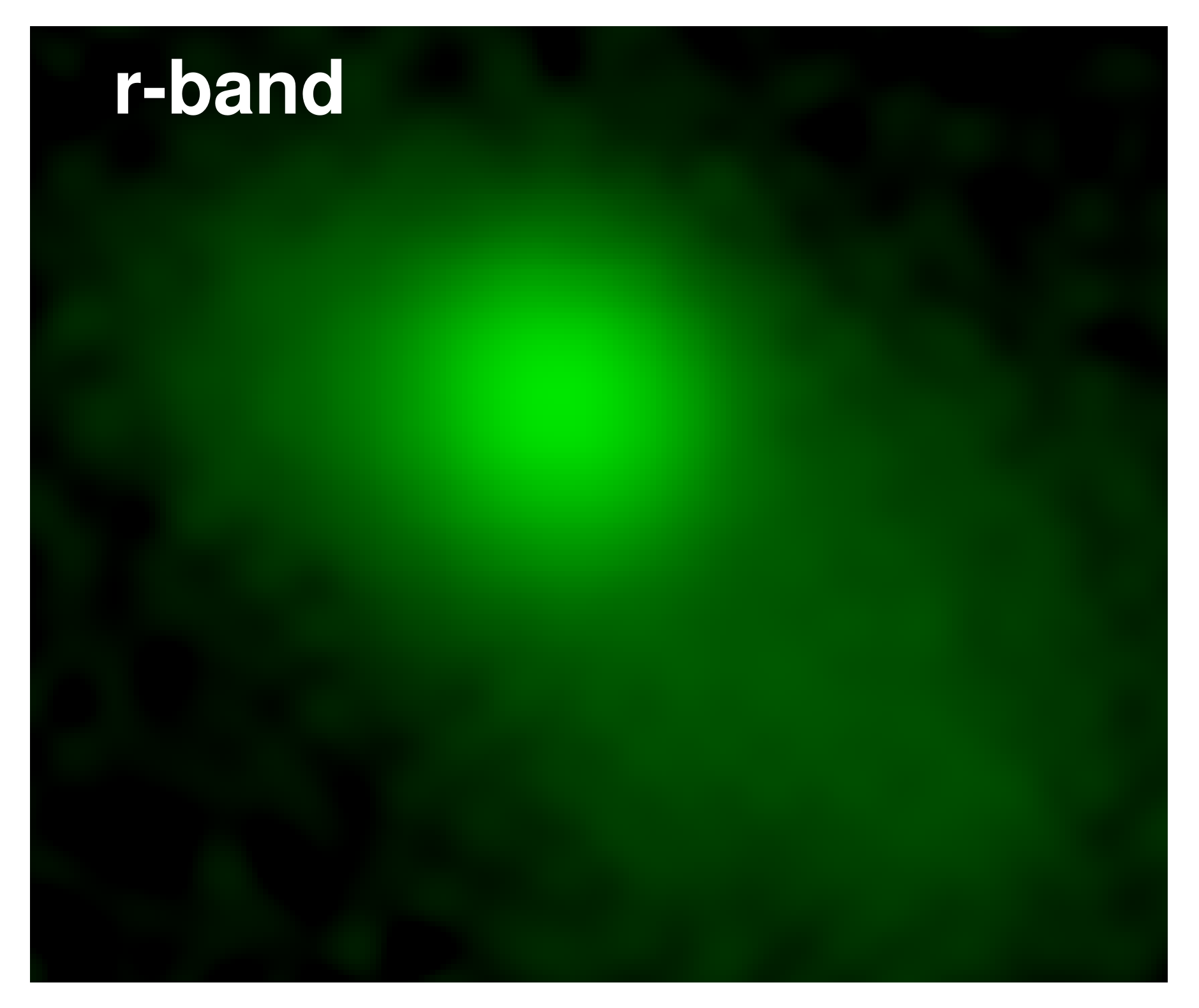}}\hfill
{\includegraphics[width=0.25\textwidth]{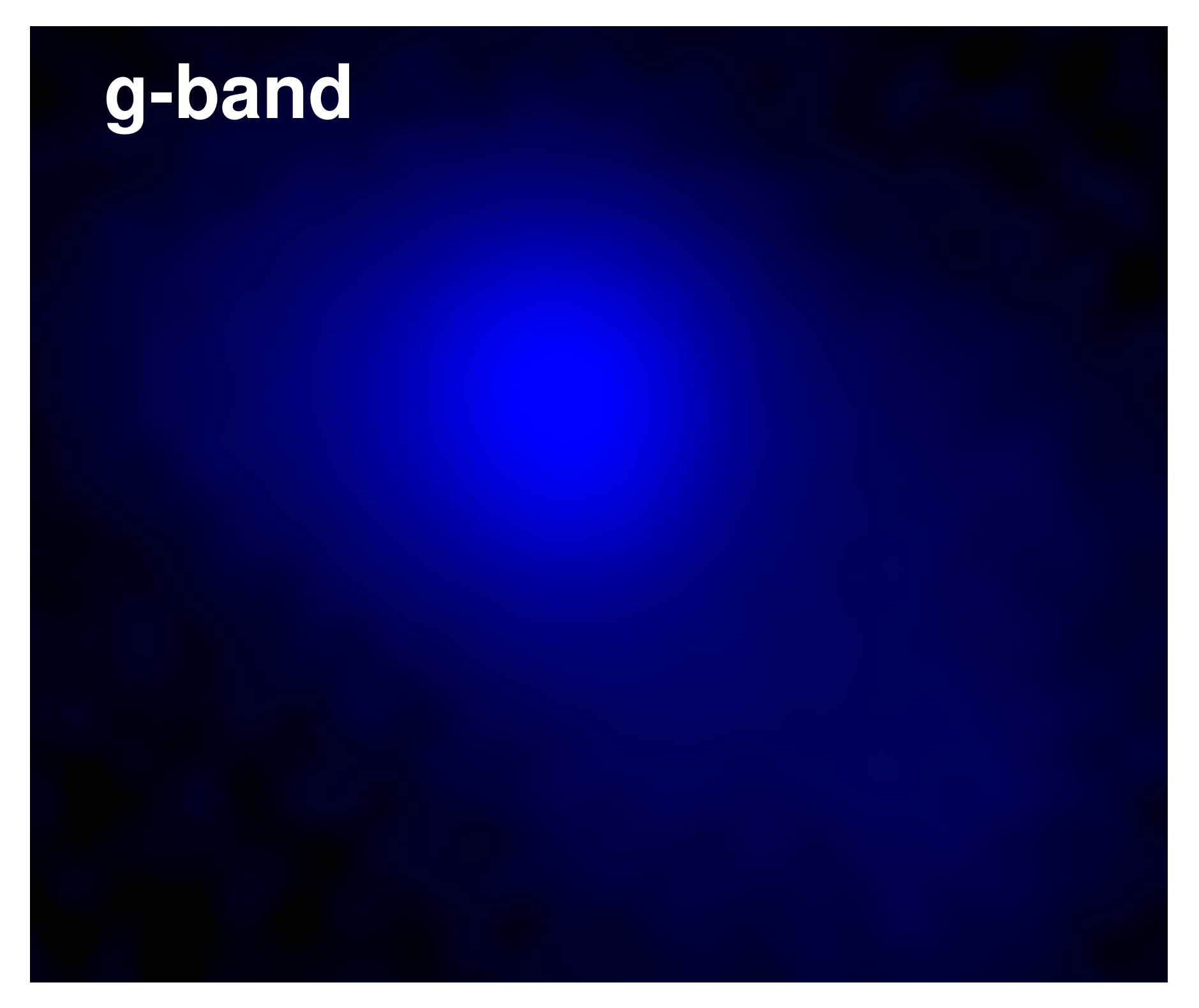}}\hfill

{\includegraphics[width=0.25\textwidth]{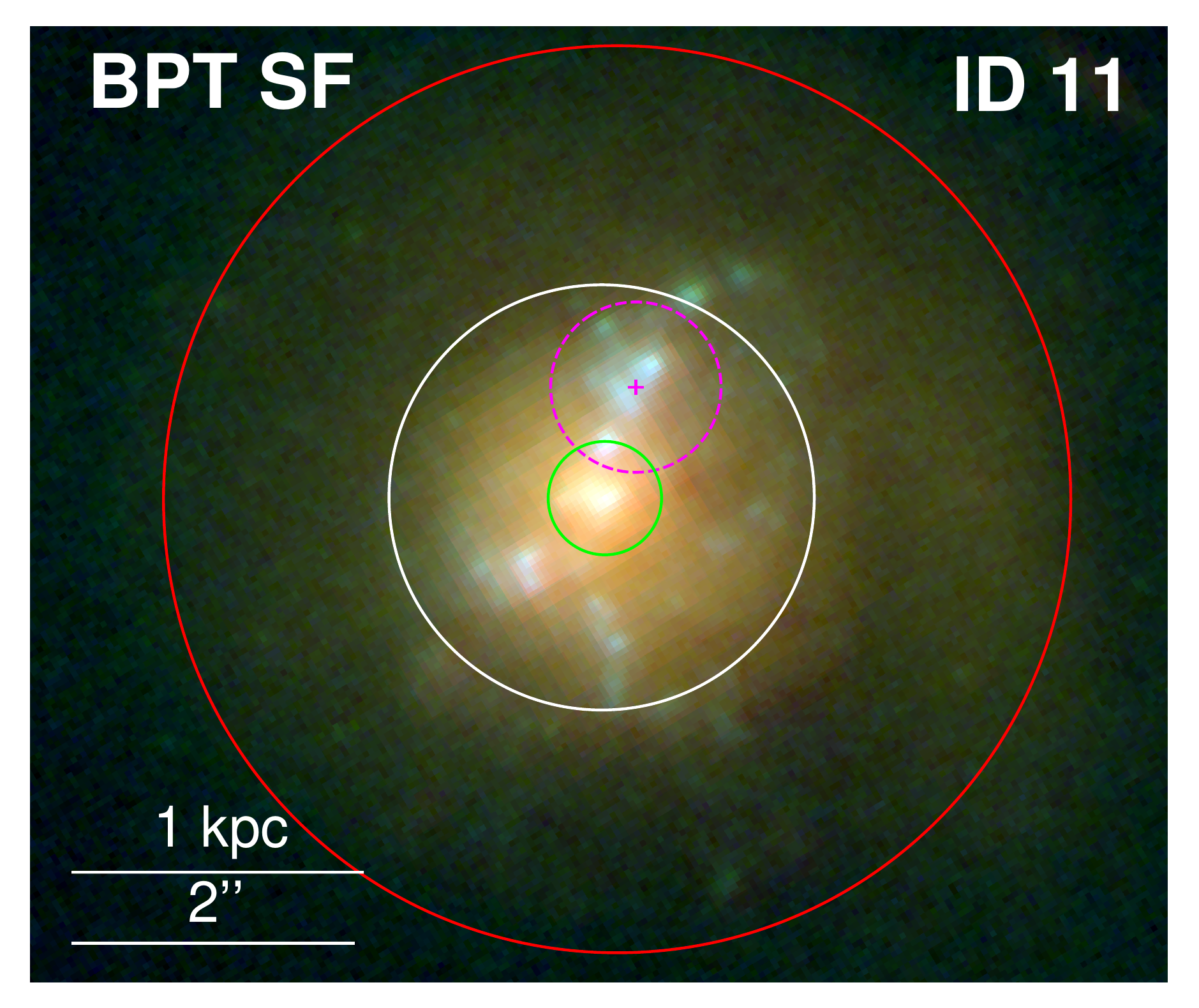}}\hfill
{\includegraphics[width=0.25\textwidth]{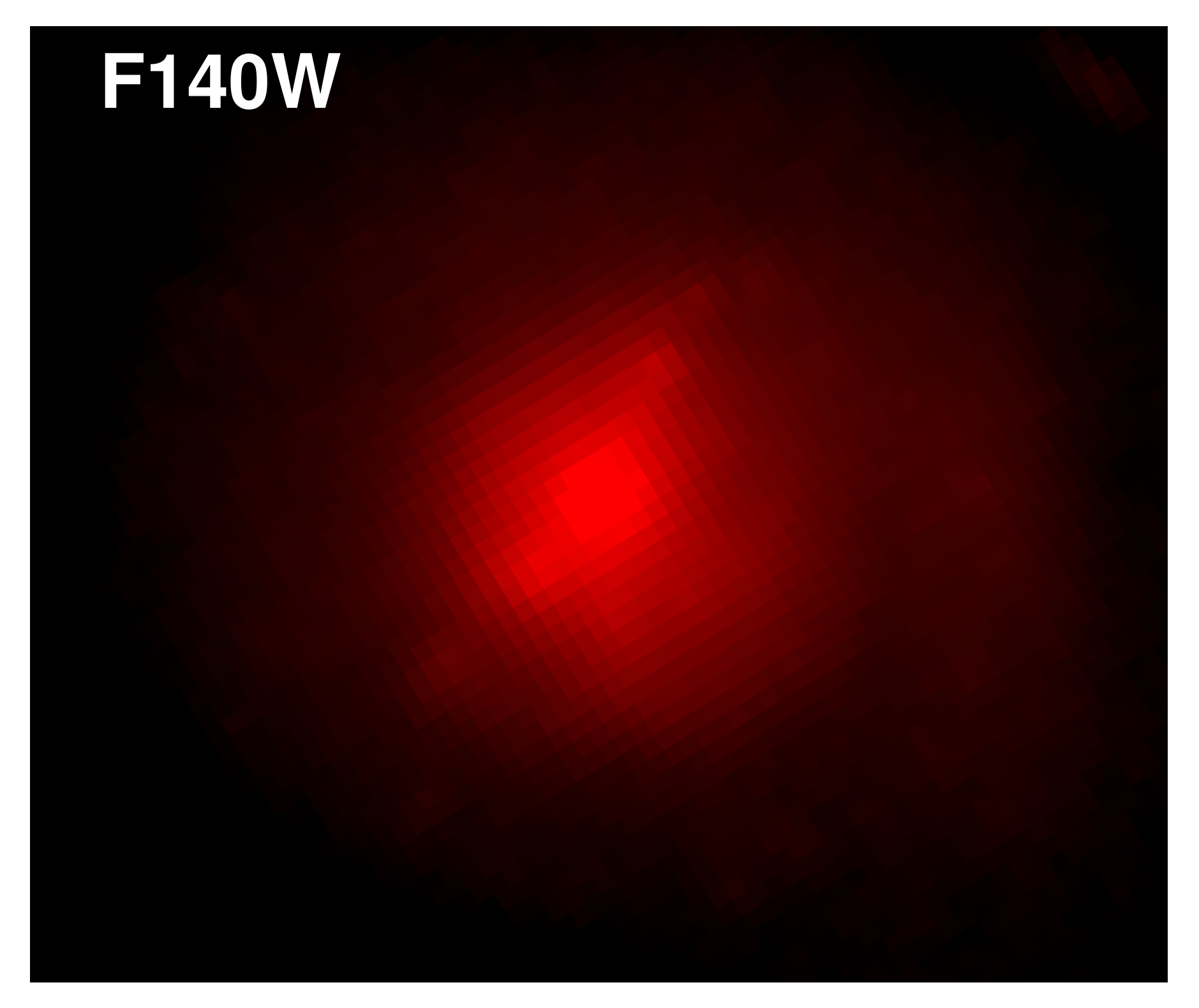}}\hfill
{\includegraphics[width=0.25\textwidth]{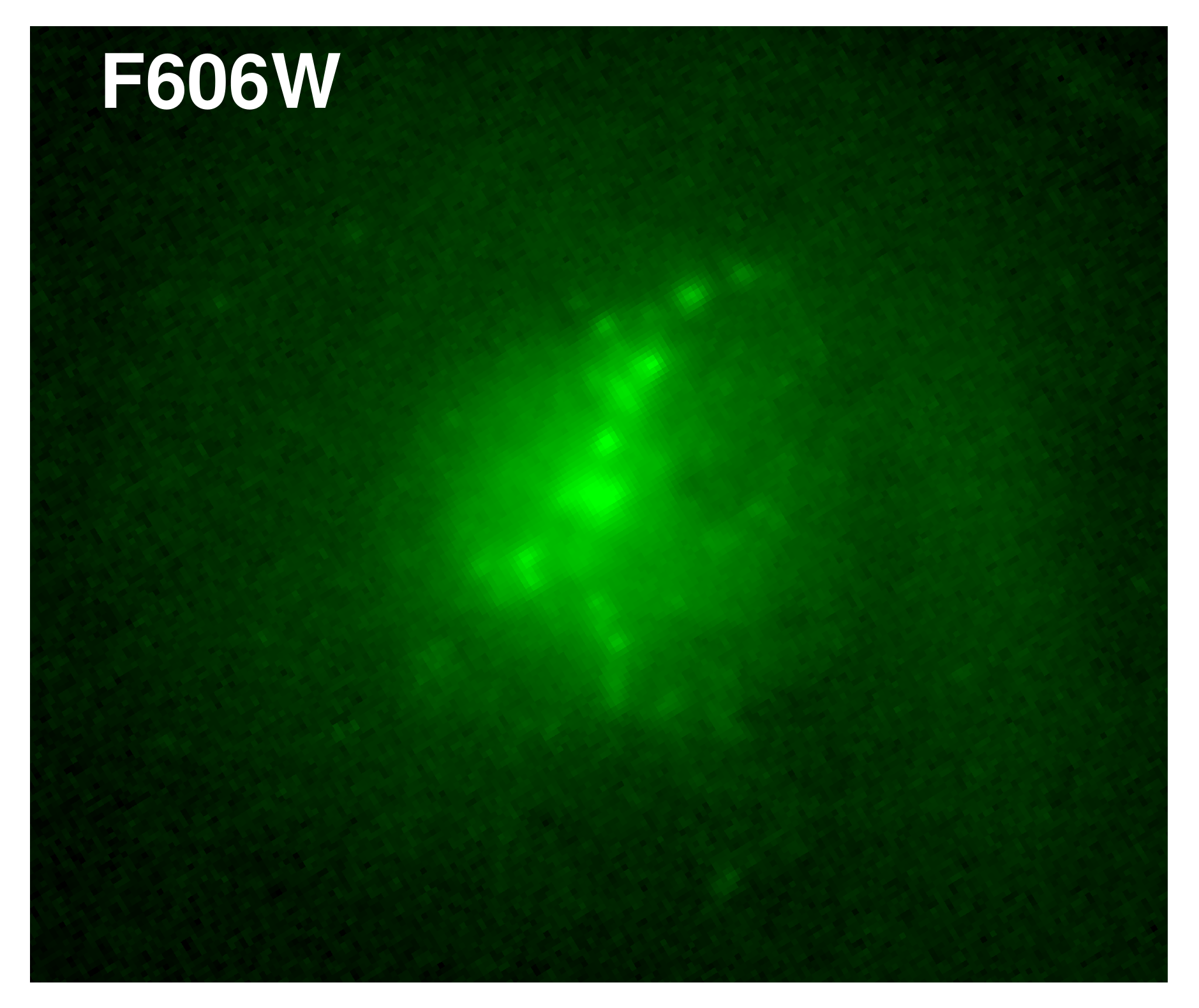}}\hfill
{\includegraphics[width=0.25\textwidth]{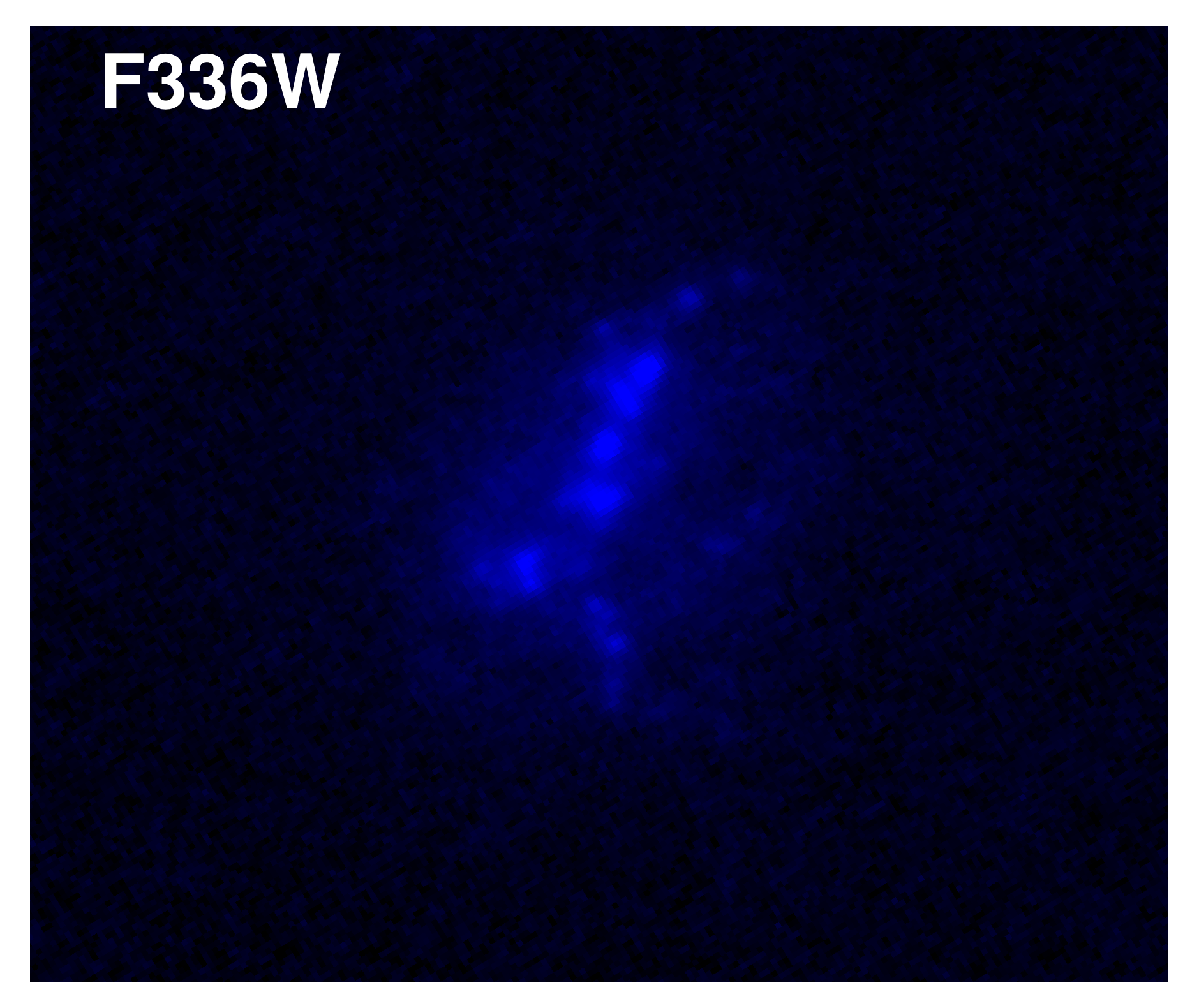}}\hfill

\caption{Same as above.
} 
\label{fig:rgb}
\end{figure*}

\section{Observations and Data Reduction} \label{sec:obstot}

\subsection{Chandra X-ray Observatory} \label{sec:xrayobs}

X-ray observations of our target galaxies were taken with \textit{Chandra} between 2014 Jan 07 and 2020 Jan 03, with exposure times ranging between 8.0 and 24.4 ks. A summary of the \textit{Chandra} observations is given in Table \ref{tab:cxo}.
In each observation, the target galaxy was placed at the center of the ACIS S3 chip. We used the \texttt{CIAO} software v4.11 \citep{fruscione06} to reprocess the data and apply calibration files (CALDB 4.8.4.1), creating new level 2 event files.

We next corrected the \textit{Chandra} astrometry by matching to SDSS catalog sources (DR12). We created a list of X-ray point sources by using the \texttt{CIAO} function \texttt{wavdetect}, a point source detection algorithm, on the S3 chip of the \textit{Chandra} image filtered from 0.5-7 keV. We excluded any \texttt{wavdetect} sources falling within 3$r_{50}$ of the galaxy, and used the \texttt{CIAO} function \texttt{wcs\_match} to match the remaining sources to optical point sources in the SDSS DR12 catalog with $i$-band magnitudes $<$ 22. We found between zero and five matching \texttt{wavdetect} and SDSS sources on the S3 chip for each observation. We only updated the astrometry if we found at least 1 match, and given the small number of matches, only applied a translation correction. This resulted in updated astrometry for six of the eleven galaxies (IDs 1, 2, 5-6, 8-9), with astrometric shifts ranging between $\pm$0.02-1.53 pixels ($0\farcs01$-$0\farcs75$). 
\begin{deluxetable}{clccc}
\tabletypesize{\footnotesize}
\tablecaption{\textit{Chandra} Observations}
\tablewidth{0pt}
\tablehead{
\colhead{ID}  & \colhead{Date observed}  & \colhead{Obs ID} & \colhead{Exp.\ time (ks)} & \colhead{$N_{\rm background}$} }
\startdata
1      & 2019 Mar 26 & 21446 & 14.9 & 0.0070 \\
2      & 2014 Dec 26 & 17033 & 15.8 & 0.0073 \\
3      & 2016 Apr 06 & 17034 & 8.0  & 0.0112 \\
4      & 2019 Mar 10 & 21447 & 24.4 & 0.0057 \\
5      & 2019 Aug 03 & 21448 & 21.8 & 0.0047 \\
6      & 2015 Feb 10 & 17037 & 10.9 & 0.0019 \\
7      & 2019 Sep 20 & 21449 & 9.9  & 0.0085 \\
8      & 2020 Jan 03 & 21450 & 20.8 & 0.0036 \\
9      & 2019 Dec 07 & 21451 & 19.8 & 0.0117 \\
10     & 2014 Jan 07 & 14908 & 10.3 & 0.0015 \\
11     & 2019 May 08 & 21452 & 12.9 & 0.0144 
\enddata
\tablecomments{$N_{\rm background}$ is the number of expected 2-10 keV $N(>S)$ background sources within $3r_{50}$, using \citet{moretti03}. }
\label{tab:cxo}
\end{deluxetable}

\subsection{Optical/{NIR} Images} \label{sec:optobs}

\textit{HST}/WFC3 images spanning ultraviolet to {NIR} wavelengths were taken between 2016 Feb 18 and 2019 Nov 02. The new data (galaxy IDs 1, 4-5, 7-9, and 11; Proposal 15607, PI: Reines) were taken in the F336W ($\sim U$-band), F606W ($\sim$ wide $V$-band), and F140W ($\sim H$-band) filters. The archival data (galaxy IDs 2-3 and 6; Proposal 13943, PI: Reines) were taken in the F275W (near-UV), F606W, and F110W ($\sim$ wide $YJ$-band) filters.   

Each galaxy was allocated one orbit, with exposure times 
of $\sim$7-9 minutes in the NIR, $\sim$11-12 minutes in the $V$-band and $\sim$12-15 minutes in the short wavelength filters. {
The images were processed using the AstroDrizzle routine in the standard \textit{HST} pipeline.} We adjust the astrometry of the \textit{HST} images to match that of the SDSS by comparing common sources between the two, resulting in astrometric shifts up to $0\farcs52$, with a median shift of $0\farcs22$.
{As galaxy ID 10 had no \textit{HST} imaging observations, we retrieved images from DECaLS in the $g$, $r$, and $z$ bands for use in Figure \ref{fig:rgb}.
}

\section{Analysis and Results} \label{sec:results}

\begin{deluxetable*}{cccccrrrr}
\tabletypesize{\footnotesize}
\tablecaption{X-ray Sources}
\tablewidth{0pt}
\tablehead{
\colhead{ID} & \colhead{R.A.} & \colhead{Decl.} & \multicolumn{2}{c}{Net Counts} & \multicolumn{2}{c}{Flux ($10^{-15}$ erg s$^{-1}$ cm$^{-2}$)} & \multicolumn{2}{c}{Luminosity (log(erg s$^{-1}$))} \\
\cmidrule(l){4-5} \cmidrule(l){6-7} \cmidrule(l){8-9}
\colhead{ } & \colhead{(deg)} & \colhead{(deg)} & \colhead{0.5-2 keV} & \colhead{2-7 keV} & \colhead{0.5-2 keV} & \colhead{2-10 keV} & \colhead{0.5-2 keV} & \colhead{2-10 keV} \\
\colhead{(1)} & \colhead{(2)} & \colhead{(3)} & \colhead{(4)} & \colhead{(5)} & \colhead{(6)} & \colhead{(7)} & \colhead{(8)} & \colhead{(9)}}
\startdata
1  	    	 & 50.602662 & 40.188872 & 6.04$^{+5.07}_{-3.46}$  & 10.58$^{+7.58}_{-4.98}$ & 4.15    & 16.21   & 39.8    & 40.4  \\
2 	    	 & 136.557572 & 56.170866 & 14.66$^{+8.30}_{-5.83}$ & 2.02$^{+3.62}_{-1.86}$  & 5.51    & 2.79    & 40.4    & 40.1  \\
3 	    	 & 148.575716 & 47.290371 & 2.07$^{+3.66}_{-1.76}$  & 2.11$^{+3.74}_{-1.83}$  & 1.69    & 5.76    & 39.6    & 40.1  \\
4\footnote{\label{minflux}No soft and/or hard band X-ray sources were detected in these galaxies. The reported fluxes and luminosities are upper limits (see Section \ref{sec:xrayemis}).} 	    	 & \nodata & \nodata & \nodata                 & \nodata                 & $<$0.74 & $<$2.03 & $<$39.6 & $<$40.0  \\
5 	    	 & 203.190100 & 26.580321 & 16.82$\pm 7.67$         & 4.09$^{+4.48}_{-3.11}$  & 6.60    & 4.18    & 40.5    & 40.3  \\
6 	    	 & 231.655677 & 6.994941 & 484.37$\pm 37.9$        & 127.75$ \pm 20.24$      & 271.89  & 255.72  & 41.9    & 41.9  \\
7\footref{minflux}    		 & \nodata & \nodata & \nodata                 & \nodata                 & $<$1.99 & $<$5.22 & $<$39.1 & $<$39.5  \\
8\footref{minflux}    		 & \nodata & \nodata & \nodata                 & \nodata                 & $<$0.96 & $<$2.42 & $<$39.3 & $<$39.7  \\
9\footref{minflux} 	    	 & \nodata & \nodata & \nodata                 & \nodata                 & $<$0.94 & $<$2.49 & $<$39.3 & $<$39.7  \\
10           & 263.755101 & 57.052294 & 13.36$^{+7.94}_{-5.50}$ & 6.48$^{+5.44}_{-3.70}$  & 6.73    & 13.58   & 40.5    & 40.8  \\
11\footref{minflux} 	    	 & 353.187593 & $-0.979194$ & \nodata                 & 3.12$^{+4.16}_{-2.48}$  & $<$1.43 & 5.40    & $<$39.2 & 39.8  
\enddata
\tablecomments{Column 1: galaxy ID. 
Column 2: right ascension of X-ray source.  
Column 3: declination of X-ray source. 
Columns 4-5: net counts after applying a 90\% aperture correction.  Error bars represent 90\% confidence intervals.
Columns 6-7: fluxes corrected for Galactic absorption.
Columns 8-9: log luminosities corrected for Galactic absorption; calculated using a photon index of $\Gamma = 1.8$.
}
\label{tab:xray}
\end{deluxetable*}

\subsection{X-ray Sources} \label{sec:xrayemis}

Using high-resolution X-ray observations from \textit{Chandra}, we search for X-ray point sources that could indicate the presence of accreting massive BHs in our target dwarf galaxies.
We first check our \textit{Chandra} images for background flares, removing time intervals where the background rate was $>3\sigma$. We then re-run \texttt{wavdetect} on images of the S3 chip filtered from 0.5-7 keV. We use wavelet scales of 1.0, 1.4, 2.0, 2.8, and 4.0, with a point spread function map of 39\% enclosed energy fraction at 2.3 keV (for hard band; 2-7 keV) and 1.56 keV (for soft band; 0.5-2 keV). We set a significance threshold of $10^{-6}$, at which we expect approximately one false source detection over the entire S3 chip. We then restrict further analysis to the \texttt{wavdetect} sources that lie within 3$r_{50}$ of the galaxy {to exclude any X-ray sources not associated with the target.}

We find source counts using circular apertures corresponding to the $90\%$ enclosed energy fraction at 4.5 keV (${\sim}2 \arcsec$ for our sources). We estimate background counts per pixel using circular annuli centered on the source with an inner radius equal to the source aperture radius and an outer radius of $12~\times $ the inner radius. We consider a source to be detected if the source counts are above the background counts in the source aperture to within a 95\% confidence level using the Bayesian methods in \cite{kraft91} for Poisson-distributed data with low source counts. All detected sources pass this test.

After subtracting off the background counts in the source aperture from the source counts, we apply a 90\% aperture correction to arrive at the net counts for each source. We detect a total of seven X-ray point sources across eleven galaxies, and report their properties in Table \ref{tab:xray}. Note that galaxy IDs 2-3 and 6 were analyzed in \cite{baldassare17}, and we find the same sources with similar luminosities.

We estimate 95\% positional uncertainties using the empirical formula from \cite{hong05} involving the \texttt{wavdetect} counts and off-axis position of the sources. 
The error bars on the net counts represent 90\% confidence intervals. If the source counts are $<10$, we take the background counts into account and follow the formalism of \cite{kraft91}. If the source counts are $\geq 10$ we use the confidence intervals from \cite{gehrels86}, which assume that the background counts are negligible. 

We calculate unabsorbed hard (2-10 keV) and soft (0.5-2 keV) X-ray fluxes using the \texttt{CIAO} function \texttt{srcflux}. We use a power-law spectral model with photon index $\Gamma = 1.8$, which is typical for low-luminosity AGN \citep{ho08,ho09} and ultraluminous X-ray sources at these energies \citep{swartz08}, and Galactic column densities from the \cite{dickey90} maps.  
The unabsorbed fluxes and corresponding luminosities are summarized in Table \ref{tab:xray}. We ignore any potential absorption intrinsic to the sources, so these values should be taken as lower limits.

Finally, we find the expected number of foreground/background hard (soft) X-ray sources to fall within 3$r_{50}$ of the galaxy. We use Equation 2 from \cite{moretti03} which, given an input flux, returns the expected number of background sources with that flux or higher that we would expect to see (per square degree). As our input fluxes, we use the minimum flux that we would classify as a source; we adopt 2 source counts, corresponding to the dimmest source we detected in each band. The resulting 2-10 keV (0.5-2 keV) minimum fluxes range from $S_{min} \sim 2\textrm{-}4.7$ ($0.7\textrm{-}2$) in units of $10^{-15}$ erg s$^{-1}$ cm$^{-2}$ for our eleven \textit{Chandra} observations. Using these fluxes, we find the expected number of hard band (soft band) background sources within 3$r_{50}$ of our galaxies range between ${\sim}$0.001-0.014 (0.001-0.011). 

 X-ray sources are detected in seven of our eleven target galaxies (at most one per galaxy) $-$ four BPT AGNs, one composite, and two star-forming galaxies. Six of the seven sources are consistent with being associated with the galaxy nucleus (see Section \ref{sec:optircntr} for further discussion). The 2-10 keV X-ray luminosities have a range of log$(L_{\rm 2-10 keV}/{\rm erg~s}^{-1}) = 39.8$--$41.9$ (see Table \ref{tab:xray}).  We use the aforementioned minimum fluxes to provide upper limits on X-ray source luminosities for the remaining four galaxies with non-detections (one BPT AGN and three star-forming galaxies).

\begin{figure}[h!]
\centering
\includegraphics[width=0.48\textwidth]{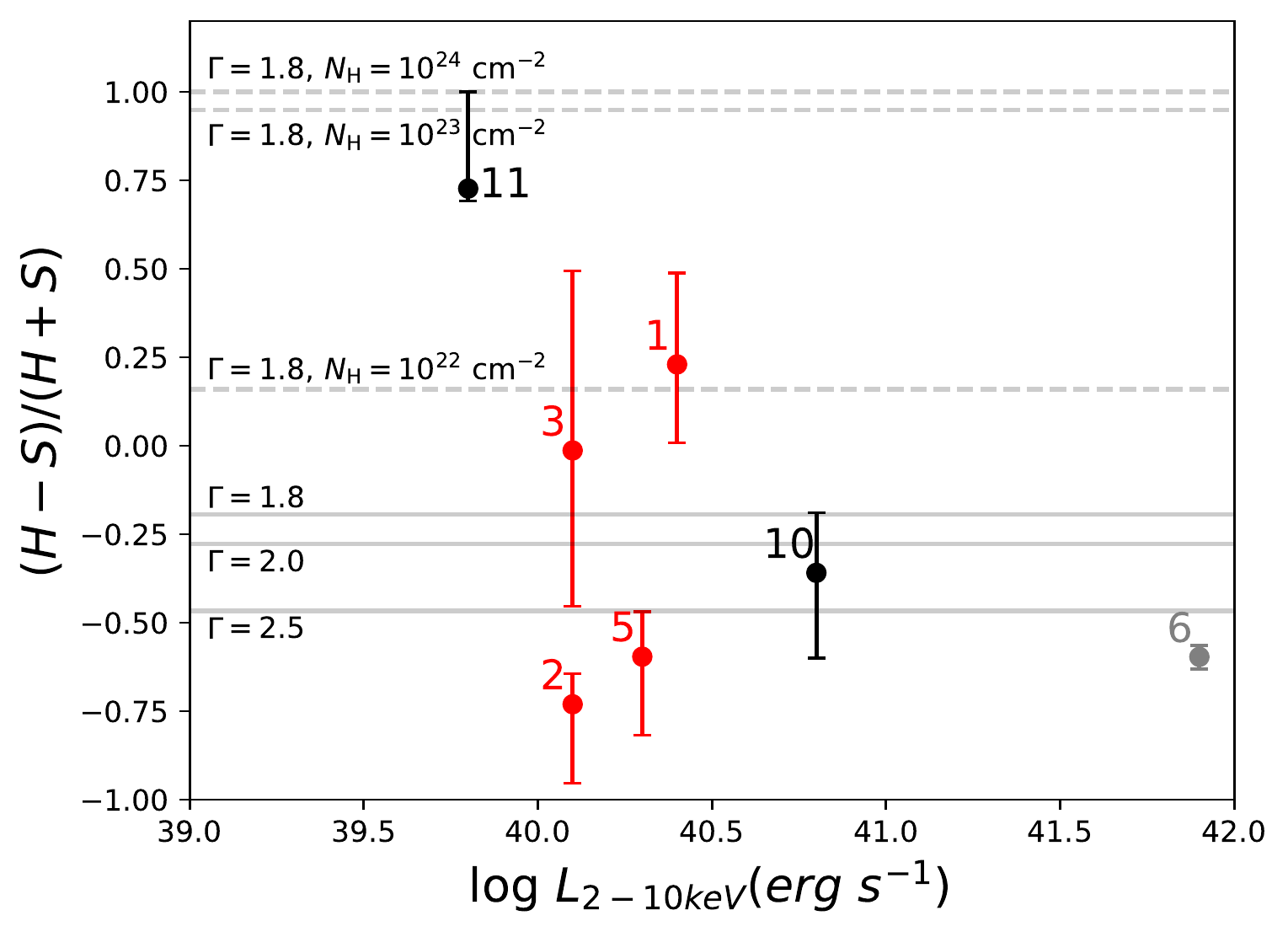}
\caption{ Hardness ratio vs.\ log 2-10 keV X-ray luminosity for the seven of our galaxies that had a detected X-ray source in at least one of the two bands. The hardness ratio was calculated using BEHR (see Section \ref{sec:xrayemis}). The error bars are the 68\% confidence intervals. Galaxies classified as AGN, composite, and star-forming are in red, grey, and black, respectively, based on the \NII/\ha~BPT diagram. Hardness ratios for unabsorbed power laws with $\Gamma=1.8$, 2.0, and 2.5 are shown as grey solid lines, while power laws with various absorptions and $\Gamma = 1.8$ are shown as dashed grey lines.
}
\label{fig:xrayhrratio}
\end{figure}

As most of our X-ray sources have a low number of counts, we estimate hardness ratios using the Bayesian Estimation of Hardness Ratios code \citep[BEHR;][]{park06}, which is useful in the Poisson regime of low counts and works even if only one of the bands has a detection. Hardness ratio here is defined as $(H-S)/(H+S)$, where $H$ and $S$ are the number of detected counts in the hard and soft X-ray bands, respectively. Here, the soft and hard bands {correspond} to the energy ranges 0.5-2 keV and 2-7 keV, respectively. We have seven galaxies with detections in at least one of the two bands, and the resulting hardness ratios are displayed in Figure \ref{fig:xrayhrratio}. We also estimate the hardness ratios using the Portable, Interactive Multi-Mission Simulator (PIMMS)\footnote{\url{https://heasarc.gsfc.nasa.gov/cgi-bin/Tools/w3pimms/w3pimms.pl}} for absorbed power laws with $\Gamma = 1.8$ and $N_{\rm H} = 10^{22-24}$ cm$^{-2}$ and for unabsorbed power laws with $\Gamma = 1.8$, 2.0, and 2.5. We plot these in Figure \ref{fig:xrayhrratio} as well.

\begin{figure}[h!]
\centering
\includegraphics[width=0.48\textwidth]{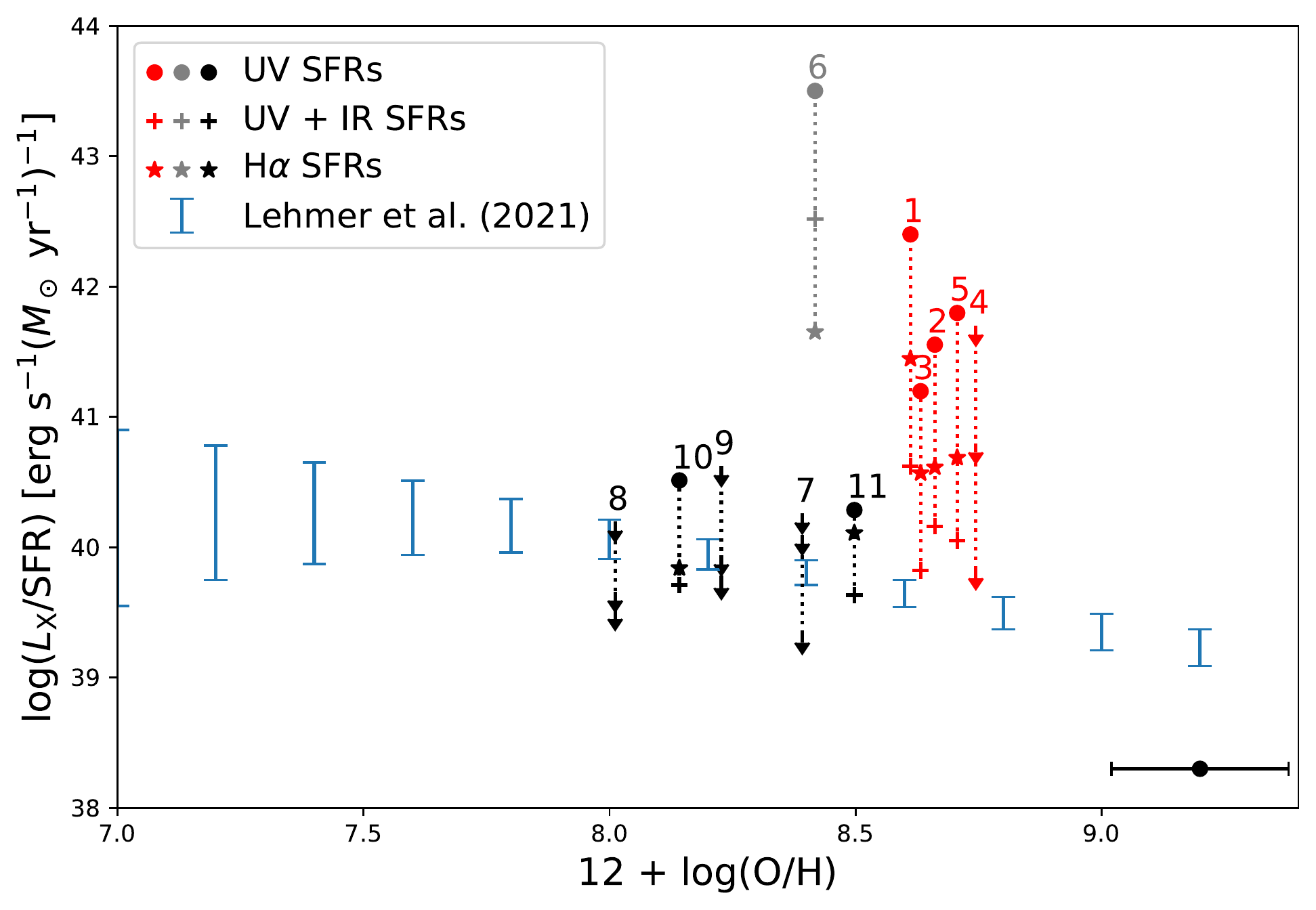}
\caption{ Log ratio of X-ray luminosity over host galaxy SFR vs.\ metallicity. 
{The crosses represent SFRs calculated with the mid-IR-corrected FUV luminosities, the dots use the uncorrected FUV luminosities (see Equation \ref{eq:sfrs}), and the stars use the \ha~derived SFRs - the three are connected by a dotted line for ease of comparison. }
SFRs based on the mid-IR-corrected FUV luminosities may be severely overestimated if the mid-IR emission is indeed dominated by an AGN as suggested by the {\it WISE} colors. {Similarly, the \ha~SFRs are also likely overestimated for the BPT AGNs/Composites since the AGN is contributing to or dominating the line ratios.} Arrows indicate upper limits on the X-ray luminosities for galaxies with no significant hard X-ray source detections (calculated from the minimum fluxes). Galaxies classified as AGN, composite, and star-forming are in red, grey, and black, respectively (using the \NII/\ha~BPT diagram). The values and error bars in blue are the expected X-ray luminosity from XRBs from \cite{lehmer21}. The ${\sim} 0.18$ uncertainty in the metallicities is shown in black in the lower right-hand corner.}
\label{fig:expxrbalt}
\end{figure}

\begin{deluxetable*}{cccccccc}
\tabletypesize{\footnotesize}
\tablecaption{Host Galaxy SFRs and Expected Luminosity from XRBs}
\tablewidth{0pt}
\tablehead{
\colhead{ID} & \colhead{FUV} & \colhead{log L(FUV)}  & \colhead{W$_{22}$} & \colhead{log L(22 $\mu$m)} & \colhead{SFR$^{\rm Cor}$} & \colhead{SFR$^{\rm Uncor}$} & \colhead{$12+$log O/H}  \\
\colhead{ } & \colhead{(mag)} & \colhead{(erg s$^{-1}$)} & \colhead{(mag)} & \colhead{(erg s$^{-1}$)} & \colhead{(\msun~yr$^{-1}$)} & \colhead{(\msun~yr$^{-1}$)} & \colhead{} \\
\colhead{(1)} & \colhead{(2)} & \colhead{(3)} & \colhead{(4)} & \colhead{(5)} & \colhead{(6)} & \colhead{(7)} & \colhead{(8)}}
\startdata
1  	    	 &  21.70  & 41.3 & 6.72  & 42.5 & 0.60 & 0.01 & 8.61 \\
2 	    	 &  20.04  & 42.5 & 6.06  & 43.3 & 3.46 & 0.14 & 8.66 \\
3 	    	 &  19.86  & 42.2 & 5.94  & 43.0 & 1.90 & 0.08 & 8.63 \\
4    		 &  23.62  & 41.1 & 7.83  & 42.6 & 0.74 & 0.01 & 8.74 \\
5    		 &  20.60  & 42.3 & 5.76  & 43.4 & 4.46 & 0.08 & 8.71 \\
6 	    	 &  20.62  & 42.1 & 7.84  & 42.4 & 0.48 & 0.05 & 8.42 \\
7 	    	 &  18.13  & 42.4 & 5.44  & 42.7 & 0.92 & 0.11 & 8.39 \\
8 	    	 &  18.99  & 42.6 & 7.05  & 42.5 & 0.76 & 0.16 & 8.01 \\
9 	    	 &  20.07  & 42.1 & 7.50  & 42.3 & 0.44 & 0.06 & 8.23 \\
10 	    	 &  16.96  & 43.7 & 4.62  & 43.9 & 15.5 & 2.44 & 8.14 \\
11           &  17.90  & 42.8 & 6.04  & 42.7 & 1.17 & 0.26 & 8.50 
\enddata
\tablecomments{Column 1: Galaxy identification number used in this paper.  
Column 2: FUV AB magnitudes from \textit{GALEX} (in the NSA).
Column 3: log FUV luminosities.
Column 4: \textit{WISE} magnitudes (W4, 22 $\mu$m). 
Column 5: log 22 $\mu$m luminosities.
Column 6: estimated SFRs from {\it GALEX} and {\it WISE} data.
Column 7: estimated SFRs using only the uncorrected FUV luminosities.
Column 8: metallicity estimated using the relation from \cite{pettini04}.
}
\label{tab:sfrs}
\end{deluxetable*}

\subsection{Expected Contribution from X-ray Binaries} \label{sec:xrbs}

It can be helpful to establish a rough baseline of how much X-ray luminosity we would expect to see coming from each galaxy in the absence of any AGN $-$ i.e. from X-ray binaries (XRBs).     
This expected X-ray luminosity scales with stellar mass for low-mass XRBs \citep{gilfanov04} and with SFR for high-mass XRBs \citep{grimm03}. Applying the relation from \cite{gilfanov04} between expected X-ray luminosity and stellar mass to our galaxies, we expect a luminosity of at most ${\sim}10^{38}$ erg s$^{-1}$ from low-mass XRBs. As this is ${\sim}1.5$ orders of magnitude smaller than the lowest observed X-ray luminosity, we conclude that low-mass XRBs are unlikely to significantly contribute to the luminosity of any of our observed X-ray sources; thus, we focus on high-mass XRBs.

We compare our observed X-ray luminosities to those expected from high-mass XRBs using the model of \cite{lehmer21}, which relates the expected XRB X-ray luminosity, SFR, and gas-phase metallicity. Figure \ref{fig:expxrbalt} shows the median values and $16-84\%$ ranges of $L_{\rm X}/{\rm SFR}$ as a function of 12 $+$ log(O/H) from their model (i.e., column 5 of table 3 in \citealt{lehmer21}).
Here, $L_{\rm X}$ refers to X-ray luminosities in the 0.5-8 keV energy band. {To this end, we re-analyze the \textit{Chandra} data for our galaxies in the 0.5-8 keV band, using the same methodology as in Section \ref{sec:xrayemis}.}
{While the 1$\sigma$ scatter for the \cite{lehmer21} relation is metallicity dependent (as evidenced by the changing 16-84\% confidence intervals shown in blue in Figure \ref{fig:expxrbalt}), our target galaxies all have $12+$ log (O/H) $> 8.0$ (see below) and the scatter in this range is less variable with a value of ${\sim}0.2$ dex.} 

In general, SFRs can be estimated using far-UV (FUV; 1528 \AA) and mid-infrared (IR; 25 $\mu$m) luminosities as follows:
\begin{equation} \label{eq:sfrs}
\begin{gathered}
 \textrm{log SFR}(\textrm{\msun~yr}^{-1}) = \textrm{log }L\textrm{(FUV)}_{\textrm{corr}} - 43.35 \\
 L \textrm{(FUV)}_{\textrm{corr}} = L\textrm{(FUV)}_{\textrm{obs}} + 3.89 L\textrm{(25 $\mu$m)}
\end{gathered}
\end{equation}
\citep{kennicutt12, hao11}. 
The \cite{hao11} sample used to derive $L({\rm FUV})_{\rm corr}$ includes normal star-forming galaxies in the local Universe drawn from the \textit{Spitzer} Infrared Nearby Galaxies Survey \citep[SINGS;][]{kennicutt03} 
and the integrated spectrophotometric survey of \cite{moustakas06}. 
Overall, the FUV and mid-IR luminosities of our target galaxies fall within the range of these samples \citep{hao11,kennicutt09}, albeit at the more sparsely populated low-luminosity end for the FUV.    
The uncertainty in the resulting SFRs is ${\sim} 0.13$ dex. 

These equations hold assuming the FUV and mid-IR emission is dominated by star formation.  Given that our galaxies have mid-IR colors similar to luminous AGNs, and some show optical evidence for AGNs as well, it is likely that the mid-IR emission is contaminated or even dominated by AGNs in some cases. Therefore we treat SFRs calculated using the relations above as upper limits.  We also calculate SFRs using only the FUV data since any potential (mid-IR selected) AGN is less likely to contribute at these short wavelengths. There could still be contamination from AGN emission, which would lead to overestimating the FUV luminosity and thus the SFR. However, this would move galaxies further upwards in Figure \ref{fig:expxrbalt}, resulting in even higher relative observed X-ray luminosities compared to what we would expect from XRBs.

We use FUV magnitudes from the \textit{Galaxy Evolution Explorer} (\textit{GALEX}) and 22 $\mu$m magnitudes from \textit{WISE} in place of 25 $\mu$m luminosities from the \textit{Infrared Astronomical Satellite} (\textit{IRAS}). The ratio between 22 $\mu$m and 25 $\mu$m flux densities is expected to be of order one \citep{jarrett13}, and while none of our galaxies have \textit{IRAS} detections, all eleven are detected by \textit{WISE}. We summarize the resulting estimates for the SFRs in Table \ref{tab:sfrs}.  The SFRs estimated from both FUV and mid-IR data (i.e., ``corrected SFRs") span a range of $0.48 \textrm{--} 15.50$ \msun~yr$^{-1}$, with a median of $0.92$ \msun~yr$^{-1}$. The SFRs estimated from the uncorrected FUV luminosities are substantially lower with a range of $0.01 \textrm{--} 2.44$ \msun~yr$^{-1}$ and a median of $0.08$ \msun~yr$^{-1}$.

{As an additional check, we also estimate SFRs based on \ha\ emission in the SDSS spectroscopic fiber using the calibration in \citet{kennicutt12}. We use the flux measurements in the NSA and correct for extinction using the Balmer decrement. It is worth noting, however, that the 3\arcsec\ fibers do not always cover the full extent of the galaxies, which could result in underestimating the SFRs.  On the other hand, AGN emission will artificially boost the \ha\ luminosities and SFRs for the BPT AGN/Composite galaxies. We find that the \ha-derived SFRs generally fall in between the FUV and FUV+mid-IR SFRs discussed above.}


We estimate metallicities via the relation from \cite{pettini04}:
\begin{equation} \label{eq:mets}
 12+\textrm{log (O/H)} = 8.90 + 0.57 \times \textrm{log \NII/\ha}.
\end{equation}
The $1\sigma$ scatter in this relation is 0.18. Using our emission line measurements from SDSS spectroscopy, the metallicities of our targets have a range of $12+$log (O/H) = 8.01-8.74, with a median of 8.50 (see Table \ref{tab:sfrs}).
As this relation is derived from \HII~regions, it may not provide accurate results when applied to galaxies with AGNs. An AGN would likely increase the \NII/\ha~flux, thus increasing the derived metallicity. This provides a possible explanation for the separation via metallicity between the BPT AGN and SF galaxies in our sample, as can be seen in Figure \ref{fig:expxrbalt}.

From Figure \ref{fig:expxrbalt}, we can also see that the four BPT AGNs and one composite galaxy with X-ray detections have X-ray luminosities that are well above the expected contributions from high-mass XRBs when adopting the SFRs derived from FUV data and excluding the mid-IR emission (the latter of which is likely dominated by the AGN). The two BPT star-forming galaxies with X-ray detections have X-ray source luminosities closer to what is expected from XRBs. {Specifically, using the FUV SFRs, the BPT AGN/Composite galaxies are ${\sim}1.6-3.7$ dex higher than expected from the relation of \cite{lehmer21}, while the BPT star-forming galaxies are ${\sim}0.1-0.7$ dex higher. In other words, the BPT AGN/Composite galaxies are at least ${\sim}8\sigma$ higher than expected, while the BPT SF galaxies are at most ${\sim}3.5\sigma$ higher.} 
While these results do not rule out the presence of AGNs in the star-forming galaxies, it certainly does not add confidence to the AGN interpretation for star-forming galaxies with anomalous mid-IR colors.  Table \ref{tab:agnindicators} summarizes the origin of the X-ray emission in our target galaxies.

The detected X-ray source luminosities are also generally consistent with ultra-luminous X-ray (ULX) sources, which are defined to be off-nuclear X-ray sources with $L_X > 10^{39}\, {\rm erg}\, {\rm s}^{-1}$ \citep[for a recent review of ULXs, see][]{kaaret17}. Though ULXs were initially considered strong candidates for typical sub-Eddington accretion onto intermediate-mass BHs \citep[e.g.,][]{sutton12}, spectroscopic and timing studies now strongly indicate super-Eddington accretion onto stellar mass sources, with several sources clearly identified as neutron stars due to the detection of X-ray pulsations \citep[e.g.,][]{bachetti14} or cyclotron absorption features \citep{brightman18}. Indeed, based on broadband X-ray spectroscopy, \cite{walton18} suggest that the ULX population as a whole might be strongly dominated by neutron star accretors. However, ULXs are rare. ULXs are perhaps viable explanations for the X-ray sources in our BPT star-forming galaxies, though invoking such a scenario is unnecessary since the X-ray luminosities of these galaxies are consistent with the expectations for their X-ray binary population (e.g., Figure \ref{fig:expxrbalt}).  We do not consider ULXs likely for the BPT AGN/Composite galaxies since the X-ray sources in these galaxies reside in prominent nuclei with additional multi-wavelength evidence for AGNs.

\begin{deluxetable*}{cccc}
\tabletypesize{\footnotesize}
\tablecaption{Multiwavelength Properties of Dwarf Galaxies with Red Mid-IR Colors}
\tablewidth{0pt}
\tablehead{
\colhead{ID}  & \colhead{BPT Classification} & \colhead{X-ray Source} & \colhead{HST Optical/{NIR} Counterpart}  }
\startdata
1   & AGN & AGN & prominent nucleus \\
2   & AGN (with broad H$\alpha$) & AGN  & prominent nucleus \\
3   & AGN (with broad H$\alpha$) & AGN  & prominent nucleus \\
4   & AGN & non-detection  & \nodata \\
5   & AGN & AGN  & prominent nucleus \\
6   & Comp. (with broad H$\alpha$) & AGN & prominent nucleus \\
7   & SF & non-detection & \nodata \\
8   & SF & non-detection & \nodata \\
9   & SF & non-detection & \nodata\\
10  & SF & XRB/ULX or AGN & {unknown (no {\it HST)}} \\
11  & SF & XRB/ULX or AGN & {blue star clusters}
\enddata
\tablecomments{All of our target galaxies were selected to have {\it WISE} mid-IR colors falling in the \citet{jarrett11} AGN selection box.  Optical emission lines are measured from SDSS spectroscopy \citep{reines13} and the classification is based on the BPT \OIII/\hbeta\ vs.\ \NII/\ha\ diagnostic diagram.  X-ray classifications are based on whether the luminosity exceeds that expected from high-mass XRBs (indicating an AGN) or not (see Section \ref{sec:xrbs}; {the BPT AGN/Composite galaxies are higher than the BPT SF galaxies by at least ${\sim}4.5\sigma$}).  {Counterparts of the X-ray sources were determined from the {\it HST} imaging shown in Figure \ref{fig:rgb}.}}
\label{tab:agnindicators}
\end{deluxetable*}

\subsection{Optical/IR Counterparts of the X-ray Sources} \label{sec:optircntr}

 Figure \ref{fig:rgb} shows the \textit{HST} images of our target dwarf galaxies with the X-ray source positions overlaid in {magenta} (the {magenta} crosses and circles mark the positions and the 95\% positional uncertainties).  We also indicate the peak of the {NIR} emission in the galaxy images using a green circle of radius 0\farcs4. The X-ray sources are almost certainly associated with the peak of the {NIR} emission for the four BPT AGNs  {(IDs 1-3,5)} and one composite galaxy  {(ID 6)} with X-ray detections.  In these cases, the positions of the X-ray sources are consistent with prominent nuclei and there are no other obvious alternative optical/{NIR} counterparts.   Of the two star-forming galaxies with detected X-ray sources, IDs 10 and 11, only ID 11 has {\it HST} imaging allowing us to determine
 optical counterparts of the X-ray source. In Figure \ref{fig:rgb} we see that the X-ray source has a larger offset from the peak of the {NIR} emission, and there are multiple blue (i.e., young) star clusters consistent with the position of the X-ray source that could host high-mass XRBs. While the X-ray source is consistent with being nuclear for ID 10, we lack HST images and are unsure whether or not this galaxy hosts star clusters close to the nucleus.
 
\begin{figure}[h!]
\centering
\includegraphics[width=0.48\textwidth]{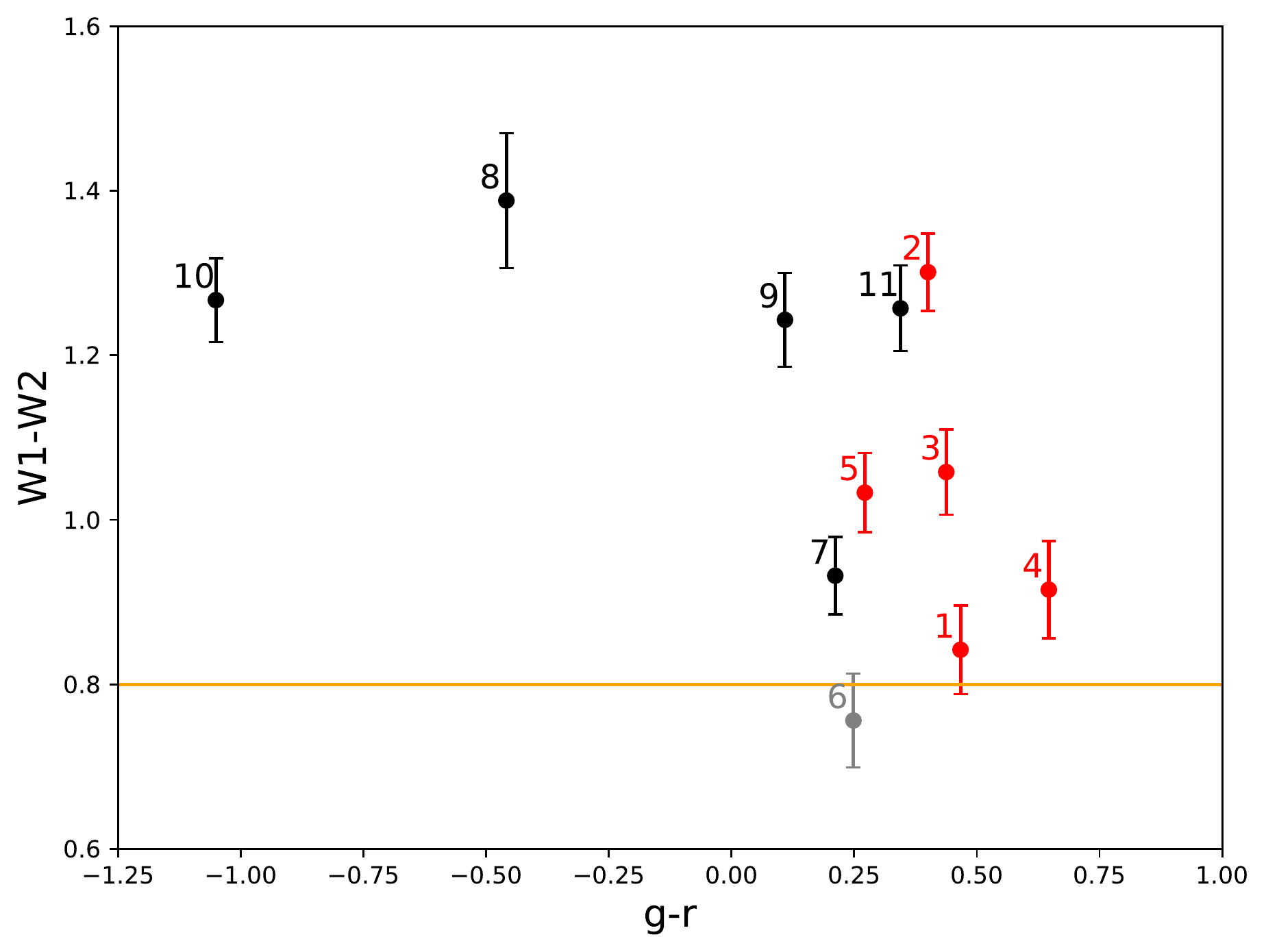}
\caption{\textit{WISE} W1-W2 color vs. $g-r$ color (from Table \ref{tab:sample}). The galaxies classified as AGN, composite, and star-forming are in red, grey, and black, respectively, based on the \NII/\ha~BPT diagram (left panel in Figure \ref{fig:bpt}). 
In orange we show the W1-W$2 > 0.8$ selection criterion from \cite{stern12}.
}
\label{fig:grwisecolor}
\end{figure}

Given that our targets were selected to have mid-IR {\it WISE} colors 
falling in the \citet{jarrett11} AGN selected box, we also show the {\it WISE} W2 angular resolution as a red circle in Figure \ref{fig:rgb}.  In all cases, {\it WISE} is capturing mid-IR emission on galaxy-wide scales and we cannot definitively determine which {\it HST} sources are physically associated with the mid-IR emission.  However, we deem it reasonable to conclude that the mid-IR emission in the BPT AGNs and the composite galaxy is in fact associated with AGNs since these galaxies have low amounts of star formation and compelling evidence for active massive BHs at multiple wavelengths. On the other hand, the mid-IR emission in the BPT star-forming galaxies could certainly be associated with young massive star clusters still partially embedded in their birth cocoons \citep[e.g.,][]{reines08}.  None of the BPT star-forming galaxies have evidence for hosting AGNs other than red mid-IR colors.  Moreover, they  {generally} have relatively blue optical colors (Figure \ref{fig:grwisecolor}; also see \citealt{hainline16}) and multiple star clusters visible in the {\it HST} images, and one  {(ID 7)} has visible extended dust lanes (see Figure \ref{fig:rgb}).  Nevertheless, we cannot rule out highly obscured AGNs in these star-forming galaxies, which could also account for the mid-IR colors.

\subsection{Multiwavelength AGN Scaling Relations} \label{sec:expxraynh}

Here we compare the observed X-ray source luminosities to expectations based on AGN scaling relations for more massive galaxies relating X-ray luminosity to mid-IR luminosity, \ha\ luminosity and \OIII\ luminosity.

\subsubsection{$L_{\rm X}$ vs. $L_{\rm IR}$} \label{sec:xraymir}

\begin{figure}
\subfloat[ \label{fig:xrayrelsa}]{\includegraphics[width=0.45\textwidth]{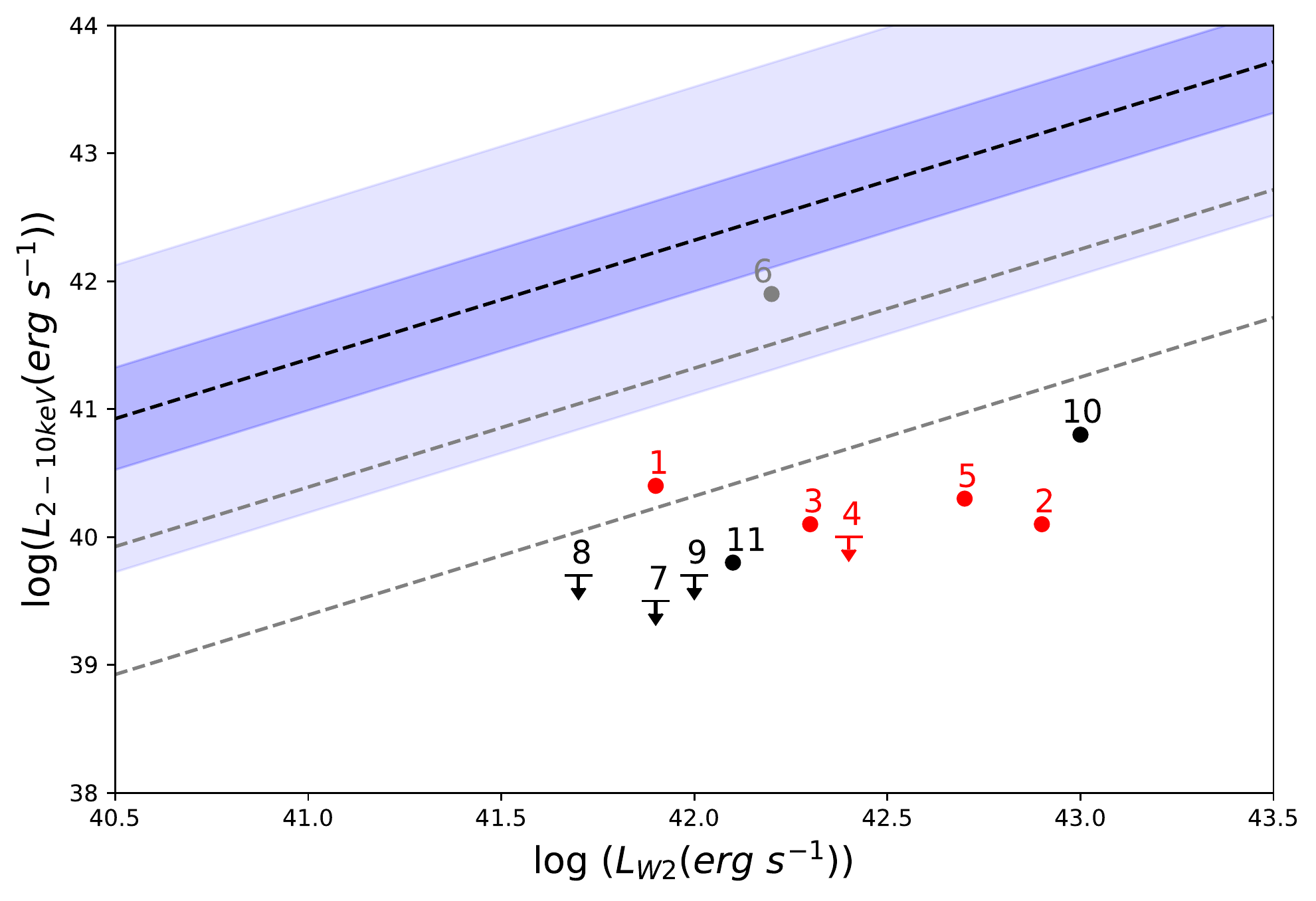}}\hfill
\subfloat[ \label{fig:xrayrelsc}]{\includegraphics[width=0.45\textwidth]{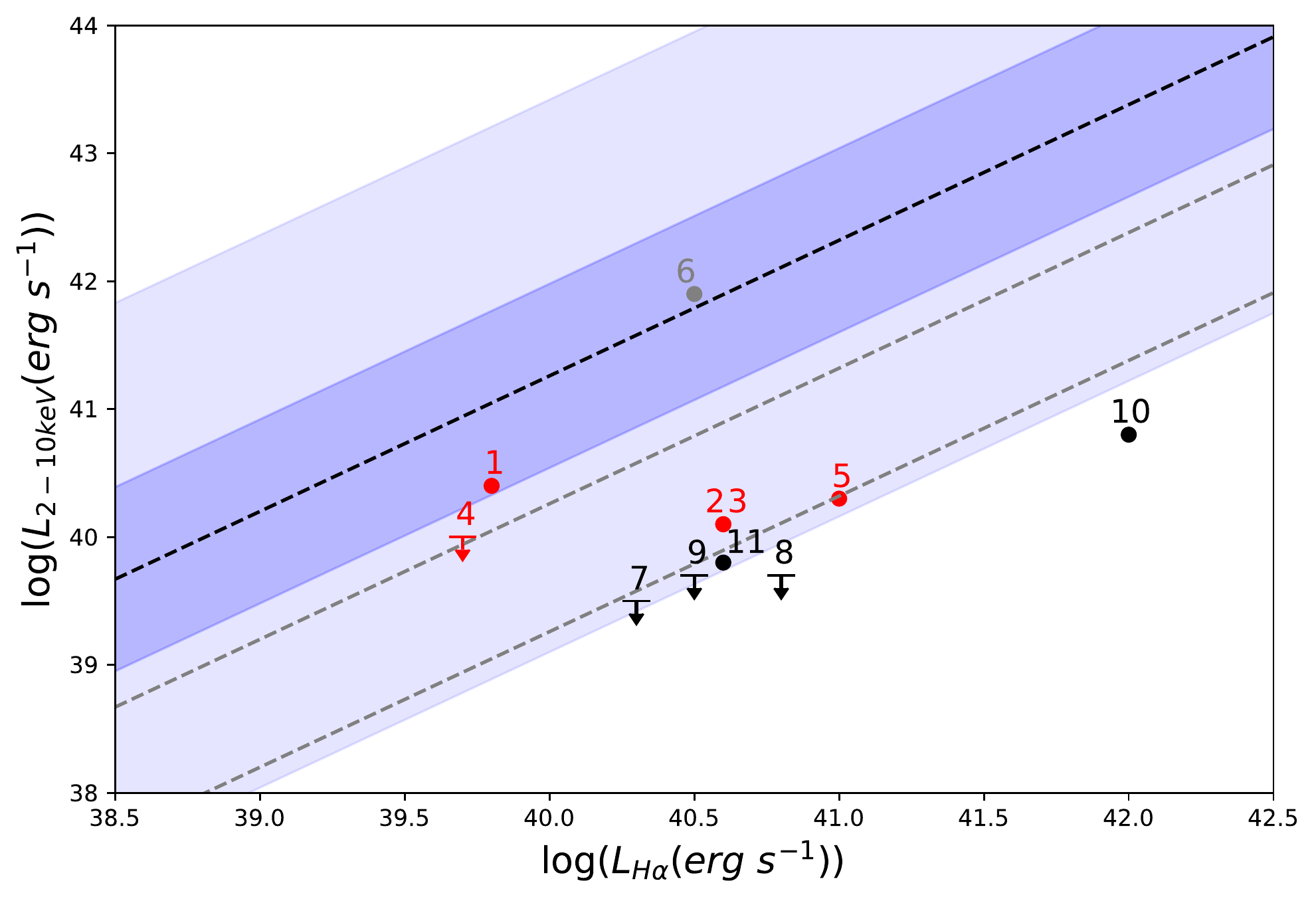}}\hfill
\subfloat[ \label{fig:xrayrelse}]{\includegraphics[width=0.45\textwidth]{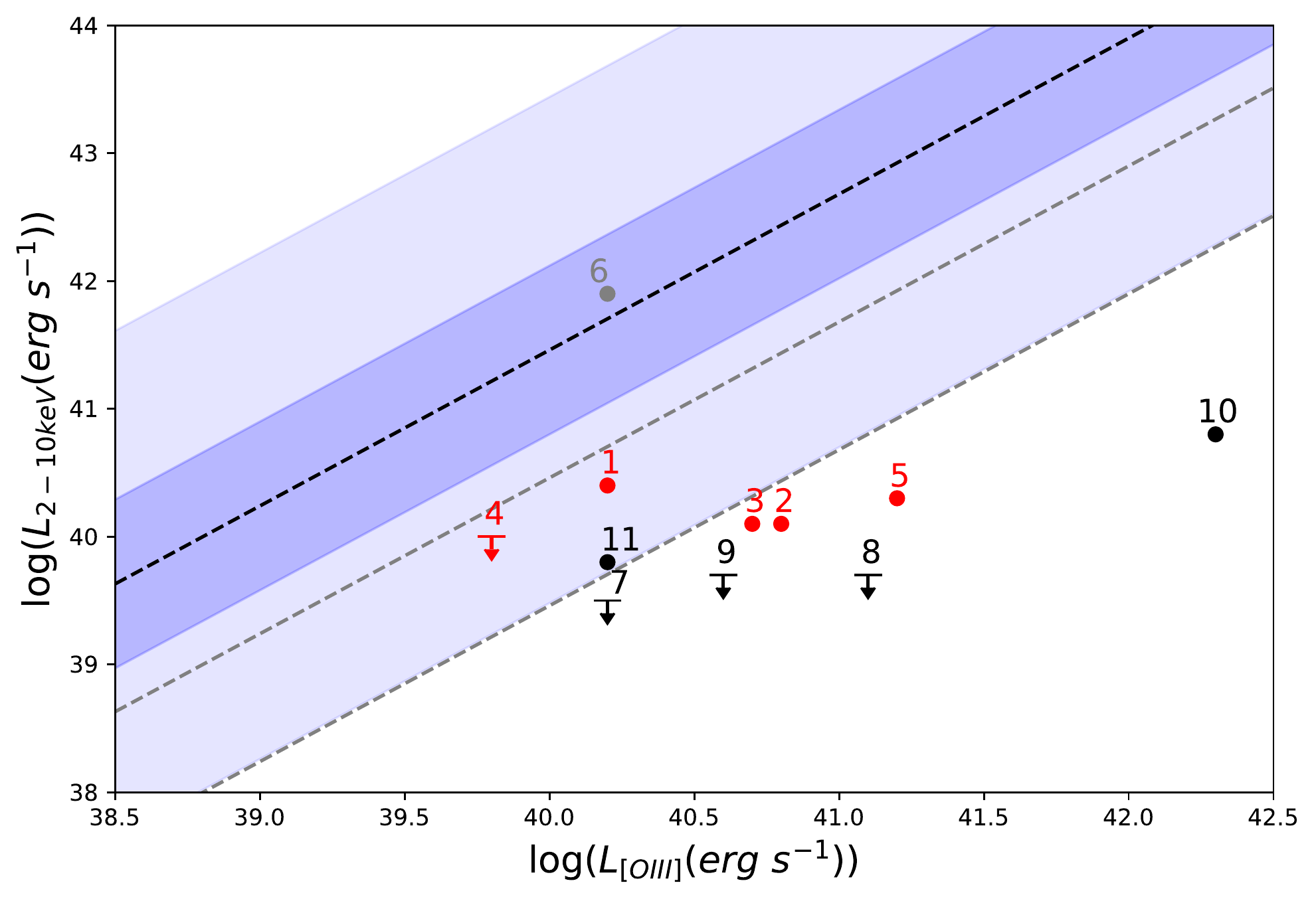}}\hfill
\caption{{Observed X-ray luminosities vs W2 (top), \ha~(middle), and \OIII~(bottom) luminosities for our galaxies. Galaxies classified as AGN, composite, and star-forming are in red, grey, and black, respectively. The arrows indicate that the X-ray luminosity comes from an upper limit (calculated from minimum fluxes - see Section \ref{sec:xrayemis}). The black dashed lines indicate the expected X-ray luminosity using the \cite{secrest15} relation (top, mid-IR; Equation \ref{eq:expxrayfrommir}) and the relations from \cite{panessa06} (middle, \ha, Equation \ref{eq:expxrayfromha}; bottom, \OIII, Equation \ref{eq:expxrayfromo3}).} The dark blue shaded areas represent the 1$\sigma$ scatters and the light blue shaded areas represent the 3$\sigma$ scatters. The dashed grey lines represent the relations shifted down by 1 and 2 dex, for ease of comparison with the plotted points.
} 
\label{fig:xrayrels}
\end{figure}

Figure \ref{fig:xrayrels} (top panel) shows the relation between $L_{\rm 2-10~keV}$ and $L_{\rm W2}$ (4.6 $\mu$m)  from \cite{secrest15}:

\begin{equation} \label{eq:expxrayfrommir}
\textrm{log}(L_{\rm 2-10~keV}) = (0.93 \pm 0.03) \cdot \textrm{log}(L_{\rm W2}) + (3.26 \pm 1.25),
\end{equation}

\noindent
where the luminosities are in units of erg s$^{-1}$ and the 1$\sigma$ scatter is 0.40 dex.
This relation was derived from a sample of 184 AGNs detected with \textit{XMM} and {\it WISE} having both X-ray and mid-IR luminosities $\gtrsim 10^{42}$ erg s$^{-1}$.  With the exception of  {ID 6} (a broad-line BPT composite), all of the observed X-ray luminosities and upper limits in our sample are $\sim 2$ dex lower than the relation between $L_{\rm 2-10~keV}$ and $L_{\rm W2}$ based on more luminous and massive AGNs. We check this result using the relation from \cite{gandhi09}, which was derived from a sample of 22 well-resolved AGNs and compares $12$ $\mu$m luminosity to 2-10 keV X-ray luminosity. For this relation we use W3 fluxes from \textit{WISE} and find similar results; the observed X-ray luminosities are lower than the relation by at least ${\sim}1.5$ dex (except ID 6). We consider three possible explanations for this discrepancy.  

First, the mid-IR luminosities could be overestimated (and the sources could be pushed to the right) if there is a significant contribution from star formation. While our \textit{Chandra} X-ray observations have high angular resolution ($\lesssim 1\arcsec$), the \textit{WISE} W2 band data has an angular resolution of $6.4\arcsec$, which in nearly all cases is roughly on par with the size of the galaxies in our sample. However, we do not expect significant contamination from star formation in the mid-IR for the BPT AGNs (IDs 1-5). In order for these sources to be consistent with the $L_{\rm X}-L_{\rm IR}$ relation in Figure \ref{fig:xrayrels}, we would have to be overestimating the mid-IR AGN luminosities by {between 2.0-3.3 dex.} 
This would imply the mid-IR emission is actually dominated by extreme star-formation, which seems highly unlikely given the optical line ratios and X-ray emission are dominated by AGNs.
The origin of the mid-IR and X-ray emission in the BPT star-forming galaxies is unclear and may not have anything to do with AGNs.

\begin{figure}[h!]
\centering
\includegraphics[width=0.48\textwidth]{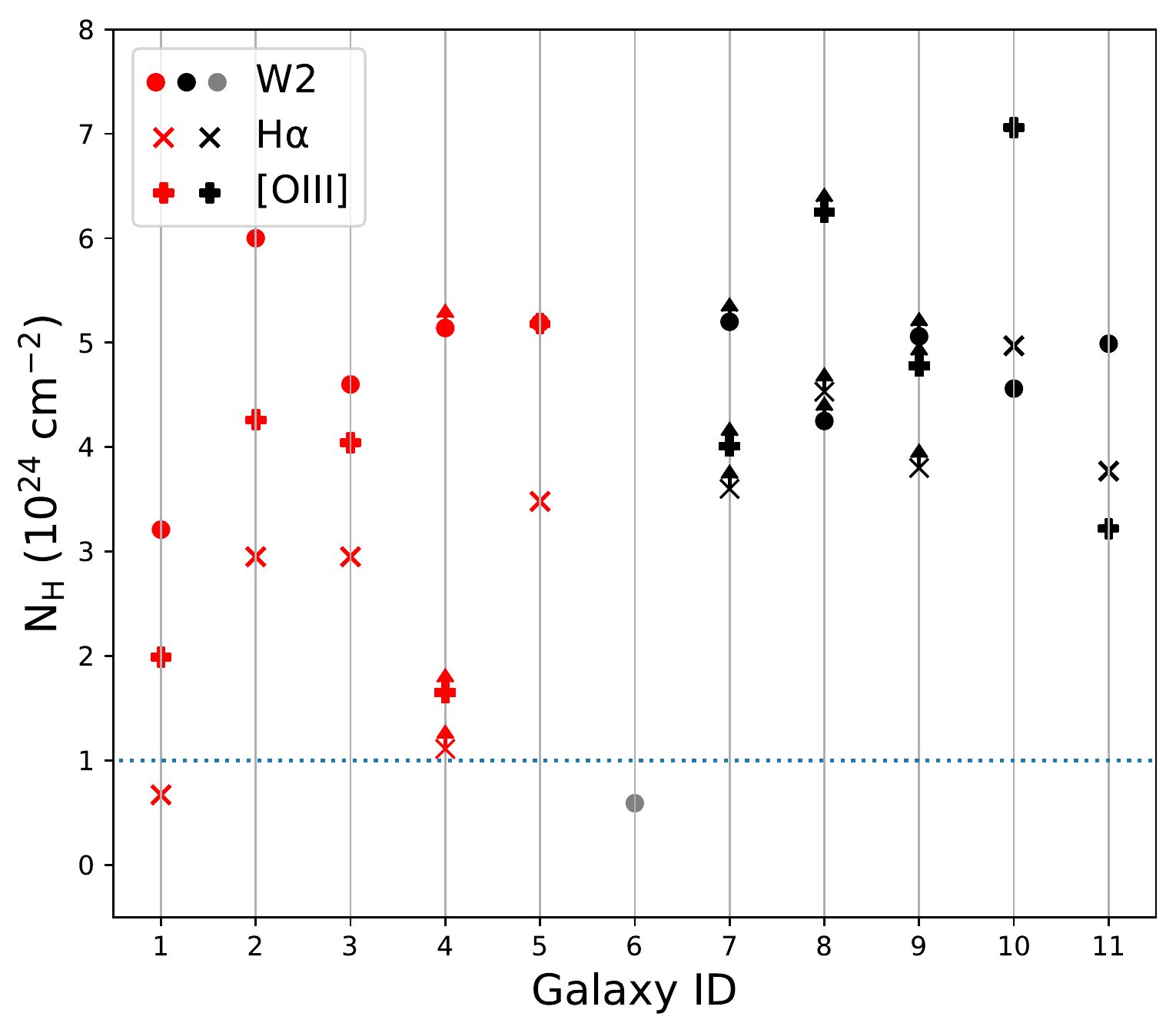}
\caption{Estimated intrinsic absorption ($N_{\rm H}$) in each galaxy using the difference between the expected and observed X-ray luminosities. Galaxies classified as AGN, composite, and star-forming are in red, grey, and black, respectively. The dots, x's, and plus signs use the expected X-ray luminosity as calculated from the W2, \ha, or \OIII~relations, respectively. The arrows indicate that the observed X-ray luminosity comes from an upper limit calculated from minimum fluxes, which translates into a lower limit for $N_{\rm H}$. The dotted blue line at $N_{\rm H} = 10^{24}$ cm$^{-2}$ separates out the plot into Compton thick (above the line) and Compton thin (below the line) regions. Note that several estimates of $N_{\rm H}$ are unavailable (e.g. the \ha~and \OIII~relations for ID 6) $-$ in these cases, the observed X-ray luminosity was higher than the expected X-ray luminosity, so an estimate was unable to be calculated. {Note also that for ID 6, \cite{baldassare17} found via spectral fitting that the models did not prefer an $N_{\rm H}$ greater than the Galactic value, suggesting little to no intrinsic absorption.}}
\label{fig:nhs}
\end{figure}

A second possibility is that the mid-IR emission is dominated by highly obscured AGNs, such that the X-rays are greatly diminished. There are examples of Compton-thick AGNs in low-mass galaxies in the literature \citep[e.g.][]{chen17,cann20}.
Under the assumption that an AGN does indeed reside in each of our target galaxies, we can estimate the column density required to suppress the X-ray emission below the $L_{\rm X}-L_{\rm IR}$ relation (assuming that intrinsic X-ray luminosity is consistent with said relation). Using PIMMS,
we find that the estimated column densities are all {(except ID 6)} high enough to qualify as Compton-thick with $N_{\rm H} > 10^{24}$ cm$^{-2}$. See Table \ref{tab:mirxray} for specific values and Figure \ref{fig:nhs} for a visual depiction. 
However, if Compton-thick AGNs resided in our target galaxies, we would expect to see hints in the X-ray emission. For example, if the AGNs were Compton-thick we would expect more photons at higher energies yielding a positive hardness ratio. Figure \ref{fig:xrayhrratio} demonstrates this is generally not the case. Three of the five BPT AGNs/composities (IDs 2,5,6) fall firmly in the soft range, and two BPT AGNs (IDs 1,3) are ambiguous. 
{While Compton-thick AGNs also typically have strong Fe K$\alpha$ emission, this may not always be the case for heavily obscured and X-ray faint AGNs \citep[see][]{chen17}. Nevertheless, using the \texttt{CIAO} function \texttt{specextract}, we check both individual and stacked X-ray spectra for the BPT AGN and Composite galaxies with X-ray detections and find no evidence for Fe K$\alpha$ emission.} Additionally, broad optical emission lines are generally not observed in obscured AGNs \citep{hickox18}, and we do see broad H$\alpha$ in three of our X-ray detected AGNs (IDs 2,3,6).    

An alternative explanation for the low X-ray luminosities is that the AGNs could be intrinsically X-ray weak \citep[e.g.,][]{dong12,baldassare17}.  It has been proposed that some AGNs in metal-poor host galaxies could simply lack the hard X-ray emission we would expect to see \citep{simmonds16}. Similar conclusions have been found regarding AGNs with weak broad lines that are not necessarily obscured \citep{denney14,laMassa15,macleod16,hickox18}. 

Finally, it is entirely possible that the $L_{\rm X}-L_{\rm IR}$ relation simply breaks down at low BH masses or is steeper than has been captured by relations calibrated at higher luminosities. Perhaps these relations cannot be properly extrapolated to dwarf galaxies, where there may be a change in the AGN spectral energy distribution and/or dust properties.

\begin{deluxetable}{cccccc}
\tabletypesize{\footnotesize}
\tablecaption{Expected $L_{\rm X}$ from $L_{\rm IR}$}
\tablewidth{0pt}
\tablehead{
\colhead{ID} & \colhead{W$_{4.6}$} & \colhead{log W$_{4.6}$}  & \colhead{log $L^{\rm Obs}_{\rm X}$} & \colhead{log $L^{\rm Exp}_{\rm X}$} & \colhead{$N_{\rm H}$} \\
 \colhead{ } & \colhead{(mag)} & \colhead{(erg s$^{-1}$)} & \colhead{(erg s$^{-1}$)} & \colhead{(erg s$^{-1}$)} & \colhead{($10^{24}$ cm$^{-2}$)}\\
\colhead{(1)} & \colhead{(2)} & \colhead{(3)} & \colhead{(4)} & \colhead{(5)} & \colhead{(6)}}
\startdata
1  	    	 & 13.17  & 41.9 &    40.4  & 42.3 &    3.21  \\
2            & 12.10  & 42.9 &    40.1  & 43.1 &    6.00  \\
3            & 12.80  & 42.3 &    40.1  & 42.6 &    4.60  \\
4    		 & 13.36  & 42.4 & $<$40.0  & 42.7 & $>$5.14  \\
5    		 & 12.48  & 42.7 &    40.3  & 43.0 &    5.19  \\
6            & 13.37  & 42.2 &    41.9  & 42.5 &    0.59  \\
7 	    	 & 12.38  & 41.9 & $<$39.5  & 42.2 & $>$5.20  \\
8 	    	 & 14.20  & 41.7 & $<$39.7  & 42.0 & $>$4.25  \\
9 	    	 & 13.28  & 42.0 & $<$39.7  & 42.3 & $>$5.06  \\
10           & 11.77  & 43.0 &    40.8  & 43.3 &    4.56  \\
11           & 12.50  & 42.1 &    39.8  & 42.4 &    4.99  
\enddata
\tablecomments{Column 1: Identification number used in this paper.  
Column 2: \textit{WISE} magnitudes (W2, 4.6 $\mu$m).
Column 3: log 4.6 $\mu$m ($W2$) luminosities.
Column 4: log observed 2-10 keV X-ray luminosities; galaxy IDs 4, 7-9 had no X-ray detections, so we instead state the calculated minimum detectable luminosities using the minimum observable fluxes (see Section \ref{sec:xrayemis}).
Column 5: expected log 2-10 keV X-ray luminosities, calculated using Equation \ref{eq:expxrayfrommir}.
Column 6: Estimated intrinsic hydrogen column density required to match our observed X-ray luminosities with those predicted by the \cite{secrest15} relation (see Equation \ref{eq:expxrayfrommir}).
}
\label{tab:mirxray}
\end{deluxetable}

\subsubsection{$L_{\rm X}$ vs.\ $L_{\rm H\alpha}$~and~$L_{\rm \OIII}$} \label{sec:xrayhao3}

Figure \ref{fig:xrayrels} (middle, bottom panels) shows the $L_{\rm 2-10~keV} - L_{\rm H\alpha}$ and $L_{\rm 2-10~keV} - L_{\rm \OIII}$ relations from \citet{panessa06}: 

\begin{equation} \label{eq:expxrayfromha}
\textrm{log}(L_{\rm 2-10~keV}) = (1.06 \pm 0.04) \cdot \textrm{log}(L_{\textrm{H}\alpha}) + (-1.14 \pm 1.78)
\end{equation}

\begin{equation} \label{eq:expxrayfromo3}
\textrm{log}(L_{\rm 2-10~keV}) = (1.22 \pm 0.06) \cdot \textrm{log}(L_{\rm \OIII}) + (-7.34 \pm 2.53),
\end{equation}

\noindent 
where the luminosities are in erg s$^{-1}$.
These relations are based on a sample of Seyferts from the Palomar optical spectroscopic survey of nearby galaxies \citep{ho95}, and the luminosities (log $L_{\rm X}/{\rm erg~s}^{-1} =38.2$-43.3) are measured in the nuclear regions of the galaxies
\citep[i.e., not the entire galaxy;][]{panessa06}. The \ha~luminosities used in \cite{panessa06} are from the combined narrow and broad (if present) components and are corrected for extinction. We also include both narrow and broad \ha\ components (when detected). While our \ha~luminosities are not corrected for extinction, this would only increase our luminosities and push them further to the right in Figure \ref{fig:xrayrels}, thereby increasing the disparity between expected and observed X-ray luminosities.

We estimate the scatter in these relations by taking the standard deviation of the differences between the observed and expected X-ray luminosities of the original sample of \cite{panessa06}, excluding galaxies that lack detection in one or more of X-ray, \ha, or \OIII~bands \citep[see Table 2 in][]{panessa06}. We find a scatter of ${\sim}0.72$ and ${\sim}0.66$ dex for the \ha~and \OIII~relations, respectively. This does not include uncertainties associated with the fitted parameters in the \citet{panessa06} relation.

Our dwarf galaxies are also shown in Figure \ref{fig:xrayrels} using our {\it Chandra} X-ray luminosities and line measurements from SDSS spectroscopy \citep{reines13} { $-$ see Tables \ref{tab:haxray} and \ref{tab:o3xray} for specific values}.
The points fall systematically below the $L_{\rm X}-L_{\rm \OIII}$ and $L_{\rm X}-L_{\rm H\alpha}$ relations, but some of the sources are within the $3\sigma$ scatter.
We note that we do not find any systematic difference between the broad-line and narrow-line AGNs with respect to these relations when considering our sample and the additional broad-line AGNs in dwarf galaxies presented in \citet{baldassare17}. 

Again, we consider various explanations for our targets generally falling below the relations. We do not expect that we are significantly overestimating the line luminosities even though the measurements come from SDSS spectra taken with a 3\arcsec-aperture (see Figure \ref{fig:rgb}). In order to match the relation, we would have to be overestimating the \ha~and \OIII\ luminosities of the BPT AGNs by 0.6-1.9 dex and 1.0-2.2 dex, respectively, and therefore star formation would have to dominate the line emission. Given that the line ratios for these galaxies indicate the emission is dominated by AGNs, we rule this scenario out.  Similar to the $L_{\rm X}-L_{\rm IR}$ relation, we conclude that the $L_{\rm X}-L_{\rm \OIII}$ and $L_{\rm X}-L_{\rm H\alpha}$ relations are {either} not applicable to AGNs with low BH masses, which may be intrinsically X-ray weak, {or the relations do not properly extrapolate to dwarf galaxies and may need to be revisited with expanded samples including more such objects.}

\begin{deluxetable*}{cccccc}
\tabletypesize{\footnotesize}
\tablecaption{Expected $L_{\rm X}$ from $L_{\rm H\alpha}$}
\tablewidth{0pt}
\tablehead{
\colhead{ID} & \colhead{F$_{\textrm{H}\alpha}$} & \colhead{log L$_{\textrm{H}\alpha}$}  & \colhead{log L$^{\rm Obs}_{\rm X}$} & \colhead{log L$^{\rm Exp}_{\rm X}$} & \colhead{$N_{\rm H}$} \\
 \colhead{ } & \colhead{($10^{-16}$ erg s$^{-1}$ cm$^{-2}$)} & \colhead{(erg s$^{-1}$)} & \colhead{(erg s$^{-1}$)} & \colhead{(erg s$^{-1}$)} & \colhead{($10^{24}$ cm$^{-2}$)}\\
\colhead{(1)} & \colhead{(2)} & \colhead{(3)} & \colhead{(4)} & \colhead{(5)} & \colhead{(6)}}
\startdata
1  	    	 & 41.77   & 39.8 &    40.4  & 41.0 &    0.67 \\
2            & 53.02   & 40.6 &    40.1  & 41.9 &    2.95 \\
3            & 164.28  & 40.6 &    40.1  & 41.9 &    2.95 \\
4    		 & 9.43    & 39.7 & $<$40.0  & 40.9 & $>$1.11 \\
5    		 & 204.76  & 41.0 &    40.3  & 42.3 &    3.48 \\
6            & 49.84   & 40.5 &    41.9  & 41.8 &    N/A\footnote{\label{hafn}Entries of N/A denote a galaxy where the observed X-ray luminosity exceeded the expected X-ray luminosity, so no $N_{\rm H}$ value could be calculated.}  \\
7 	    	 & 315.78  & 40.3 & $<$39.5  & 41.6 & $>$3.60 \\
8 	    	 & 329.23  & 40.8 & $<$39.7  & 42.1 & $>$4.53 \\
9 	    	 & 177.75  & 40.5 & $<$39.7  & 41.8 & $>$3.80 \\
10           & 2343.00 & 42.0 &    40.8  & 43.4 &    4.97 \\
11           & 368.45  & 40.6 &    39.8  & 41.9 &    3.77 
\enddata
\tablecomments{Column 1: Identification number used in this paper.  
Column 2: \ha~flux.
Column 3: log \ha~ luminosity. This include both narrow and broad (if present) components.
Column 4: log observed 2-10 keV X-ray luminosity.
Column 5: expected log 2-10 keV X-ray luminosity, calculated using Equation \ref{eq:expxrayfromha}.
Column 6: Estimated intrinsic hydrogen column density required to match our observed X-ray luminosities with those predicted by the \cite{panessa06} relation (see Equation \ref{eq:expxrayfromha}).
}
\label{tab:haxray}
\end{deluxetable*}
\begin{deluxetable*}{cccccc}
\tabletypesize{\footnotesize}
\tablecaption{Expected $L_{\rm X}$ from $L_{\rm \OIII}$}
\tablewidth{0pt}
\tablehead{
\colhead{ID} & \colhead{F$_{\rm \OIII}$} & \colhead{log L$_{\rm \OIII}$}  & \colhead{log L$^{\rm Obs}_{\rm X}$} & \colhead{log L$^{\rm Exp}_{\rm X}$} & \colhead{$N_{\rm H}$} \\
 \colhead{ } & \colhead{($10^{-16}$ erg s$^{-1}$ cm$^{-2}$)} & \colhead{(erg s$^{-1}$)} & \colhead{(erg s$^{-1}$)} & \colhead{(erg s$^{-1}$)} & \colhead{($10^{24}$ cm$^{-2}$)} \\
\colhead{(1)} & \colhead{(2)} & \colhead{(3)} & \colhead{(4)} & \colhead{(5)} & \colhead{(6)}}
\startdata
1  	    	 & 111.86  & 40.2 &    40.4  & 41.7 &    1.99 \\
2            & 139.83  & 40.8 &    40.1  & 42.4 &    4.26 \\
3            & 243.27  & 40.7 &    40.1  & 42.3 &    4.04 \\
4    		 & 11.97   & 39.8 & $<$40.0  & 41.2 & $>$1.65 \\
5    		 & 390.51  & 41.2 &    40.3  & 43.0 &    5.18 \\
6            & 51.16   & 40.2 &    41.9  & 41.7 &    N/A\footnote{\label{o3fn}Entries of N/A denote a galaxy where the observed X-ray luminosity exceeded the expected X-ray luminosity, so no $N_{\rm H}$ value could be calculated.}  \\
7 	    	 & 275.72  & 40.2 & $<$39.5  & 41.7 & $>$4.01 \\
8 	    	 & 630.52  & 41.1 & $<$39.7  & 42.8 & $>$6.25 \\
9 	    	 & 219.95  & 40.6 & $<$39.7  & 42.2 & $>$4.78 \\
10           & 3969.37 & 42.3 &    40.8  & 44.2 &    7.06 \\
11           & 133.58  & 40.2 &    39.8  & 41.7 &    3.22 
\enddata
\tablecomments{Column 1: Identification number used in this paper.  
Column 2: \OIII~flux.
Column 3: log \OIII~ luminosity.
Column 4: log observed 2-10 keV X-ray luminosity.
Column 5: expected log 2-10 keV X-ray luminosity, calculated using Equation \ref{eq:expxrayfromo3}.
Column 6: Estimated intrinsic hydrogen column density required to match our observed X-ray luminosities with those predicted by the \cite{panessa06} relation (see Equation \ref{eq:expxrayfromo3}).
}
\label{tab:o3xray}
\end{deluxetable*}

\section{Summary and Conclusions}

We have presented high-resolution \textit{Chandra} observations of eleven dwarf galaxies having mid-IR selected candidate AGNs culled from the sample of \citet{hainline16}. Multi-band \textit{HST} observations are also presented for ten of these galaxies. Based on optical SDSS spectroscopy \citep{reines13}, five of our target galaxies are classified as BPT AGNs (two of which have broad lines), five as star-forming, and one as Composite (which also has broad lines). 
Using a suite of multiwavelength diagnostics, we investigate whether mid-IR color-color AGN selection \citep[e.g.][]{jarrett11} is effective when applied to low-mass galaxies.  Our primary findings are summarized below (also see Table \ref{tab:agnindicators}).

\begin{enumerate}
    \item We detect seven X-ray point sources across our sample of eleven dwarf galaxies, with luminosities in the range log$(L_{\rm 2-10 keV}/{\rm erg~s}^{-1}) = 39.8$--$41.9$. Five of these sources are in BPT AGN or Composite galaxies, and two are in BPT star-forming galaxies. 

\item There is strong evidence that the X-ray sources detected in the five optically-selected BPT AGN and Composite galaxies (IDs 1,2,3,5,6) are indeed accreting massive BHs in the nuclei of their host galaxies. The X-ray luminosities are well above the expected contribution from XRBs, and the positions of the X-ray sources are consistent with prominent nuclei, the peak of the {NIR} emission in the {\it HST} images, and the centroid of the mid-IR emission from {\it WISE} (see Figure \ref{fig:rgb}). It is thus reasonable to attribute the detected mid-IR emission from {\it WISE} to these optically-selected AGNs.

\item The same cannot be said of the BPT star-forming galaxies. Only two of five of these galaxies have X-ray detections
(IDs 10,11) and the X-ray luminosities are not so high as to rule out high-mass XRBs.  Additionally, {\it HST} imaging of ID 11 indicates the X-ray source is associated with blue (i.e.\ young) star clusters that are plausible hosts of an XRB/ULX.  The X-ray source in this galaxy is also significantly offset from the peak of the {NIR} emission and the centroid of the mid-IR emission (see Figure \ref{fig:rgb}), strongly suggesting different origins for the X-ray and mid-IR sources. While the X-ray source in ID 10 is consistent with the nucleus of the galaxy, without HST imaging we cannot rule out the possibility of star clusters close to the nucleus, and as such cannot reliably determine the optical counterpart of the X-ray source.

    \item We compare the observed X-ray luminosities to those expected from AGN scaling relations using mid-IR, \ha, and \OIII~luminosities. In nearly all cases, we find that the observed X-ray emission falls below the relations 
    (see Figure \ref{fig:xrayrels}), with the largest discrepancies for the $L_{\rm X}-L_{\rm IR}$ relation.  We consider various explanations and hypothesize that the sources are either intrinsically X-ray weak, or that the scaling relations break down in {(or cannot be reliably extrapolated to)} the regime of low-mass galaxies/BHs (see Sections \ref{sec:xraymir} and \ref{sec:xrayhao3}).
\end{enumerate}

The work of \citet{hainline16} demonstrated that using a single {\it WISE} $W1-W2$ color cut to select AGN candidates in dwarf galaxies led to severe contamination from young starbursts.  Here we show that using a more stringent color-color selection \citep[i.e.,][]{jarrett11} still leads to unreliable results for optically-selected star-forming dwarf galaxies. {While the WISE color-color selection for our sample of BPT AGN and Composite galaxies seems to be reliable, this mid-IR selection method misses many optically selected AGN and Composite galaxies \citep[see Figure 1 in][]{hainline16}.}  We also demonstrate that $\sim80-90\%$ of the secure mid-IR selected AGNs in our sample (with optical and X-ray evidence for AGNs) have lower than expected X-ray luminosities when compared to multiwavelength scaling relations based on more massive/luminous systems, suggesting these mid-IR selected AGNs are either {highly obscured, intrinsically X-ray weak, or that the scaling relations break down when applied to low-mass galaxies such as we have here.}

Recently, there have been additional papers analyzing the efficacy of these mid-IR selection techniques. \cite{satyapal18} use photoionization and stellar population synthesis models to model AGNs and starbursts, finding that in some extreme cases these starbursts can have mid-IR colors that would classify them as AGNs using a one-band ($W1-W2$) color cut \citep[e.g.,][]{stern12}. They also find that abnormally high ionization parameters or gas densities would be required for a starburst to be classified as an AGN using the two-band color cut of \cite{jarrett11}, and that these conditions are inconsistent with current observations of star-forming galaxies.  Our observational findings presented here suggest such extreme conditions do exist in at least some star-forming dwarf galaxies, or that other physical processes are causing them to fall in the \citet{jarrett11} AGN selection box.

Mid-IR AGN selection in dwarfs also has issues due to the relatively poor resolution of {\it WISE} compared to optical surveys, potentially resulting in contamination due to overlapping sources.
\cite{lupi20} re-analyzed the sample of \cite{kaviraj19}, who combined optical and infrared data from the Hyper Suprime-Cam \citep[HSC;][]{aihara17} and \textit{WISE}, respectively, in order to search for AGN in dwarf galaxies. \citet{kaviraj19} find an AGN occupation fraction of $10-30\%$ in their sample of ${\sim}800$ galaxies, which is much larger than surveys at other wavelengths. The re-analysis by \cite{lupi20} takes into account resolution effects and source overlapping, as HSC has a resolution of ${\sim}0\farcs6$ compared to \textit{WISE} with ${\sim}6$\arcsec. \cite{lupi20} match HSC sources to \textit{WISE} sources (and apply an additional signal-to-noise ratio cut for the W3 emission), arriving at a sample of ${\sim}500$ dwarf galaxies. All but 15 of these dwarfs are in groups, with multiple HSC sources associated with one \textit{WISE} source. Assuming the {\it WISE} source is associated with a dwarf, as opposed to a more luminous galaxy in the group that is also consistent with the {\it WISE} source, leads to errors in the AGN fraction in dwarf galaxies. \citet{lupi20} account for this effect and find 
an AGN occupation fraction of ${\sim}0.4\%$, consistent with other results.

While mid-IR AGN selection in dwarf galaxies at the angular resolution of {\it WISE} appears to be fraught with problems, mid-IR prospects are likely to improve in the near future.  The {\it James Webb Space Telescope (JWST)} will yield significantly higher resolution images in the infrared and facilitate the study of IR coronal emission lines to help identify elusive AGNs \citep{cann18,satyapal20}.


\acknowledgements
{We thank the anonymous referee for their helpful comments.} Support for this work was provided by NASA through \textit{Chandra} Award Number GO9-20094X issued by the \textit{Chandra X-ray Observatory Center}, which is operated by the Smithsonian Astrophysical Observatory for and on behalf of the NASA under contract NAS8-03060.
Based on observations with the NASA/ESA \textit{Hubble Space Telescope} obtained from MAST at the Space Telescope Science Institute, which is operated by the Association of Universities for Research in Astronomy, Incorporated, under NASA contract NAS5-26555. Support for Program number HST-GO-15607.001-A was provided through a grant from the STScI under NASA contract NAS5-26555.
{AER also acknowledges support for this work provided by NASA through EPSCoR grant number 80NSSC20M0231. The work of DS was carried out at the Jet Propulsion Laboratory, California Institute of Technology, under a contract with NASA.} 

{
The Legacy Surveys consist of three individual and complementary projects: the Dark Energy Camera Legacy Survey (DECaLS; Proposal ID \#2014B-0404; PIs: David Schlegel and Arjun Dey), the Beijing-Arizona Sky Survey (BASS; NOAO Prop. ID \#2015A-0801; PIs: Zhou Xu and Xiaohui Fan), and the Mayall z-band Legacy Survey (MzLS; Prop. ID \#2016A-0453; PI: Arjun Dey). DECaLS, BASS and MzLS together include data obtained, respectively, at the Blanco telescope, Cerro Tololo Inter-American Observatory, NSF’s NOIRLab; the Bok telescope, Steward Observatory, University of Arizona; and the Mayall telescope, Kitt Peak National Observatory, NOIRLab. The Legacy Surveys project is honored to be permitted to conduct astronomical research on Iolkam Du’ag (Kitt Peak), a mountain with particular significance to the Tohono O’odham Nation.

NOIRLab is operated by the Association of Universities for Research in Astronomy (AURA) under a cooperative agreement with the National Science Foundation.

This project used data obtained with the Dark Energy Camera (DECam), which was constructed by the Dark Energy Survey (DES) collaboration. Funding for the DES Projects has been provided by the U.S. Department of Energy, the U.S. National Science Foundation, the Ministry of Science and Education of Spain, the Science and Technology Facilities Council of the United Kingdom, the Higher Education Funding Council for England, the National Center for Supercomputing Applications at the University of Illinois at Urbana-Champaign, the Kavli Institute of Cosmological Physics at the University of Chicago, Center for Cosmology and Astro-Particle Physics at the Ohio State University, the Mitchell Institute for Fundamental Physics and Astronomy at Texas A\&M University, Financiadora de Estudos e Projetos, Fundacao Carlos Chagas Filho de Amparo, Financiadora de Estudos e Projetos, Fundacao Carlos Chagas Filho de Amparo a Pesquisa do Estado do Rio de Janeiro, Conselho Nacional de Desenvolvimento Cientifico e Tecnologico and the Ministerio da Ciencia, Tecnologia e Inovacao, the Deutsche Forschungsgemeinschaft and the Collaborating Institutions in the Dark Energy Survey. The Collaborating Institutions are Argonne National Laboratory, the University of California at Santa Cruz, the University of Cambridge, Centro de Investigaciones Energeticas, Medioambientales y Tecnologicas-Madrid, the University of Chicago, University College London, the DES-Brazil Consortium, the University of Edinburgh, the Eidgenossische Technische Hochschule (ETH) Zurich, Fermi National Accelerator Laboratory, the University of Illinois at Urbana-Champaign, the Institut de Ciencies de l’Espai (IEEC/CSIC), the Institut de Fisica d’Altes Energies, Lawrence Berkeley National Laboratory, the Ludwig Maximilians Universitat Munchen and the associated Excellence Cluster Universe, the University of Michigan, NSF’s NOIRLab, the University of Nottingham, the Ohio State University, the University of Pennsylvania, the University of Portsmouth, SLAC National Accelerator Laboratory, Stanford University, the University of Sussex, and Texas A\&M University.

BASS is a key project of the Telescope Access Program (TAP), which has been funded by the National Astronomical Observatories of China, the Chinese Academy of Sciences (the Strategic Priority Research Program “The Emergence of Cosmological Structures” Grant \# XDB09000000), and the Special Fund for Astronomy from the Ministry of Finance. The BASS is also supported by the External Cooperation Program of Chinese Academy of Sciences (Grant \# 114A11KYSB20160057), and Chinese National Natural Science Foundation (Grant \# 11433005).

The Legacy Survey team makes use of data products from the Near-Earth Object Wide-field Infrared Survey Explorer (NEOWISE), which is a project of the Jet Propulsion Laboratory/California Institute of Technology. NEOWISE is funded by the National Aeronautics and Space Administration.

The Legacy Surveys imaging of the DESI footprint is supported by the Director, Office of Science, Office of High Energy Physics of the U.S. Department of Energy under Contract No. DE-AC02-05CH1123, by the National Energy Research Scientific Computing Center, a DOE Office of Science User Facility under the same contract; and by the U.S. National Science Foundation, Division of Astronomical Sciences under Contract No. AST-0950945 to NOAO.}


\bibliography{ref}

\end{document}